\documentclass[letterpaper,12pt,peerreviewca,draftcls]{IEEEtran}
\usepackage{csm16}
\usepackage[margin=1in]{geometry}
\usepackage{amsmath} 
\usepackage{amssymb}
\usepackage{cite}
\usepackage{url}
\usepackage{graphicx,xcolor}
\usepackage{verbatim}
\makeatletter
\newcommand{\verbatimfont}[1]{\def\verbatim@font{#1}}%
\makeatother
\verbatimfont{\ttfamily\small}

\usepackage{tikz}
\usepackage{pgfplots}
\usetikzlibrary{decorations.markings}
\usetikzlibrary{patterns}
\usetikzlibrary{arrows}
\usetikzlibrary{shapes}
\usetikzlibrary{fit}
 \usetikzlibrary{calc}
\usetikzlibrary{decorations.pathreplacing}
\usetikzlibrary{arrows,positioning} 
\usetikzlibrary{pgfplots.groupplots}

\usepackage{algorithm}
\usepackage[noend]{algpseudocode}
\usepackage{tabulary}

\newcommand{\bi}{\begin{itemize}}\newcommand{\ei}{\end{itemize}}
\newcommand{\be}{\begin{equation}}\newcommand{\ee}{\end{equation}}
\newcommand{\bee}{\begin{enumerate}}\newcommand{\eee}{\end{enumerate}}
\newcommand{\bea}{\begin{eqnarray}}\newcommand{\eea}{\end{eqnarray}}
\newcommand{\beas}{\begin{eqnarray*}}\newcommand{\eeas}{\end{eqnarray*}}
\newcommand{\bc}{\begin{center}}\newcommand{\ec}{\end{center}}
\usepackage[left,pagewise]{lineno} 
\usepackage[english]{babel} 
\usepackage{blindtext}

\newcommand{\R}{\mathbb{R}}
\renewcommand{\S}{\mathcal{S}}

\newcommand{\K}{\kappa}

\newcommand{\W}{\mathcal{W}}

\newcommand{\U}{{U}}
\newcommand{\X}{{X}}

\newcommand{\ub}{u_{\rm b}} 
\newcommand{\ud}{u_{\rm des}} 
\newcommand{\ua}{u_{\rm act}} 
\newcommand{\xcand}{x_{\rm cand}} 
\newcommand{\xcur}{x_{\rm curr}} 
\newcommand{\Cb}{\mathcal{C}_{\rm b}} 
\newcommand{\Cs}{\mathcal{C}_{\rm S}} 
\newcommand{\Ca}{\mathcal{C}_{\rm A}} 

\newcommand{\ie}{\emph{i.e. }}
\newcommand{\eg}{\emph{e.g. }}
\newcommand{\Rn}{\mathbb{R}^n}
\newcommand{\Rm}{\mathbb{R}^m}

\makeatletter
\newsavebox{\@brx}
\newcommand{\llangle}[1][]{\savebox{\@brx}{\(\m@th{#1\langle}\)}%
  \mathopen{\copy\@brx\kern-0.5\wd\@brx\usebox{\@brx}}}
\newcommand{\rrangle}[1][]{\savebox{\@brx}{\(\m@th{#1\rangle}\)}%
  \mathclose{\copy\@brx\kern-0.5\wd\@brx\usebox{\@brx}}}
\newcommand{\corn}[1]{\llangle #1 \rrangle}
\makeatother

\DeclareMathOperator*{\argmin}{arg\, min} 

\newcommand{\revised}[1]{\textcolor{black}{#1}}

\newcommand{\rev}[1]{\textcolor{black}{#1}}

\newenvironment{example}{\paragraph{Example}}{\hfill$\Diamond$}

\title{Run Time Assurance \\for Safety-Critical Systems\\
\Large An Introduction to Safety Filtering Approaches \\ for Complex Control Systems}
\author{Kerianne L. Hobbs, Mark L. Mote, Matthew C. L. Abate, \\ Samuel D.  Coogan, and Eric M. Feron \\
	POC: K.\ Hobbs (kerianne.hobbs@us.af.mil)\\ \today \\ Approved for public release: distribution unlimited. Case Number AFRL-2021-2407.}

\newif\ifPDF \ifx\pdfoutput\undefined\PDFfalse \else\ifnum\pdfoutput > 0\PDFtrue \else\PDFfalse \fi \fi
\ifPDF 
\usepackage[pdftex, plainpages = false, colorlinks=true, linkcolor=black, citecolor = green!50!blue, urlcolor = blue, filecolor=black, pagebackref=false, hypertexnames=false,  pdfpagelabels ]{hyperref}
\fi

\begin{document}
\maketitle
\CSMsetup


More than three miles above the Arizona desert, an F-16 student pilot experienced a gravity-induced loss of consciousness (GLOC), passing out while turning at nearly 9Gs (nine times the force of gravity) flying over 400 knots (over 460 miles per hour). With its pilot unconscious, the aircraft turn devolved into a dive, dropping from over 17,000 feet to less than 8,000 feet in altitude in less than 10 seconds. An auditory warning in the cockpit called out to the pilot ``altitude, altitude" just before he crossed through 11,000 feet, switching to a command to ``pull up" around 8,000 feet. Meanwhile, the student's instructor was watching the event unfold from his own aircraft. As the student's aircraft passed through 12,500 feet, the instructor called over the radio ``two recover," commanding the student (``two") to end the dive. As the student's aircraft passed through 11,000 feet the instructor's ``two recover!" came with increased urgency. At 9,000 feet, and with terror rising in his voice the instructor yelled ``TWO RECOVER!" Fortunately, at the same time as the instructor's third panicked radio call, a new Run Time Assurance (RTA) system kicked in to automatically recover the aircraft. The Automatic Ground Collision Avoidance System (Auto GCAS), an RTA system integrated on the jets less than two years earlier in the Fall of 2014, detected that the aircraft was about to collide, commanded a roll to wings level and pull up maneuver, and recovered the aircraft less than 3,000 feet above the ground. The event described here occurred in May 2016.  A video from the event was declassified and publicly released in September 2016 and the footage may be found at \rev{\cite{F16SaveVideo}}. While Auto GCAS monitored the behavior of a safety-critical cyber-physical system with a human providing the primary control functions, the same concept is gaining attention in the autonomy community looking to assure safety while integrating complex and intelligent control system designs.

RTA Systems are online verification mechanisms that filter an unverified \emph{primary} controller output to ensure system safety. The primary control may come from a human operator, an advanced control approach, or an autonomous control approach that cannot be verified to the same level as simpler control systems designs. The critical feature of RTA systems is their ability to alter unsafe control inputs explicitly to assure safety. In many cases, RTA systems can functionally be described as containing a \textit{monitor} that watches the state of the system and output of a \textit{primary controller}, and a \textit{backup controller} that replaces or modifies control input when necessary to assure safety. \rev{Note that RTA and the controllers within the architecture go by many different names, as described in the sidebar titled ``RTA Aliases." RTA designs specifically to bound the behavior of neural network control systems used as the primary controller are discussed in the sidebar titled ``Shielded Learning."} 
An important quality of an RTA system is that the assurance mechanism is constructed in a way that is entirely agnostic to the underlying structure of the primary controller. 
By effectively decoupling the enforcement of safety constraints from performance-related objectives, RTA offers a number of useful advantages over traditional (offline) verification. \rev{Another way to think of RTA is to consider designing controllers so that there is always a plan B, as discussed in a sidebar titled ``The Case for Plan B."}

Verification, validation, assurance, and certification methods present the largest barriers to operational use of autonomous control in safety critical systems, such as passenger aircraft, vehicles, medical devices, and nuclear power plants. For example, the commercial aviation domain requires showing stringent safety constraint satisfaction for any failure that is more likely than ``extremely improbable," defined as an event that occurs with a probability of $10^{-9}$ \cite{circular1988ac}, (more than one in a billion flight hours). While verification, validation, assurance, and certification are related, there are slight variations in their meaning. \rev{Additionally, these are also confused with concepts described in the ``Safety, Reliability, and Security" sidebar.} \textit{Verification} is an activity that determines whether a system meets requirements \cite{DAUverification}, in effect answering the question ``Did we build the system right?" \textit{Validation} assesses whether a system meets the needs of the end user \cite{DAUvalidation}, answering the question ``Did we build the right system?" \textit{Model validation} refers to evaluating how well a model represents reality. \textit{Assurance} is justified confidence that the system functions as intended with limited vulnerability to uncertainty, hazards, and threats based on evidence generated through development activities \cite{Nichols2019Assurance}. \textit{Certification} determines whether a system conforms to a set of criteria or standards for a class of similar systems \cite{FAAAirworthiness,MILHDBK516C,MILSTD882E,ARP4761,ARP4754A,DO-178C,DO-254,DO-333}. Verification, validation, certification and assurance of safety critical dynamical systems requires the development of techniques that incorporate comprehensive analysis based on rigorously developed specifications as well as techniques that continuously evaluate system behavior at run time. 

A significant benefit of RTA is that it alleviates the need to design the primary controller in a way that conforms to traditional safety standards, which may not be directly applicable to the primary control design. 
Practical consequences of this are that RTA provides a means of testing new control algorithms on hardware platforms without compromising safety potential as well as a \textit{near-term certification path} for autonomous controllers and safety-critical human-controlled systems. 
Additionally, the verification of an assurance mechanism is generally simpler than verification of a performance-based controller as it
does not require consideration of the potentially complex performance-related objectives. 
That is, safety verification does not become more difficult as the primary controller grows more complex, and the verification process does not need to be repeated when changes are made to the primary controller. Note that RTA is envisioned to increase safety confidence, not provide a substitute for having some level of confidence in the safety of a primary controller. \rev{Additionally, as described in the sidebar ``To RTA or not to RTA? That is the Question," there are times when RTA is not appropriate.}

RTA is emerging as the dominant approach for enforcing safety in real world autonomous systems and semi-autonomous systems. 
Though not always directly associated with the term \emph{run time assurance}, notable autonomous systems applications are seen in road vehicles \cite{hu2019lane, xu2017correctness}, bipedal walking \cite{gurriet2018towards, harib2018feedback}, air traffic control \cite{Ariadne_Olatunde2021}, fixed wing aircraft collision avoidance \cite{squires2018constructive, squires2019composition, squires2020safety, schouwenaars2005implementation, schouwenaars2006safe} and flight envelope protection \cite{hanson2009capability}, VTOL aircraft \cite{ wang2017safe_quads}\cite{singletary2020safety}, spacecraft collision avoidance \cite{Mote2021Natural}, manipulator arms \cite{singletary2019online}, and propulsion \cite{wong2014towards} to name a few.

The rise in popularity of RTA can be attributed to  many factors, including: (i) the emergence of mobile autonomous systems operating in safety critical environments, \emph{e.g.} away from very specialized and \revised{remote} applications to the vicinity of humans; (ii) the increasing complexity of the primary control system in these systems making traditional verification techniques prohibitively difficult; (iii) the growing desire to quickly update complex system software without compromising safety or repeating a costly verification process; (iv) the ability to perform fast computation online, \emph{e.g.}, integrating trajectories or reachable sets online has only recently become a tractable task for non-trivial systems; (v) the emergence of control barrier function (CBF) and active set invariance filtering (ASIF) methods that give smooth and \emph{minimally invasive} modifications to the desired actions. 
Additionally, RTA seems to provide an attractive solution to bounding the behavior of machine learning and artificial intelligence (AI) systems. 
While some notable efforts have been made in the direction of generating and verifying provably safe deep neural network controllers \cite{katz2017reluplex,tran2020nnv,bak2020improved,gokulanathan2020simplifying}, the complexity and dynamic nature of these systems often precludes the ability to generate rigorous safety guarantees on the primary controller. The RTA paradigm enables an alternate path, without a need to compromise on safety. Reinforcement learning (RL) applications on safety-critical systems often fall into the domain of safe RL, described as a process of policy learning that maximizes expected return while ensuring system performance and/or respecting safety constraints \cite{garcia2015comprehensive}. The safe RL community generally takes one of two approaches: modifying the optimality criterion to reduce risk or using an RTA mechanism during the training process, often referred to as a shield \cite{alshiekh2018safe}.

This article provides a tutorial on developing RTA systems. First a description of the basic architecture,  modeling framework, and fundamental definitions are introduced. Second, categories and properties of RTA systems are identified, and systems engineering and human-machine interaction considerations beyond the system dynamics are discussed. Third, a particular category of RTA systems called the Simplex architecture is described. Fourth, a second category of RTA systems called active set invariance filters is described.  Fifth, implicit and explicit variations of both the Simplex architecture and active set invariance filter RTA approaches are described for the canonical double integrator system. Sixth, approaches for RTA system performance under uncertainty is described. Seventh, verification approaches for RTA algorithms and architectures are presented. Additional topics, applications, definitions, examples, and supplemental information are found in sidebars throughout.





\section{Run Time Assurance for Safety-Critical Cyber-Physical Systems}
Cyber-physical systems feature computing devices that interact with the physical world via actuators and sensors \cite{Alur2015}, and such systems are  \emph{safety-critical} when  failure would result in loss of life, damage to property, or other unacceptable damages such as environmental harm \cite{knight2002safety}.
An RTA architecture acts as an online-verification and enforcement tool for cyber-physical systems that guarantees certain system-level safety requirements are met at run time. In this section, the RTA problem formulation, appropriate RTA employment in the control loop, and other important design considerations are discussed.  


\subsection{The Run Time Assurance Architecture}

\begin{figure}[] 
    \centering
    \includegraphics[width=1\textwidth]{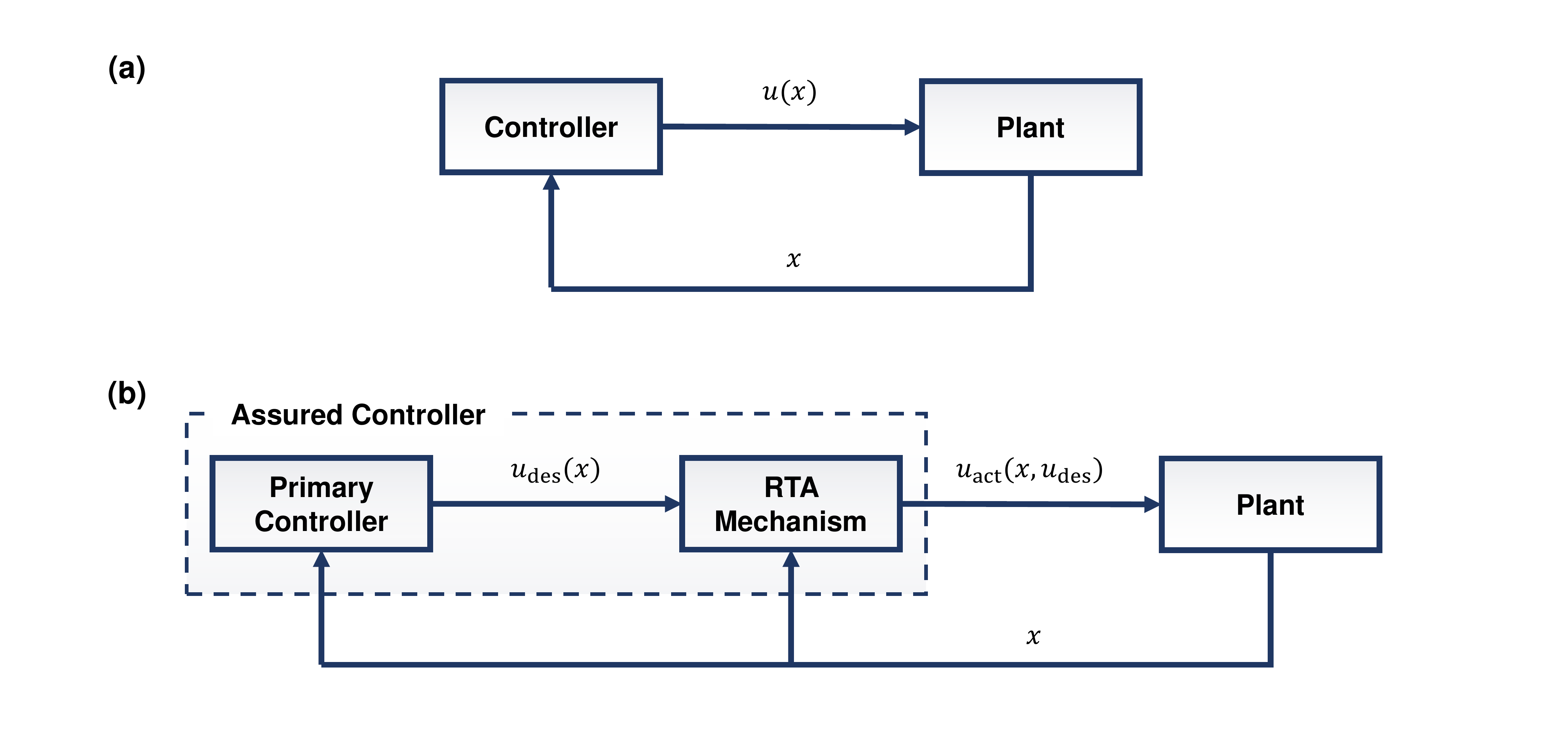}
    \caption{\rev{To illustrate where run time assurance sits within the control system architecture, }(a) \rev{depicts a p}rototypical feedback control system architecture, \rev{and} (b) shows the insertion of a run time assurance mechanism. \rev{The primary controller (e.g. human operator, an advanced control approach, or an autonomous control approach) is performance-driven and outputs a desired input $\ud$. The run time assurance mechanism maps the state $x\in\R^n$ and a desired input $\ud\in\R^m$ to an actual input $\ua\in\R^m$, with the objectives of (i) enforcing state constraints, and (ii) outputing $\ua$ as {close} as possible to $\ud$. Run time assurance can be thought of as a safety filter on the performance controller output.}}
    \label{fig: RTA_architecture}
\end{figure}

RTA presents an approach to control design that allows a designer to sidestep the common trade-off between performance and safety.   
The central idea behind RTA is to {decouple} the task of enforcing safety constraints from all other objectives of the controller. 
This is achieved by splitting the control system into two components: a performance-driven \emph{primary} controller, and a safety-driven \emph{RTA mechanism}. 
The RTA mechanism preempts the desired inputs from the primary controller when necessary for ensuring the safety of the system, and otherwise lets the input pass through unaltered. 
This approach is associated with the feedback control architecture shown in Figure \ref{fig: RTA_architecture}b. 
The output of the performance-driven controller is referred to as the \emph{desired} input and assigned the variable $\ud$, and the output of the RTA mechanism is referred to as the \emph{actual} input (or assured input) and is assigned the variable $\ua$. 
At its core, an RTA mechanism is a mapping from a state $x\in\R^n$ and a desired input $\ud\in\R^m$ to an actual input $\ua\in\R^m$, and has the objectives of (i) enforcing safety of the system as defined by state constraints, and (ii) having $\ua$ as {close} as possible to $\ud$,
 so long as it does not conflict with the first objective. A common choice for measuring how \emph{close} the two signals are is the norm of the difference between the two variables, however, this measurement may also take other forms, such as the integral of the deviation. 
An RTA mechanism is said to be \emph{assured} when the output $\ua$ is always verifiably safe, as defined in later sections. Moreover, an assured RTA mechanism guarantees safety for the entire system, and for all time, and this is true regardless of the chosen primary controller. \rev{Although not considered an RTA architecture under the definitions in this paper, a related concept for an alternative safety architecture is described in the sidebar titled ``Reference and Command Governors."}




\subsection{Modelling Cyber-Physical Systems }


Mathematical models that provide an abstraction of real world system behavior may be defined at varying levels of complexity based on the contrasting needs of accurately predicting system behavior and conducting mathematical analysis. This article focuses mainly on cyber-physical systems modeled as \emph{dynamical systems}; hereafter the variable $x\in \X\subseteq\Rn$  denotes the state vector of the system model and the variable $u\in \U \subset \Rm$  denotes a bounded control input, where $\X$ is the set of all possible system states and the set $\U$ is the \emph{admissible} set of controls, determined \emph{a priori} by the actuation constraints of the real-world system.  \emph{Continuous-time} system models appear as systems of ordinary differential equations as in 
\begin{equation} \label{eq:matt}
    \dot{x} = F_{\rm c}(x, u),
\end{equation}
and \emph{discrete-time} system models appear as state-update maps as in 
\begin{equation}\label{eq:mark}
    {x}^+ = F_{\rm d}(x, u). 
\end{equation}
While the computational elements responsible for the control of a CPS live in a discrete world, the physical laws of motion tend to take the form of ordinary differential equations. 
In many cases, systems of the form \eqref{eq:mark} are obtained from discretizing equations of the form \eqref{eq:matt} over the controller update period. 

A natural generalization is to consider \emph{hybrid} systems that combine continuous-time and discrete-time elements \cite{goebel2012hybrid, tabuada2009verification}. In fact, as seen below, some RTA mechanisms are naturally interpreted as inducing a hybrid system in closed-loop, even when the open-loop system is of the form \eqref{eq:matt}. For the purposes of this article, attention is restricted to to systems with dynamics governed by \eqref{eq:matt} or \eqref{eq:mark} and will not formally characterize hybrid systems, but we note instances where RTA tools have been specifically extended to such systems in the literature.

Throughout this article, it is discussed how one may verify or assure system models of the form \eqref{eq:matt} or \eqref{eq:mark} against safety constraints.
However, \eqref{eq:matt} and \eqref{eq:mark} are only models, and 
real-world systems generally deviate from their models. When this deviation is significant, 
safety guarantees derived from the models are invalid. One method for addressing this limitation is to instead consider a \emph{nondeterministic model} that explicitly considers uncertainties that account for the deviation between the real-world system and the \emph{deterministic models} presented in \eqref{eq:matt} or \eqref{eq:mark}.  This article first focuses on the deterministic system models \eqref{eq:matt} and \eqref{eq:mark} and discusses extensions to nondeterministic models in the Section ``Assurance in the Presence of Uncertainty."  

\subsection{Admissible Control of Cyber-Physical Systems}
The main focus of this article is in devising \emph{safe} controllers for cyber-physical systems modeled as in \eqref{eq:matt} or \eqref{eq:mark}. Safety control schemes must conform to the boundaries of admissible control. That is, a control law can only be realized on a physical system if its outputs are bounded to a set of signals that respect the actuation constraints of that system.
A feedback \emph{control law} $u$ denotes a mapping to $\Rm$ from either the state domain or an augmentation of the state domain. 
For instance, in the context of RTA, many primary control laws will take the form $u: X \to \Rm$, while RTA control laws will generally take the form $u:\X\times\R^m\to\R^m$, and either can be extended to include explicit dependence on time. 
%
Control laws are said to be \emph{admissible}
when their range is the admissible input set $U$, 
\eg $u:X\to U$ or $u:X\times\R^m\to U$.

Any control law $u$ can be made admissible via composition with a saturation (also called clamping) function; a \emph{saturation function} $\sigma:\Rm \to U$ is such that $\sigma(u)$ is approximately equal to $u$ when $u$ is in the interior of $\U$ and $\sigma(u)$ is approximately the projection of $u$ onto the boundary of $\U$ when $u$ is outside of $\U$.  Here, there are two main approaches: (i) \emph{hard clamping}, whereby the saturation function $\sigma(u)=u$ for all $u\in\U$ 
and $\sigma(u)$ is a projection onto the boundary $\partial U$, and (ii) \emph{smooth clamping}, whereby $\sigma$ takes the form of a differentiable function that such that $\sigma(u)\approx u$ for $u\in \U$.  For example, for the case of scalar input and $U=[-1,1]\subset \mathbb{R}$, then $\sigma_1(u)=\max(-1,\min(1,u))$ is a hard clamping  function while  $\sigma_2({u}) = \rm{tanh}({u})$, $\sigma_3({u}) = \frac{2}{\pi}{\rm{tan}(\frac{\pi}{2}{u})}$,  $\sigma_4({u})={u}(1+{u}^2)^{-1/2}$, and $\sigma_5({u}) = {\rm tanh}({u}+ 0.5 {u}^3)$ are examples of smooth clamping functions.




\subsection{Safety and Invariance of Cyber-Physical Systems}

In a colloquial context, safety means freedom from harm or danger. For safety critical systems, safety is freedom from conditions that would, in a worst-case environment, lead to an unacceptable loss, such as a loss of control, physical damage to the system under control, loss of human life, human injury, property damage, environmental pollution, or failure of the mission \cite{leveson2011engineering,schierman2015runtime}. 
\revised{In the context of a dynamical system, safety is a characterization of the state of the system and its evolution.} 
While alternate approaches to forming safety specifications exist (\eg temporal logic), this research focuses specifically on set invariance requirements derived from static properties on the state.
Specifically, safety properties are specified with state constraints, and safety relates to whether a particular initial condition will lead to the safety constraints being satisfied for all time. 

In this article, \emph{safety constraints} are defined with
inequality constraints on the state. In particular, we consider  $\varphi_i:\X \to\mathbb{R}$ for $i\in \{1, \dots, M\}$, where $M$ is the number of safety constraints and it is required that always $\varphi_i(x) \geq 0$ for each $i$.  
The set of states that satisfy all of the safety constraints is referred to as the \emph{constraint set} (sometimes called the  \emph{allowable set} or the constraint space) and is given by 
\begin{equation}\label{eq: allowable set}
    \Ca:=\{x\in \X \, | \, \varphi_i(x) \geq 0, i\in\{1,\dots, M\} \}.
\end{equation}

\begin{example}
    Consider a wingman aircraft with position $(x_{\rm W}, y_{\rm W})$ flying co-altitude in formation with a lead aircraft with position $(x_{\rm L}, y_{\rm L})$ and the requirement is that the two aircraft never go within 30 meters of each other. A constraint set over the states $x=[ x_{\rm L},  y_{\rm L},\dot{x}_{\rm L},\dot{y}_{\rm L}, x_{\rm W}, y_{\rm W},\dot{x}_{\rm W}, \dot{y}_{\rm W} ]^T$ is given by: 
\begin{equation}
    \Ca = \{{x}\in \mathbb{R}^8 \,|\, \sqrt{(x_{\rm L}-x_{\rm W})^2+(y_{\rm L}-y_{\rm W})^2}-30\geq0\}.
\end{equation} 
\end{example} 

Importantly, there may exist states in $\Ca$ that obey the safety constraints at a given time, but that inevitably lead to violations in the future. 
For example, in the aircraft case it is possible for the wingman aircraft to have just more than 30 m separation with the lead aircraft initially, but be traveling too fast to  
be able turn, climb, or descend to avoid a collision. 
Though the state falls in the constraint set $\Ca$ at a given time, it is not ``safe"  as it will inevitably lead to a departure from the set. 
A meaningful definition of safety must encode additional information relating to whether the safety constraints will continue to be satisfied for all time with a particular control law and subject to a particular set of dynamics and actuation constraints.


 

Informally, a system is safe if the state \revised{belongs to} $\Ca$ for all time. 
This concept is formalized with the notion of \emph{set invariance}:
A state $x(t_0)$ is safe with respect to a closed-loop dynamical system if that state lies in a forward invariant subset of the constraint set, \ie if: 
\begin{equation}
    x(t_0)\in\Cs \,\, \implies  \,\, x(t)\in\Cs \,\,\,\,  \  \forall \, t\geq t_0
\end{equation}
where  $\mathcal{C}_{\rm S}\subseteq\mathcal{C}_{\rm A}$.
In this case, $\Cs$ is said to be a \emph{safe set}.
Furthermore, 
control laws are said to be safe when they render some nonempty subset of $\Ca$ forward invariant.
It is generally desirable that a control law be designed such that the safe set is as large as possible. 
However, it is generally not possible to find an admissible control law that will render $\Ca$ itself forward invariant.
The distinction between $\Cs$ and $\Ca$ is necessary as the system is subject to actuation and dynamics constraints, meaning that the control signal only has a limited amount of influence over the evolution of the state.

\begin{figure}[htb]
\begin{center}
\includegraphics[width=.97\textwidth]{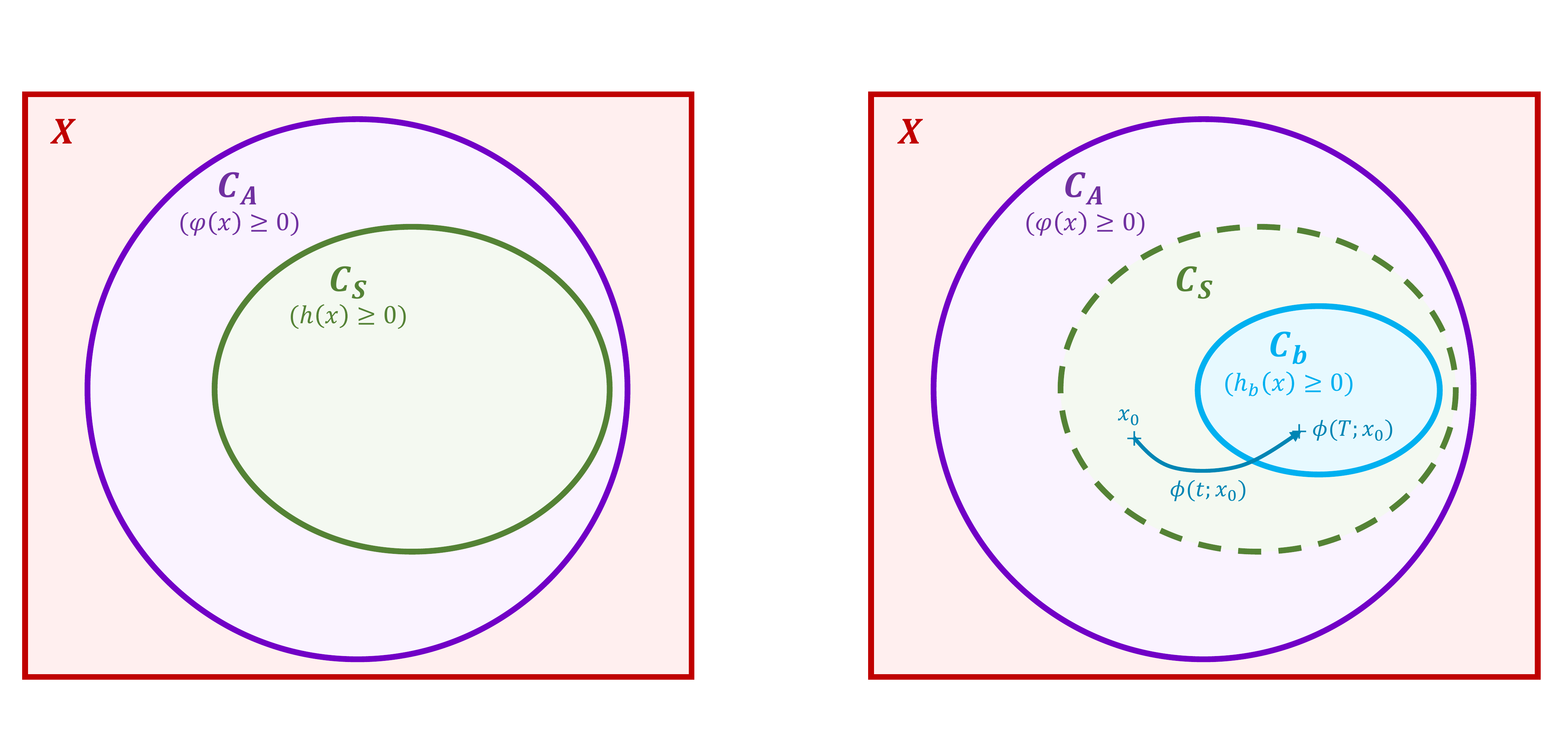}
\caption{Safe set $\Cs$ and constraint set $\Ca$ topology for explicit (left) and implicit (right) cases. \rev{Here $X$ represents the set of all states, the constraint set $\Ca$ is the set of states considered safe when there are no controller limitations, and the safe set $\Cs$ is a subset representing the set of states that are safe under the limits of the control system. For example, a car approaching a wall is in the constraint set $\Ca$ so long as it does not hit the wall. However, if the car is two feet from the wall and moving at 100 miles per hour, the it can be in the constraint set $\Ca$ but not in the safe set $\Cs$. To be in $\Cs$, the car must be able to stop before hitting the wall. In cases where it is hard to define $\Cs$, but a predefined backup control is known (e.g. car slamming on the brakes), an implicit backup set $\Cb$ may be used instead of $\Cs$.}} 
\label{fig: ImplicitExplicit_Set_Topology} 
\end{center}
\end{figure}

It is important to note that forward invariance, and by extension safety, are properties of the closed-loop system and are not defined in absence of a controller. The existence, size, and shape of $\Cs$ are all dependent on the controller.
The question of whether it is possible to find a control law that will render a particular set invariant is addressed with the concept of control invariance. 
A set $\mathcal{S}\subseteq \X$ is said to be \emph{control invariant} (also called viable) if there exists an admissible control law 
that renders $\mathcal{S}$ forward invariant. 
The largest control invariant subset of $\Ca$ is known as the \emph{viability kernel} \cite{aubin2011viability}. 
Intuitively, the viability kernel is the largest achievable safe set, and it forms the boundary between the states for which it is and is not possible to find a control input that will keep the system safe for all time. 
A fact that will become important in later sections is that control invariant sets are forward invariant under controllers that take the form of certain optimization-based procedures. 

\subsubsection{Identifying Safe Sets}

Computing  invariant sets has been the subject of a wide variety of research. 
A common approach to identifying an invariant region is to find a Lyapunov function. Since stability is a stronger condition than invariance, any stable region under a particular controller must also be invariant under that controller.
Moreover, invariance of a set under an admissible control law $u:\X\to\U$, also implies control invariance under $U$.  
Invariant regions may also be computed by identifying barrier certificates \cite{prajna2004_safety_verification_of_hydrid}. 
While identifying Lyapunov functions or barrier certificates is typically a difficult task, they may be computed automatically for polynomial dynamics using sum-of-squares (SOS) optimization \cite{anderson2015advances, xu2017correctness, prajna2007framework} at moderate to high computational costs. 
Other approaches include using discretized solutions the the Hamilton-Jacobi equations \cite{mitchell2005time}, and sampling \cite{gillula2014sampling}. 
Safe sets may also be computed for discrete time systems using approaches in computational geometry \cite{bertsekas1971minimax}. These approaches typically rely on convexity of the set $\Ca$ and linearity of the dynamics for the purposes of implementing set operations. See \cite{gurriet2020applied, mitchell2015summary, borrelli2017predictive} for a more extensive overview of techniques for invariant set computation.

\begin{example}
    Consider the double integrator system---\eg a spacecraft in linear motion--- described by
\begin{equation}
    \begin{split}
    \label{eq: double_integrator_syst}
    \dot{x}_1&=x_2\\
    \dot{x}_2&=u
\end{split}
\end{equation}
where the state vector $x=[x_1,x_2]^T\in\mathbb{R}^2$ is composed of a position state $x_1$, and a velocity state $x_2$, and $u\in[-1,1]$ is the acceleration due to thrust.
The safety constraint 
    $\varphi(x)= -x_1 \geq 0$
is imposed on the system, reflecting the requirement that the system avoid collision with an object located at $x_1=0$\revised{, and extending to $x_1\geq 0$}. The constraint set is \begin{equation} \label{eq: double_integrator_Ca}
    \Ca=\{x \in\mathbb{R}^2 \, | \, -x_1 \geq 0 \}.
\end{equation}
The largest control invariant subset (\ie the viability kernel) of $\mathcal{C}_{\rm A}$ is $\mathcal{C}_{\rm S} = \{ x\in\mathbb{R}^2 \, | \, h(x)\geq 0\} $  where 
\begin{equation}\label{e:doubleintegratorviability}
\begin{split}
    h(x) &=
    \begin{cases}
    -2 x_1 - x_2^2  & \text{if}\quad x_2 > 0  \\
    -x_1 &\text{if}\quad x_2\leq0 .\\ 
    \end{cases} \\
\end{split}
\end{equation} 
The unsafe states in the constraint set $\mathcal{C}_{\rm A}\setminus\mathcal{C}_{\rm S}$ represent states for which a future collision is inevitable. 
Intuitively, $x_2=\sqrt{-2x_1}$ represents the maximum safe velocity at $x_1$. 
If this approach speed is exceeded, then there will not be sufficient distance to stop before a collision occurs. 
This result is made apparent by considering the flow of the system under a recovery maneuver $u=-1$, which applies maximum thrust force away from the obstacle. 
It can be seen that $\mathcal{C}_{\rm S}$ is a forward invariant set under this control law.
Figure \ref{fig: double_integrator_sets} shows a depiction of the described sets, and the flow under various inputs. 

\begin{figure}[htb]
\begin{center}
\includegraphics[width=0.5\textwidth]{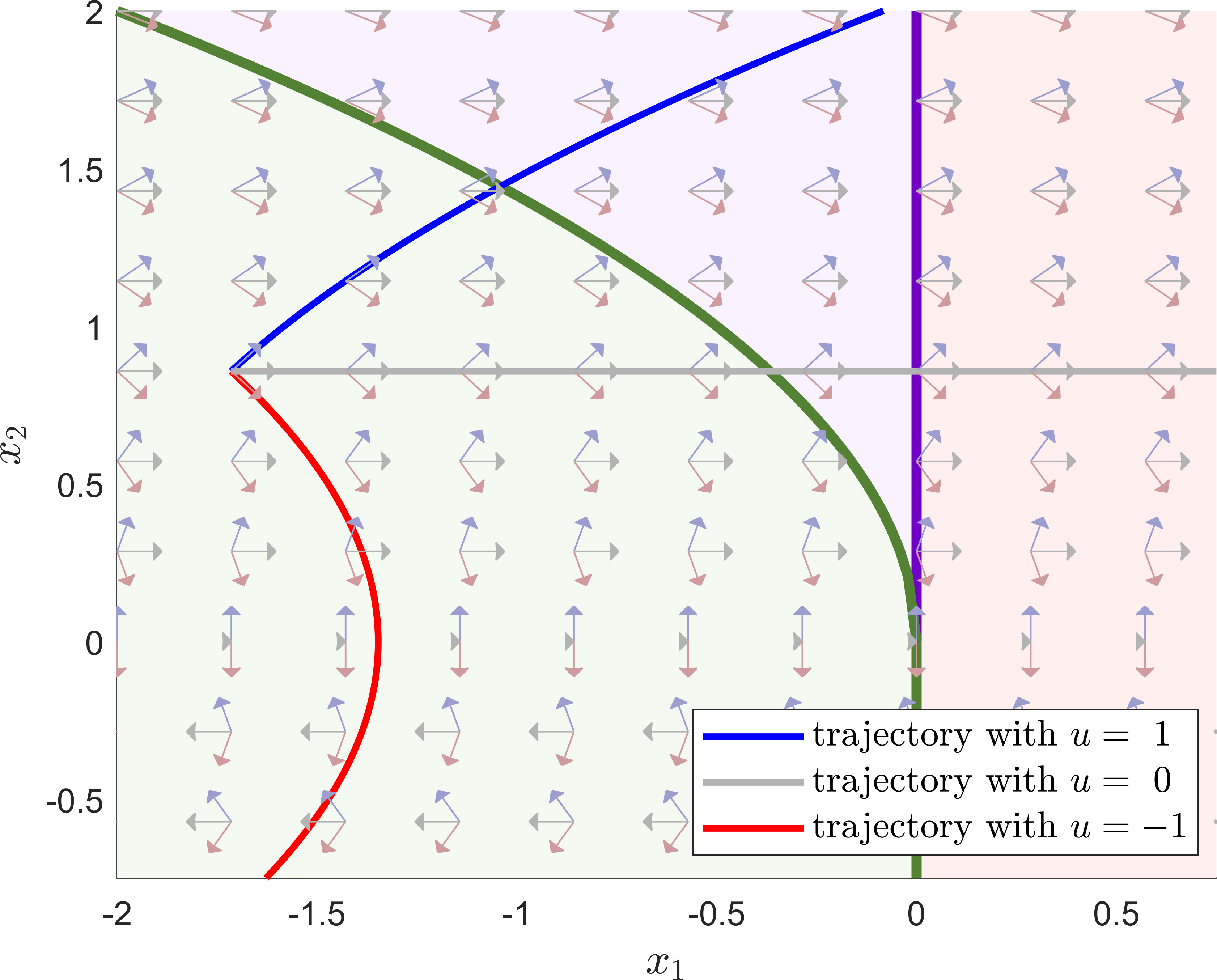}
\caption{\rev{A car moving straight towards a wall can be modeled as a double integrator system.} \rev{This shows a p}hase plot for double integrator system with $u=-1,0,1$\rev{, where $x_1$ is position, $x_2$ is velocity, $u=1$ represents full forward acceleration, $u=0$ represents continuing at a constant velocity, and $u = -1$, represents full acceleration away from the wall}. The viability kernel \rev{$\Cs$} is shaded green \rev{and represents the safe states where it is possible to apply full negative acceleration to avoid hitting the wall.} Collision states are shaded red \rev{ and represent states in which the car has hit or driven past the wall $X\setminus\mathcal{C}_{\rm A}$.} \rev{T}he \emph{inevitable} collision states \rev{$\mathcal{C}_{\rm A}\setminus\mathcal{C}_{\rm S}$} are shaded purple \rev{and represent cases where the car has not yet hit the wall, but is traveling too fast towards the wall to be able to stop in time}.} 
\label{fig: double_integrator_sets} 
\end{center}
\end{figure}
\end{example}

 
 \subsubsection{Implicit and Explicit Definitions of the Safe Set}

Given a constraint set $\Ca$ and an admissible control law $u:\X\to\U$, the question arises of how one can determine if a given state $x\in\X$ is safe. 
In many cases, the set of safe states $\Cs$ can be identified \emph{explicitly} with a functional representation; 
\emph{e.g.}
\begin{equation} \label{eq: explicitly_defined_safe_set}
    \mathcal{C}_{\rm S} = \{x\in\X \, | \, h(x)\geq 0 \},
\end{equation} 
with $h:\X\to\mathbb{R}$.
Checking whether $x\in\Cs$ is then equivalent to checking whether $h(x)\geq0$.
The safe set boundary may be determined in a number of ways, such as through Lyapunov arguments or reachability analysis. 
Additionally, for continuous-time systems as in \eqref{eq:matt}, Nagumo's theorem states that, under appropriate regularity assumptions on $f$, $u$, and $h$, a necessary and sufficient condition for invariance of a candidate set $\Cs$ as in \eqref{eq: explicitly_defined_safe_set} is  $\frac{\partial h}{\partial x}(x)^T F_c(x,u(x))\geq 0$ for all $x$ such that $h(x)=0$ \cite{nagumo1942lage}, and thus 
Nagumo's theorem can be used to verify that a candidate safe set is invariant. 
Moreover, as discussed in the Section ``Assurance in the Presence of Uncertainty," it is possible to consider nondeterministic effects through the identification of sets that are \emph{robustly} forward invariant, that is, that are forward invariant under any realization of the nondeterminism. 

\begin{figure*}[t!]
    \centering
    \begin{subfigure}[b]{0.5\textwidth}
\begin{center}
\includegraphics[width=0.97\textwidth]{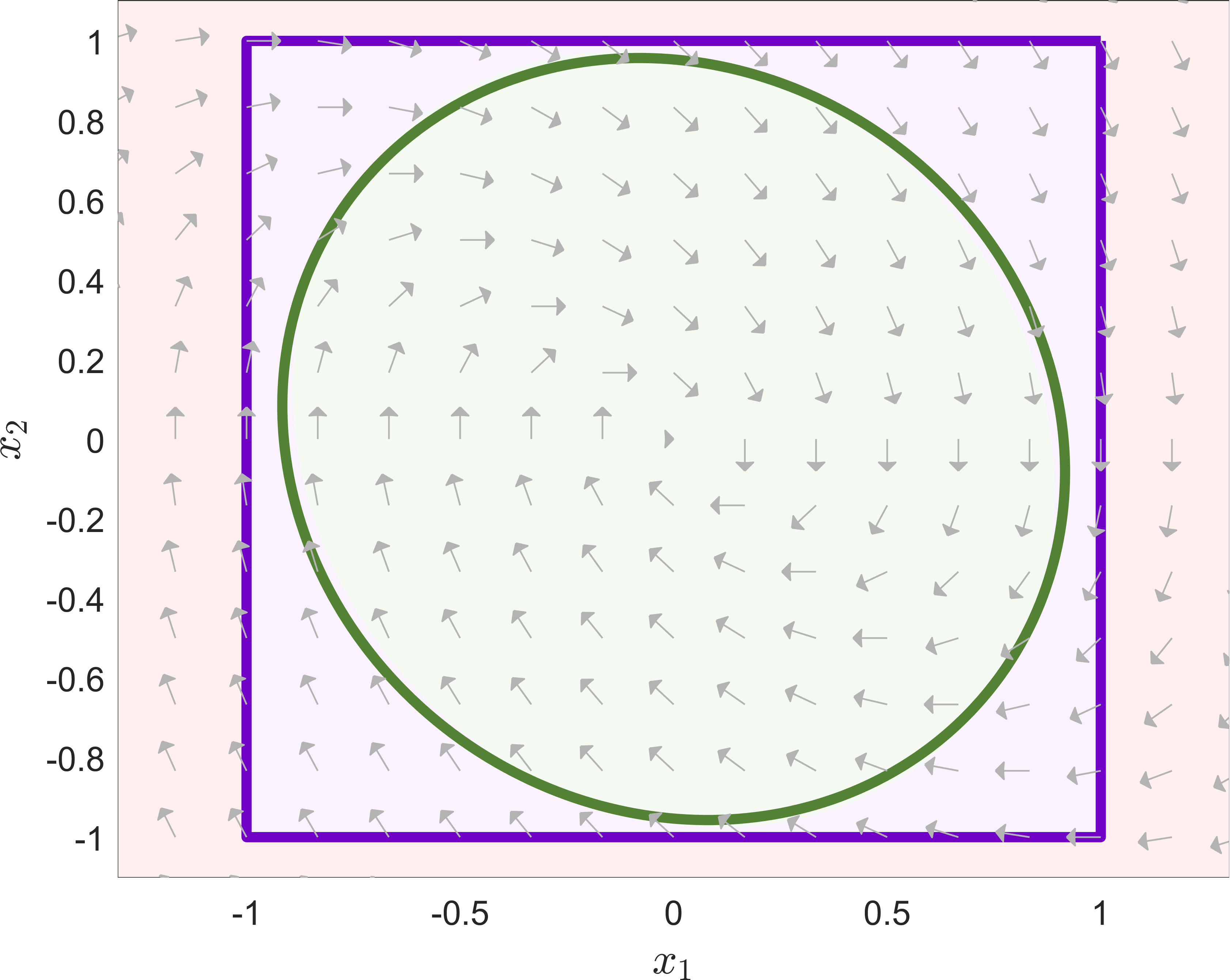}
\caption{Explicit safe set $\Cs$ (green)} 
\label{fig: mass_spring_damper_explicit} 
\end{center}
    \end{subfigure}%
    ~ 
    \begin{subfigure}[b]{0.5\textwidth}
\begin{center}
\includegraphics[width=0.97\textwidth]{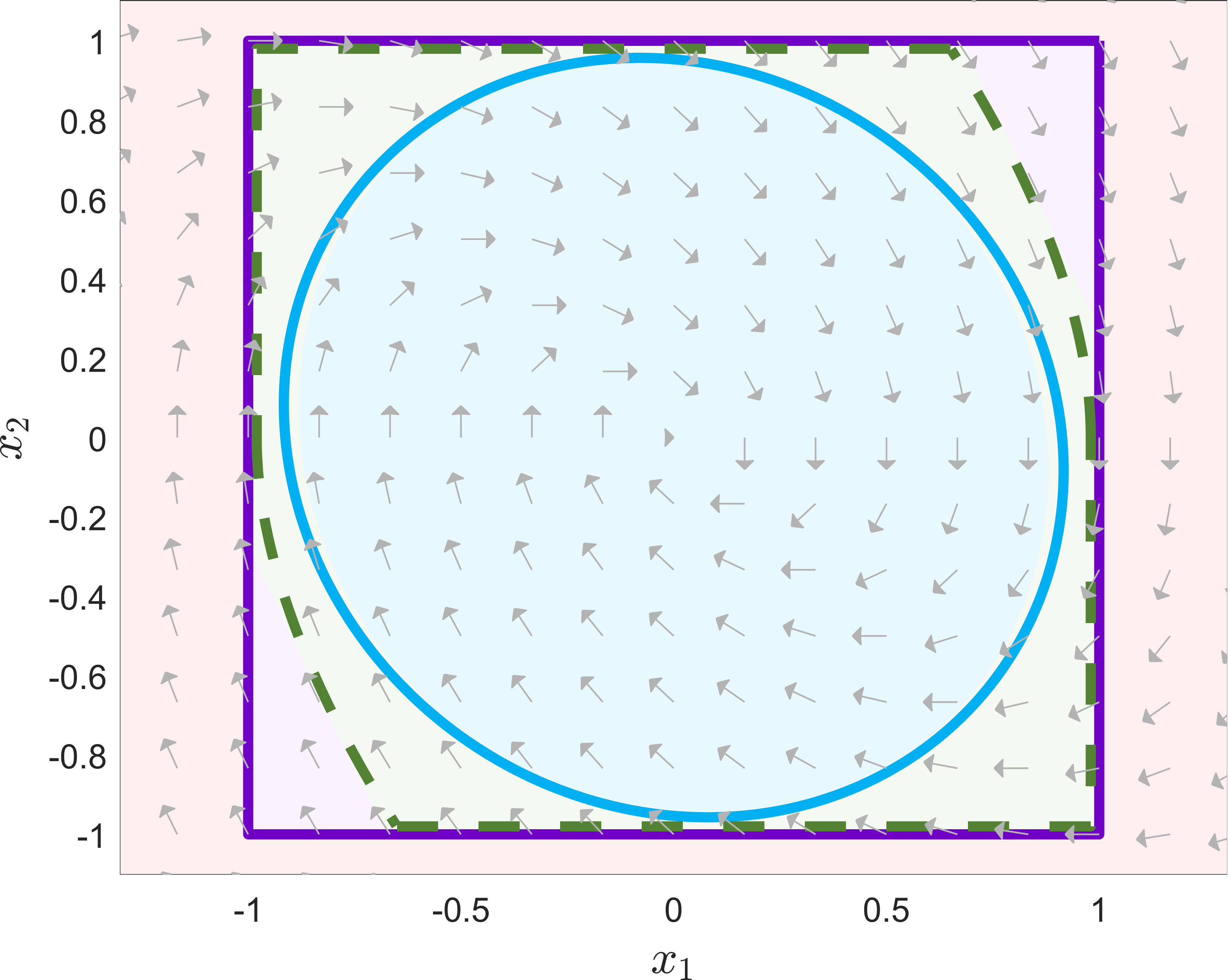}
\caption{Implicit safe set $\Cs$ (green) and backup set $\Cb$ (blue)} 
\label{fig: mass_spring_damper_implicit} 
\end{center}
    \end{subfigure}
    \caption{Phase plots for \rev{a mass-spring-damper} system \eqref{eq: mass_spring_damper}, \rev{modeled as a double integrator with the constraint set $\Ca = \{x\in\mathbb{R}^2 \, | \, 1-\Vert x \Vert_\infty \geq 0\}$, where} with the complement of the constraint space $\Ca$ \rev{is} shaded red, and $\Ca\backslash\Cs$ \rev{is} shaded purple.}
    \label{fig: mass_spring_damper_set_comparison}
\end{figure*}

\begin{example} \label{example: explicit_mass_spring_damper}
    Consider the double integrator system \eqref{eq: double_integrator_syst}, now paired with the constraint set $\Ca = \{x\in\mathbb{R}^2 \, | \, 1-\Vert x \Vert_\infty \geq 0\}$.
    Suppose that a control law is constructed such that for all $x\in\Ca$,  $u=-x_1-x_2$. 
    Note that actuation constraints can be ignored in $\Ca$ if a hard clamping function is used, and that for all $x\in\Ca$,
    $u(x)\in \U$. 
    In this case, 
    the closed-loop dynamics are
    \begin{equation} \label{eq: mass_spring_damper}
        \dot{x} = \begin{bmatrix}
        0 & \,1 \\ -1 & -1
    \end{bmatrix} x,
    \end{equation}
    for all states in $\Ca$.
    A valid safe set can be constructed by identifying a region of attraction contained in $\Ca$. For instance, note that 
    \begin{equation} \label{eq: mass_spring_lyapunov}
    V(x)=1.2x_1^2 + 0.2x_1x_2 + 1.1x_2^2
    \end{equation}
    is a Lyapunov
    function for the closed-loop system and the level curve $V(x)=1$ is contained in $\Ca$. Thus, 
    $\Cs=\{x\in\mathbb{R}^2\,|\, h(x)\geq 0 \}$, where $h(x)=1-V(x)$
    defines an invariant set contained in $\Ca$. 
    The sets $\Cs$ and $\Ca$ are depicted along with the system flow in Figure~\ref{fig: mass_spring_damper_explicit}. Note that while $\Cs$ is invariant, it is not the largest forward invariant set contained in $\Ca$. It is said to be conservative.
\end{example}


Explicit identification of forward invariant subsets is typically obtained only at the expense of conservatism. 
In particular, the methods used to identify forward invariant sets are not generally scalable to  complex and high-dimensional systems, and the safe sets obtained with these methods may greatly underapproximate the largest safe set obtainable.
However, an explicit (functional) representation of the safe set boundary is not always necessary. 
One may \emph{implicitly} define $\Cs$ in terms of the closed-loop trajectories under the control law. For example, consider a \emph{backup} control law $\ub:\X\to U$, and 
let $\phi^{\ub}(t;x)$ represent the state reached after starting at $x\in\X$ and applying $\ub$ for $t$ units of time. Then the set
\begin{equation} 
    \mathcal{C}_{\rm S} = \{x\in\X \, | \, \forall t\geq 0,\,\, \phi^{\ub}(t;x)\in\mathcal{C}_{\rm A} \} 
\end{equation}
is by definition an invariant set under $\ub$, and it is entirely  contained in $\Ca$. 
The utility of this definition arises from the observation that it is possible to check whether \emph{individual states} are safe simply by integrating the dynamics forward from those states. 
For example, if $\Ca=\{x\in\X\,|\,\varphi(x)\geq 0\}$  
then the following are equivalent: 
\begin{equation}
    (i)\,\, \forall t\geq 0, \,\, \phi^{\ub}(t;x)\in\mathcal{C}_{\rm A}, \qquad (ii)\,\, \inf_{t\in [0,\infty)} \varphi(\phi^{\ub}(t;x))\geq 0
\end{equation}
It is interesting to note that in special cases where the infimum can be solved in closed form, the solution can be used to define an explicit safe set, as shown in \revised{(\ref{e:doubleintegratorviability}) and in }the following example. 

\begin{figure}[htb]
\begin{center}
\includegraphics[width=0.5\textwidth]{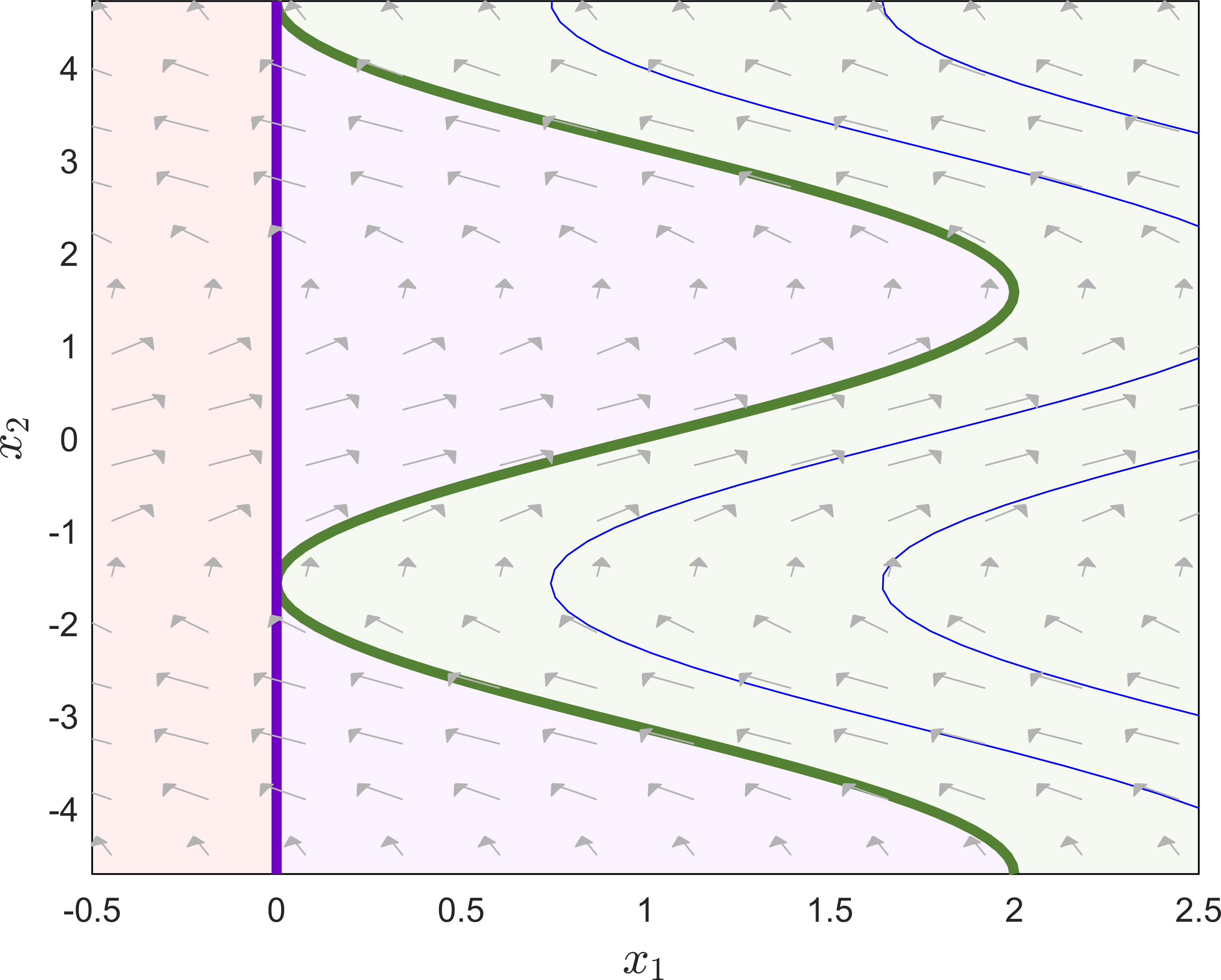}
\caption{Phase plot for system \eqref{eq: constant_speed_unicycle_dynamics} under $\ub=1$. The safe set $\Cs$ is shaded green, the complement of $\Ca$ is shaded red, and $\Ca\backslash\Cs$ is shaded purple.} 
\label{fig: unicycle_sets} 
\end{center}
\end{figure}

\begin{example}
    Consider the system, 
    \begin{equation} \label{eq: constant_speed_unicycle_dynamics}
        \begin{bmatrix}
        \dot{x}_1 \\ \dot{x}_2 
        \end{bmatrix}
        = 
        \begin{bmatrix}
        \cos(x_2) \\ u
        \end{bmatrix}
    \end{equation}
    with $u\in[-1,1]$, the constraint set $\Ca = \{x\in\R^2 \, | \, x_1\geq 0 \}$ and the admissible control law $\ub=1$.  
    The solution for the flow from initial state $x_0=[x_{0,1},x_{0,2}]^T$ is given by
    \begin{equation}
        \begin{bmatrix}
        \phi^{\ub}_1(t;x_0) \\ 
        \phi^{\ub}_2(t;x_0)
        \end{bmatrix}
        =
        \begin{bmatrix}
        x_{0,1}-\sin(x_{0,2})+\sin(x_{0,2}+t) \\ 
        t+x_{0,2}
        \end{bmatrix}
    \end{equation}
    and $x_0$ is safe if $\inf_{t\in [0,\infty)} \phi^{\ub}_1(x_0,t)  = x_{0,1}-\sin(x_{0,2}) - 1 \geq 0$. 
    Using this information, a safe set under this maneuver can be defined explicitly as
    \begin{equation}
        \Cs=\{x\in\R^2 \, | \,  x_{1}-\sin(x_{2}) - 1 \geq 0 \}.
    \end{equation}
    The sets $\Ca, \, \Cs$ are depicted in Figure~\ref{fig: unicycle_sets} along with the flow of the system under $\ub=1$. 
\end{example}

Importantly, the system flow $\phi^{\ub}(t;x)$ can readily be evaluated over a \emph{finite} time horizon $\mathcal{T}= [0,T]$ by numerically integrating the dynamics. 
In this case, safety of a trajectory can be shown by ensuring that (i) the trajectory lies in $\Ca$ over $\mathcal{T}$ and (ii) the endpoint of the finite trajectory $T$ lies in a known invariant subset $\Cb\subseteq \Ca$. 
That is, 
\begin{equation} \label{eq: Safe_Backward_Image}
    \Cs = \{x\in\X \, | \, \forall t\in\mathcal{T},\,\, \phi^{\ub}(t;x)\in\mathcal{C}_{\rm A}\, \wedge \, \phi^{\ub}(T;x)\in\mathcal{C}_{\rm b} \}.
\end{equation}
In this context, a safe terminal set $\mathcal{C}_{\rm b}\subseteq\Ca$ is referred to as a \emph{backup set}, and the set \eqref{eq: Safe_Backward_Image} is the \emph{safe backward image} (SBI) of $\Cb$.
Note that numerical integration over a continuous interval $[0,T]$ requires that the trajectory be approximated with a finite amount of points sampled along this interval; \eg $\mathcal{T}=\{t_1,...,t_K\}\subset [0,T]$. 
Testing with the discrete set of points does not strictly prove invariance as it does not ensure that the trajectory will not leave between the sampled points. While the approximation is sufficient for most practical purposes, this may be addressed with a finer time discretization or by deriving bounds for an inner approximation of the constraint set as in \cite{gurriet2020scalable}.

While a single case is presented here, various forms of implicit safety may arise under modified assumptions. For instance, \cite{gurriet2019scalable} considers the case where $\Cb$ is not forward invariant, and guarantees are  in finite time.
As discussed in the Section ``Assurance in the Presence of Uncertainty," this concept may be extended to nondeterministic systems by approximating the forward reachable sets or invariant tubes around the recovery trajectories.
The topology of the constraint and safe sets for both implicit and explicit cases is illustrated in Figure~\ref{fig: ImplicitExplicit_Set_Topology}.

\begin{example}
    Consider the system \eqref{eq: mass_spring_damper} along with the associated constraint set $\Ca=\{x\in\R^2\,|\,\varphi(x)\geq 0\}$ with $\varphi(x)=1-\Vert x \Vert_\infty$. We now consider the safe set from the previous example as the backup set. That is, consider the backup set $\Cb = \{x\in\R^2\,|\, h_{\rm b}(x) \geq 0 \,\} $ with $h_{\rm b}(x)=1-V(x)$ and  $V(x)$ described by \eqref{eq: mass_spring_lyapunov}. 
    The flow for this system can be evaluated exactly as 
    \begin{equation}
        \phi^{\ub}(t;x)= e^{A t}x
    \end{equation}
where $A$ is the system matrix from \eqref{eq: mass_spring_damper}. 
Consider the backup time horizon $\mathcal{T}=[0,T]$. A state $x\in\R^2$ is in the safe backward image if the following conditions are met
\begin{equation}
    (i)\,\, \forall t\in\mathcal{T},\quad \varphi( e^{A t}x ) \geq0, \qquad (ii)\,\, h_{\rm b}( e^{A T}x)\geq 0. 
\end{equation}
The sets for this case are depicted in Figure~\ref{fig: mass_spring_damper_implicit}. 
\end{example}


\subsubsection{Identifying Safe Backup Sets}

Backup sets $\Cb$ are safe terminal sets that act as \emph{seeds} of safety. As seen in Figure~\ref{fig: mass_spring_damper_set_comparison} the backup set may be used to implicitly identify a  larger safe region via trajectories under a backup controller. 
This is useful because small invariant sets are generally much easier to find than large ones.
Furthermore, a finite-time trajectory can be used to show safety over an infinite horizon when it is shown that it the trajectory safely reaches such a set. 
This concept has been frequently explored in the context of trajectory optimization and \emph{Model-Predictive Control} (MPC) \cite{mayne2013apologia, schouwenaars2006safe}. In the literature related to MPC, the backup set is often referred to as a \emph{terminal feasible invariant set}.  
Backup sets \emph{are} safe sets and may be computed as any safe subset of $\Ca$ using the methods described above.  
However, it is common to construct these from domain knowledge of safe configurations of the particular system. 

For an example of a safe backup set, consider that ground vehicles and vertical takeoff and landing (VTOL) air vehicles may bring themselves to rest in a safe location. 
In this case, the role of $\ub$ is to generate a safe stopping maneuver; \eg \cite{how17_agressive_3d_avoidance_quads}. 
Individual rest states are invariant points in the state space, and sets of adjacent rest states form invariant surfaces or volumes. 
In many cases it is not possible or practical to bring a vehicle to rest, and a backup set may instead be identified from safe periodic trajectories. 
For example, fixed wing aircraft may compute safe loiter circles \cite{Ariadne_Olatunde2021,schouwenaars2005implementation, schouwenaars2004receding, schouwenaars2006safe} or spacecraft may compute a set of safe orbits \cite{mote2021_NMT}. 
In many cases, the invariant set is a region without volume, and only an infinitesimal perturbation is required to cause a departure from the set. 
In practice, it is desirable to show that such surfaces are locally attractive under some backup control law, or to explicitly identify a larger invariant region around the initial set. 
For example, \cite{mote2021_NMT} develops a control law that stabilizes the a spacecraft to a set of safe orbits. 



\begin{example}
    Consider the damped linear system 
    \begin{equation}
        \ddot{x}= -\dot{x} + u
    \end{equation}
    with the constraint set given by $\Ca = \{x\in\R \, | \, x\geq 0 \} $. A safe backup set is given by the subset of $\Ca$ with zero velocity: $\Cb = \{ x\in \R \, | \, x\geq 0, \dot{x}=0 \} $. Note that $\Cb\subset\Ca$ and that $\Cb$ is invariant when $u=0$. 
\end{example}

\begin{example}
    Consider two spacecraft in planar orbit around a central body: (i) a  \emph{target} spacecraft, which is in a fixed circular orbit with period $\tau$, and (ii) a \emph{chaser} spacecraft of mass $m$.  In this setting, the dynamics of the chaser spacecraft are given by the Clohessy-Wiltshire-Hill equations, as developed in \cite{clohessy1960terminal}:
\begin{equation}\label{eq: CWH_dynamics}
    \begin{split}
        \dot{x}_1 &= x_3 \\
        \dot{x}_2 &= x_4 \\
        \dot{x}_3 &= 3 n^2 x_1 + 2 n x_4 + \tfrac{1}{m} u_1 \\
        \dot{x}_4 &= -2n x_3 + \tfrac{1}{m} u_2 \\
        \dot{x}_5 &= -|u_1|-|u_2|
    \end{split}
\end{equation}
with state $x\in \R^5$ and control input $u \in [-u_{\rm max},u_{\rm max}]^2$.  In this setting,  $x_1,\, x_2$ denote the relative distances between the spacecraft, $x_3,\, x_4$ denote the relative velocities, and $x_5$ denotes a fuel state. Additionally, $n = \frac{2\pi}{\tau}$ \cite{clohessy1960terminal}. The constraint set is defined by the constraints that the chaser must not run out of fuel and must stay at a distance greater than $R_{\rm min}$ from the target spacecraft. The constraint set is $\Ca=\{x\in\R^5 \, | \, x_1^2+x_2^2 -R_{\rm min}^2 \geq 0,\, x_5 \geq 0 \} $.  

\textit{Backup Set from Invariant Points:} Invariant {points} exist as zero velocity states in the $x_2$-$x_5$-plane, hence $\{x\in\R^5\,|\,x_1 =x_3 =x_4 =0\}$ is invariant for $u\equiv0$. A safety set is obtained from this set as ${\Cb}_{,1}=\{x\in\R^5 \,|\, x_1, x_2, x_3=0, x_5\geq0, |x_2|\geq R_{\rm min}\}$. 

\textit{Backup Set from Periodic Trajectories:} It can be shown that the chaser spacecraft is on an elliptical orbit around the target spacecraft whenever $u=0$ and $x_3=\frac{n}{2}x_2, \, x_4= -2n x_1$, with each individual orbit occupying the position states $x_1^2+\frac{1}{4}x_2^2=b^2$ where $b\geq 0$ is the semi-minor axis of the orbit. 
Hence, the linear subspace $\{x\in\R^5\,|\, x_3=\frac{n}{2}x_2, \, x_4= -2n x_1\}$ is an invariant manifold composed of all such closed orbits. 
A backup set can be constructed from this subspace by considering range of orbits that fall into the constraint set: 
$
    {\Cb}_{,2}=\{x\in\R^5 \,|\, x_3=\frac{n}{2}x_2, \, x_4= -2n x_1, \, x_1^2+\frac{1}{4}x_2^2 \geq R_{\rm min}^2 \}
$
\end{example}

\section{Properties of Run Time Assurance Systems} 
RTA has evolved over the last several decades stemming from research in control theory, computer science, electrical, mechanical, and aerospace engineering.
In this section, each of several basic classifications and properties of RTA mechanisms are defined, concluding with RTA-human interaction considerations. RTA approaches can be classified as: explicit or implicit, zero or first order, and latched or unlatched. Explicit approaches precisely define a specific safe set, while implicit RTA approaches define recovery trajectories under a predefined backup control law. Zeroth order RTA methods do not use gradients of the dynamics and are often associated with RTA systems that use methods described in the Section  ``The Simplex Architecture," while first order methods take advantage of gradient computations and are used in methods such as those described in the Section ``Active Set Invariance Filtering." Latched RTA systems switch to a backup controller and remain under its control until a specified release condition is met, while unlatched implementations return to the primary control as soon as the system returns to a safe set. In addition to these classifications, RTA systems feature safety and performance properties described in this section, including innocuity, viability, nuisance freedom, and integrity monitoring. Finally, this section discusses important concepts with respect to RTA-human interaction including: variable risk tolerance, ability to turn the RTA off, transparency, and missed detection/false alarm considerations.

\subsection{Implicit and Explicit RTA Approaches} \label{sec: implicit_vs_explicit_approaches}

Fundamentally, an RTA mechanism is a control law that renders a subset of the constraint space forward invariant, activates a recovery response near the boundary of that set, and that passes through the desired input from the primary controller everywhere else.  
The set made safe under the RTA is referred to as the \emph{safe operational region} of the RTA system, and it is denoted by $\Cs$. 
In cases where the recovery response relies on switching to a backup control law $\ub$, the safe operational region is a 
subset of the constraint set $\Ca$ that is forward invariant under that control law.
Approaches of this type are considered in the section entitled ``The Simplex Architecture." 
Alternatively, in cases where the recovery response is determined as the solution to an optimization program, and safety is enforced with the constraints of that program, then the safe operational region is a control invariant subset of $\Ca$.
This is because the optimization program searches over all feasible control actions --- \ie control invariant sets are forward invariant under certain optimization procedures.
Approaches of this nature are considered in the section ``Active Set Invariance Filtering." 

As with the safe sets discussed previously, the safe operational region can be identified explicitly or implicitly.
\emph{Implicit} (or trajectory-based) approaches compute finite-time trajectories  under a backup controller $\{\phi^{\ub}(t;x)\in \X \,|\, t\in \mathcal{T}\}$ \emph{online} (at run time) and use the information to decide when or how to intervene. 
\emph{Explicit} (or region-based) approaches may or may not switch to a backup controller, but do not rely on simulating trajectories of the backup controller online. 
Typically, explicit approaches identify trusted regions \emph{offline}
via
the construction of either a large forward invariant or control invariant set. 
That is, implicit RTA approaches are those that utilize an online \emph{look-ahead} and will bound the system to an implicitly defined safe set such as \eqref{eq: Safe_Backward_Image} while explicit RTA approaches determine safe states \emph{a priori} and will bound the system to an explicitly defined safe set such as \eqref{eq: explicitly_defined_safe_set}.
This concept is demonstrated in the following example.

\begin{example}
    Consider the mass-spring-damper system described by \eqref{eq: mass_spring_damper} with the constraint space $\Ca = \{x\in\mathbb{R}^2 \, | \, 1-\Vert x \Vert_\infty \geq 0\}$.
    An RTA mechanism can be constructed by requiring that the backup controller be applied near the boundary of a set $\Cs$, where $\Cs$ is safe under the backup controller. 
    Let the RTA mechanism be, 
    \begin{equation} \label{eq: switching_RTA_for_mass_spring_damper_example}
    \begin{split}
        u(x) &=
        \begin{cases}
        u_{\rm b}  & \text{if}\quad \vartheta(x)  \\
        u_{\rm des} &\text{otherwise} .\\ 
        \end{cases} \\
    \end{split}
    \end{equation}
    where $\ud$ is the desired control input and $\vartheta(x)$ represents the activation condition (or switching condition). Considered below are implicit and explicit activation conditions.
    
    \textit{Explicit Condition}
    
    Let $\Cs =\{x\in\R^2\,\,|\,\,h(x)\geq 0\}$ with $h(x)=1-V(x)$ and $V(x)$ being the Lyapunov function \eqref{eq: mass_spring_lyapunov}. The set $\Cs$ is said to be trusted as it is safe under $\ub$. 
    The activation condition
    \begin{equation}
        \vartheta(x): h(x)\geq \varepsilon
    \end{equation}
    ensures safety of $\Cs$ with respect to $\ua$ by activating $\ub$ near the boundary where $\varepsilon>0$ defines the size of the RTA activation region. 
    Since, $V(x)$ defines a Lyapunov function for the closed-loop dynamics under $\ub$, it is known that
    $h(x)$ is increasing whenever $\ub$ is applied. 
    Furthermore, since $u_{\rm b}$ is activated before the boundary $h(x)=0$, \eqref{eq: switching_RTA_for_mass_spring_damper_example} enforces that $h(x)$ is always positive, meaning that the system will be forward invariant in $\Cs\subset\Ca$. 
    
    \textit{Implicit Condition}
    
    Let $\Cb =\{x\in\R^2 \,\,|\,\,h(x)\geq 0\}$ with $h(x)=1-V(x)$ and $V(x)$ being the Lyapunov function \eqref{eq: mass_spring_lyapunov}, then a condition for invariance in the safe backward image of $\Cb$ is
    \begin{equation}
    \vartheta(x):\,\, [\forall t\in[0,T],\quad \varphi( e^{A t}x ) \geq \varepsilon_1] \,\,\wedge\,\, [h_{\rm b}( e^{A T}x)\geq \varepsilon_2] 
\end{equation}
where $\varepsilon_1, \varepsilon_2 > 0$. The safe backward image in this case defines the safe operational region for the RTA. It is depicted as the green region in Figure \ref{fig: mass_spring_damper_implicit}.  
\end{example}

There are important tradeoffs associated with implicit and explicit approaches.
While identifying invariant sets explicitly offline generally reduces the online computational burden of the algorithm, the task may become intractable for complex and high-dimensional systems. 
In such cases, implicit, trajectory-based approaches become favorable.
The key advantage to assessing safety 
through trajectories online is that, in the simplest case, it only requires a finite-time simulation of the recovery maneuver, which is generally a tractable task.
A secondary benefit is that it handles dynamic environments and changes in the model better; \ie rather than needing to recompute an invariant subset of $\Ca$, one only needs to update the simulation parameters.
However, an important practical consideration is that trajectory-based approaches rely on the ability to accurately forecast recovery trajectories, and the predicted behavior may be increasingly sensitive to noise (\ie less accurate) as the length of the backup trajectory is increased.

\begin{figure}[htb]
\begin{center}
\includegraphics[width=1\textwidth]{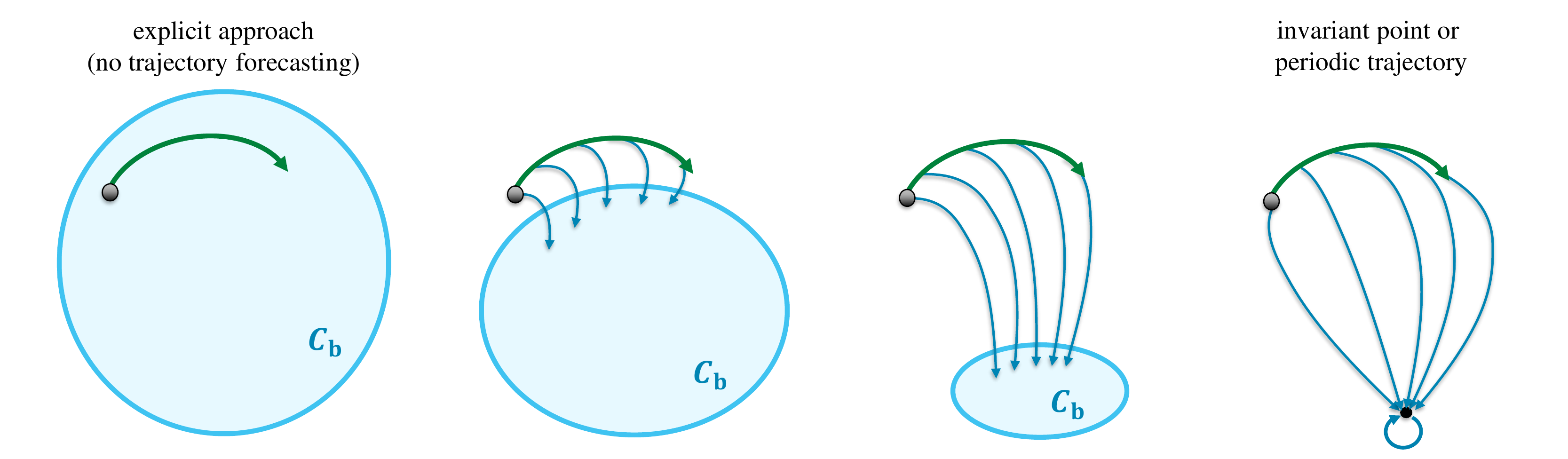}
\caption{Notional depiction of the tradeoff between a longer integration horizon and a larger backup set. \rev{The thick green arcing arrow represents the desired trajectory of the primary controller, the teal region represents a backup safe set $\Cb$, and the thin teal arcing arrows represent trajectories to the backup set. Large explicit backup sets $\Cb$, reduce the need for online computation of trajectories to the backup set, but may be intractable for complex and high-dimensional systems. Computing online finite-time simulation of a recovery maneuver may be more tractable in some cases, but is more vulnerable to noise or model inaccuracies.}}
    \label{fig: implicit_explicit_tradeoff}
\end{center}
\end{figure}

It is interesting to note that since a backup set $\Cb$ is an explicitly defined invariant subset of the constraint set $\Ca$, an implicit approach to safety will typically involve a combination of (i) explicitly identifying an invariant set $\Cb$ offline \rev{and} (ii) finding a safe trajectory 
to this set online. 
In this sense, the explicit approach can be viewed as a special case of the implicit approach where the backup trajectory consists of only a single point. 
Furthermore, any explicitly defined safe set can be used as part of an implicit approach. 
The implicit approach allows for the problem to be solved by taking full advantage of both offline and online resources.
Specifically, finding a larger backup set offline reduces the online computational burden by reducing the length of the trajectory required to obtain the same operational region; likewise, integrating backup trajectories over a longer horizon  reduces the need to identify large safe sets offline. 
An depiction of this tradeoff is illustrated in Figure~\ref{fig: implicit_explicit_tradeoff}.
Generally, for simple systems, the problem can be solved offline, and as systems become more complex, one must rely more and more on online simulations.

\subsection{Zero-Order and First-Order Methods}

Filtering approaches in RTA may be classified into zero-order and first-order methods.
Zero-order algorithms are derivative free. They are typically, though not necessarily,  associated with the \emph{Simplex} architecture and choosing from a discrete set of possible control signals. 
For example, the RTA mechanism may choose between applying the input from the primary control law $\ud$ or a backup control law $\ub$ at any time.   
In this case, one may observe the effects of chatter; \ie variations in the output of the RTA mechanism are not smooth with respect to variations in the initial state or desired input. 
This may have practical implications to real world systems. For instance, the action may cause excessive wear to the actuators, on-board components may experience interruption to their operations, or passengers may observe the intervention as being more abrupt, resulting in a less pleasant experience (\eg consider an automobile that only intervenes through applying the brakes at maximum capacity). 
These effects may be handled with various heuristics; \eg hysteresis, or interpolating between $\ub$ and $\ud$ as the state approaches the boundary of the safe operational region. Alternatively, one may turn to a first order method. 

First order methods require  computing derivatives, and are typically associated with methods described in the ``Active Set Invariance Filtering" section. 
In this case, the gradient information is used in a \emph{barrier condition}, which constrains the set of available inputs to a subset of admissible and safe inputs. 
An optimization framework may be used to choose the input within this set according to some cost function.
A common cost specification requires that the RTA choose the control that is closest (in a 2-norm sense) to the desired input. 
RTA filters constructed in this way will have smooth variations in the output signal $\ua$. 
While this eliminates the effect of chatter, it introduces added complexity into the problem and generally requires greater computational resources. \rev{The Robotarium is an example explicit ASIF RTA implementation is discussed further in the sidebar titled ``The Robotarium: An RTA Enabled Remote-Access Swarm
Robotics Testbed"}


\begin{table}[hbt!]
\begin{center}
\caption{\label{t: classification_of_RTA_algs} Classification of RTA Applications in the Literature}
\resizebox{\textwidth}{!}{%
\begin{tabular}{|c|c|c|} \hline
                    & Implicit                 & Explicit \\\hline
\,      & spacecraft collision avoidance \cite{Mote2021Natural} & waypoint tracking cyber-physical system \cite{bak11} \\  
\,      & fixed wing aircraft ground avoidance (Auto-GCAS) \cite{swihart2011automatic,griffin2012automatic} & advanced flight control systems \cite{aiello2010run}  \\
Simplex      & fixed wing aircraft obstacle avoidance \cite{schouwenaars2006safe, schouwenaars2005implementation,burns2012advanced}  & lane keeping \cite{Benoit2010_simplex_lane_keeping} \\
\, & safety against LTL specifications \cite{hobbs2020elicitation, abate2019monitor} & aircraft conflict resolution \cite{tomlin1998conflict, tomlin2000game, tomlin2001safety} \\
\,      & collision avoidance for high speed quadrotor navigation \cite{kumar16_high_speed_naviagtion, how17_agressive_3d_avoidance_quads} & decentralized collision avoidance for quadrotors \cite{gillula2011applications, hoffmann2008decentralized} \\
\,      & collision avoidance during lane-changing maneuvers \cite{lin2014active}  &  ground robot navigation around obstacles \cite{muthukumaran2019hybrid, phan2017collision }  \\ \hline 
\,      & bipedal walking/ exoskeletons \cite{gurriet2019towards_exoskeletons} & bipedal walking/ exoskeletons \cite{nguyen2015safety_cbf_bipedal, nguyen20163d_bipedal} \\
\, & robotic manipulator arms \cite{singletary2019online} & lane keeping \cite{xu2017correctness, hu2019lane} \\
ASIF & mobile inverted pendulum (Segway) \cite{gurriet2019scalable} &  mobile inverted pendulum (Segway)  \cite{gurriet2018ASIF} \\
\, & rapid aerial exploration of unknown environments \cite{singletary2020safety} &  robotic grasping \cite{cortez2019control}  \\
\, & multi-robot systems \cite{chen2020guaranteed} & swarm robotics (Robotarium) \cite{pickem2016safe, pickem2017robotarium, wilson2020robotarium, robotarium_2021} \\
\, & fixed wing aircraft collision avoidance \cite{chen2021_backup_CBFs} & motorized rehabilitative
cycling system\cite{Isaly2020Zeroing} \\ \hline 

\end{tabular}}
\end{center}
\end{table}

\subsection{Latched and Unlatched Implementations}
RTA systems based on those described in the Section  ``The Simplex Architecture" activate a recovery maneuver when the monitor detects that one of the boundary violation conditions has become active. 
At this point, the decision logic must specify when to return control to the primary controller. 
In an \emph{unlatched} implementation, the decision logic releases control of the recovery controller as soon as the boundary violation condition becomes inactive. 
For example, an unlatched RTA implementation that switches from the primary control law $\ud$ to the backup control law $\ub$ whenever a condition $\vartheta$ is satisfied will take the form of a hybrid system as in \eqref{eq: switching_RTA_for_mass_spring_damper_example}. 
Alternatively, a \emph{latched} implementation will hold its activation state until a separate release condition is met. 
The release condition may for example consist of following the backup trajectory for some fixed amount of time, bringing the system to a particular configuration, or having a human operator return control after assessing the problem. Auto GCAS is an example of a temporary latched system that holds the maneuver until the aircraft is on a collision free course\cite{swihart2011automatic}. Other implementations of latched RTAs may require a system reset before switching back, such as a computer reset supervised by a human operator. Note that this requires that the decision logic remember the current activation state.

\subsection{Properties and Systems Engineering Considerations for Run Time Assurance Design}

While it is convenient to treat safety as a binary ``safe" or ``unsafe" property, safety is often a complex weighting of risk among primary and secondary safety constraints. This section provides a few definitions of design considerations for incorporating RTA solutions into the larger control system. In practice, RTA system design ultimately relies on domain-specific knowledge to achieve acceptable performance. However, there are common safety and performance properties of good RTA designs.
In order to keep the system safe, the RTA mechanism must select control inputs that are \emph{innocuous} and \emph{viable}. In addition to safety, the performance of the RTA may be evaluated by its adherence to a \textit{nuisance free} property.

\subsubsection{Innocuity}

 Ideally the entire state of the system is contained in one vector $x$ and the entire constraint set can be explicitly defined in a set $\Ca$. However, in practice this is rarely the case. Instead, RTA decisions are not solely based on system dynamics, but must factor in the states of interacting subsystems, human interactions, fault management, or interlock condition management. \textit{Innocuity} is the property that each component of the design has no unacceptable impact on the larger system design \cite{holloway2019understanding}. Innocuity of RTA systems implies the selected control action does not have unintended behavior, such as causing a violation of a secondary safety constraints. An interlock is a set of two mutually exclusive states such as mechanisms that prevent elevator doors from opening when the elevator is in motion, or that prevent the elevator from moving when the doors are open. This relationship between RTA-level and system-level constraints may result in a combination of RTA system modes (described as a finite state machine), if-statements, and dynamics considerations \cite{hobbs2021formal}. 

For example, in the development of Auto GCAS, the high-level requirements in order of precedence were to do no harm, do not interfere, and prevent collisions \cite{swihart2011automatic, swihart2011design}. The ``do no harm" and ``do not interfere" requirements in Auto GCAS are examples of innocuity properties. These requirements don't directly define the RTA response, but describe special considerations to prevent unintended consequences of the RTA design. This requirement ordering might seem counter-intuitive, as one might expect preventing collisions to be the top priority of a collision avoidance system. However, this order of requirements proved critical in acceptance of the system \cite{lyons2016trust}. To give a little more insight, selected generalized, conceptual requirements for Auto GCAS are presented below and divided here in terms of ``do no harm," ``do not interfere," and ``prevent collisions." The ``prevent collisions" requirements are typical of what is generally considered an RTA requirement. However, the ``do no harm" and ``do not interfere" requirements describe how the RTA system should operate in the context of the aircraft control system that interacts with human operators, experiences failures, and has a defined set of interlock conditions.

\noindent\textbf{Do No Harm}
\begin{itemize}
\item The automatic recovery shall not cause harm to pilot, aircraft, or components.
\item The automatic recovery maneuver shall not place the aircraft in an uncontrollable state.
\item The pilot shall be able to interrupt a maneuver.
\item The pilot shall be able to manually engage a maneuver.
\item Subsystem monitors shall determine if a failure exists and be provided to Auto GCAS.
\item Auto GCAS shall not activate an automatic recovery maneuver if a failure of a subsystem supporting Auto GCAS exists.
\item Auto GCAS shall leave the failed mode state if a failure does not exist.
\item The automatic recovery maneuver shall not activate during aerial refueling.
\item The automatic recovery maneuver shall not activate when the aircraft has an excessively high angle of attack.
\item The automatic recovery maneuver shall not activate when the aircraft velocity is too low.
\item Auto GCAS shall inform the pilot if the aircraft speed is too low for the automatic recovery maneuver.
\end{itemize}

\noindent\textbf{Do Not Interfere}
\begin{itemize}
\item The automatic recovery maneuver shall not activate during landings.
\item The pilot shall be able to turn the system off.
\item The pilot shall be able to select the protection level.
\item Auto GCAS shall notify the pilot when the automatic recovery maneuver initiates and terminates.
\item The system shall record data about each automatic recovery maneuver activation.
\item The Digital Terrain Elevation Data used by the system shall have a resolution and accuracy that supports safe and nuisance free operation. 
\end{itemize}

\noindent\textbf{Prevent Collisions}
\begin{itemize}
\item The system shall conduct an automatic recovery maneuver when the projected recovery trajectory intersects the terrain profile contour with buffers.
\item The system shall terminate an automatic recovery maneuver as soon as it is determined that the aircraft will clear the terrain threat.
\item The recovery maneuver shall consist of a roll to wings level and a pull up.
\end{itemize}

Systems engineering, computing hardware availability, and human-machine interaction all require practical considerations that play important roles in imparting secondary safety constraints on RTA designs. 
From a computing hardware standpoint, assuming availability of high performance computing may not be valid. In some domains, design constraints such as size, weight, and power coupled with vibration and radiation tolerance limit the processing and memory of computing hardware onboard the dynamical system. In these cases simpler RTA solutions may be favored over optimal solutions. An example of this occurs in aviation, where RTA designs are expected to be backwards compatible with older aircraft, and simple approaches may be to climb, descend, or turn right or left at a specified rate \cite{olson2015airborne}. Additionally, when the automated backup control of the RTA systems impact human operators, humans trust an easily understood maneuver consistent with human training, expectations, and preferences  \cite{lee2004trust}. For Auto GCAS, familiarity of the roll to wings level and pull maneuver and its consistency with pilot training and behavior resulted in strong positive perceptions of the system \cite{lyons2016trust}.

\subsubsection{Viability} 
\textit{Viability} ensures the selected control action does not eliminate availability of a safe action downstream, \emph{i.e.} the RTA mechanism will always be able to identify a safe input in the future. This concept is formalized in Figure \ref{fig: double_integrator_sets}, where the green area under the curve describes maximum velocity for any distance from the obstacle. 
One way to ensure viability in an implicit RTA design approach is to define a limited set of pre-determined recovery actions. Simple, pre-defined recovery responses can be pre-verified for use, reduce online computation burden, and make the RTA system more easily understood and trusted by a human operator \cite{lee2004trust}. In Auto GCAS, rather than optimizing a trajectory for each specific collision avoidance scenario, a predefined set of verified maneuvers was created. In F-16 Auto GCAS, a single roll to wings level and 5g pull maneuver is conducted \cite{griffin2012automatic}. In the Automatic Air Collision Avoidance System (Auto ACAS), nine  different maneuvers are considered \cite{turner2012automatic}. In patented \cite{skoog2017ground} and experimental Auto GCAS designs for small UAVs \cite{sorokowski2015small}, lesser capability aircraft \cite{carpenter2019simulation}, and cargo-class aircraft \cite{suplisson2015optimal}, three or more possible maneuvers may be considered. In the case of multiple maneuvers, one popular approach that combines viability with nuisance freedom is to engage the last available maneuver right before it is too late.


\subsubsection{Nuisance Freedom}
In a \textit{nuisance-free} RTA system, the constraint set $\Ca$ and corresponding operating envelope $\Cs$ of the primary controller is as large as feasible. Other ways to describe this property is that the RTA should be minimally invasive and should not interfere with the primary function of control system $\ud$ unless absolutely necessary. In Auto GCAS, a collision avoidance maneuver is considered nuisance-free if it occurs after an aware pilot would have maneuvered to avoid a collision. To formally define nuisance-free operations, a \textit{time available} metric described zero time available as the point where an activation would just barely scrape the ground, and increasingly positive time available would result in maneuvers at greater altitudes above the terrain. To assign a value to an acceptable time available activation range, a 1995 study measured when pilots felt an aggressive recovery should be activated to avoid a collision \cite{swihart2011automatic, swihart2011design}. Pilots flew towards the ground at a variety of dive angles, bank angles, air speeds, and load factors and activated a recovery maneuver when his or her comfort threshold was met. After each run, the pilots rated the timing of the recovery initiation, their anxiety level during the maneuver, and the precision/aggressiveness of the maneuver for each run. Based on this information a 1.5 second time available design criteria was identified, and collision avoidance maneuvers that activated with less than 1.5 seconds time available were considered nuisance-free \cite{swihart2011automatic, swihart2011design}.

More generally, the number of RTA activations, magnitude of the response, activation timing, and size of the safe set can be use to compare nuisance criteria across different RTAs.
 \begin{itemize}
        \item In many scenarios, the RTA should not come on often because activations interrupt the primary mission. An \textit{RTA activations} metric tracks the number of seconds or discrete time steps that a recovery maneuver is in control. 
        This metric allows comparing different RTA approaches as it provides a measure of relative conservatism of various RTA solutions in the presence of the same, possibly faulty primary controller.
        \item The RTA should ideally not impart a large change in control, measurable via a  \textit{control magnitude} metric.
        \item An \textit{RTA invasiveness} metric describes whether the RTA activates appropriately when near the safety constraint boundary, while refraining from activating when far from the allowable boundary.  A variety of application-specific measurements of this may be applied, such as the time available metric in Auto GCAS.
        \item An \textit{RTA safe set size}  measures of the volume of the safe set $\Cs$. The larger the volume, the more space the primary controller has to operate optimally, and the less invasive the RTA is. Different RTA approaches may be more or less conservative, which impacts the size of $\Cs$.
\end{itemize}


\subsubsection{Run Time Assurance Integrity Monitoring}

Recognizing and accommodating failures that could impact an RTA system is a critical element of RTA design. The failures could come from inside the RTA system or from externally provided information that the RTA uses to make a response decision. Possible software checks include monitoring for missing ``heartbeat" signals from components and ``reasonableness checks" that ensure incoming data are within a reasonable range of values \cite{sorokowski2015small}. For example, a failure of the inertial navigation system should not allow Auto GCAS to roll inverted and pull down into the ground instead of pulling up to avoid a collision \cite{swihart2011automatic}. Auto GCAS uses System Wide Integrity Management (SWIM) hardware and software tests used to mitigate single-point-of-failure risks \cite{turner2012automatic,burns2011auto, swihart2011automatic}.  

\subsection{Human Interaction with Run Time Assurance Systems}

When designing RTA systems, it is important to consider the interaction of the system with human operators. While the topic of human-machine interaction can swell quickly, this article focuses on a few key design considerations in RTA.

\subsubsection{Variable Risk Tolerance and an ``Off Switch"}
One of the challenges of developing an RTA system is the need to balance nuisance-freedom with the risk tolerance of a particular task or operator. Adding the ability to tune risk corresponding to safety buffers or other factors facilitates flexibility and acceptability of the RTA in operational use. For example, Auto GCAS features two pilot selectable modes that enable variable risk tolerance: a ``norm" mode  with adequate safety buffers for most cases, and a ``min" mode which minimizes buffer size for nuisance-free low-level flying at the trade-off of reduced protection \cite{swihart2011automatic, swihart2011design}. In addition, the ability to turn the system off was an important feature that engendered pilot trust \cite{lyons2016trust}.

\subsubsection{Transparency and Trust}
The primary basis of trust in automation comes from dependability and predictability \cite{madhavan2007similarities}. However additional factors with a measurable impact on human trust include the past performance ($\eg$ system pedigree or heritage), simplified and understandable performance, system intent, and reliability \cite{lee2004trust}. In Auto GCAS transparency comes through pilot training before operations and through visual indicators on the aircraft's head-up-display (HUD) \cite{lyons2016trust,ho2017trust}. 
A discussion of the difference between safety, security, and reliability is provided in the sidebar titled ``Safety, Reliability, and Security." Reliability is important for human trust in RTA systems as humans initially assume that machines are perfect and any errors rapidly deteriorate trust \cite{hoff2015trust}. Knowledge of Auto GCAS's 98\% reliability increased test pilot trust of the system \cite{lyons2016trust}.

\subsubsection{Missed Detections and False Alarms}
 
It is important to consider an acceptable rate of missed detections and false alarms in the design of an RTA system. \textit{Missed detections} are a failure of the monitor to detect an imminent safety violation. For example, if Auto GCAS fails to detect an imminent collision with terrain, it could lead to loss of life and the aircraft. \textit{False alarms} occur when the RTA monitor believes the system is entering an unsafe state, when in reality no safety violation is imminent. False alarms erode confidence in the RTA system. For example, during psychological study on the development and deployment of Auto GCAS, one pilot indicated that a single false fly-up (automatically maneuvering to avoid the ground when there was no imminent collision) would likely cause pilots to turn the system off and lose the protection it provided \cite{swihart2011automatic, swihart2011design}. Developing pilot trust in Auto GCAS hinged on its nuisance avoidance criteria, \ie keeping the number of false alarms very low \cite{ho2017trust}.

In statistical hypothesis testing, missed detections and false alarms are Type I and Type II errors. Hypothesis testing is a formal technique to verify that a system meets a specification or requirement \cite{Milton2003Probability}. A hypothesis can have two possible outcomes: a null hypothesis $H_0$, such as a collision is not imminent, and an alternative hypothesis $H_1$, such as a collision is imminent. As summarized in Table \ref{t:HypothesisTest}, a Type I error is a false alarm ($\ie$ activate $\ub$ when not near the constraint boundary $\partial\Ca$), while a Type II error is a missed detection ($\ie$ failing to activate $\ub$ near $\partial\Ca$, resulting in a departure from the safe set. When considering a collision avoidance system for example, a Type I error occurs when a collision avoidance system maneuvers to avoid a collision that would not have happened, and a Type II error occurs when a collision avoidance system does not maneuver to prevent a collision.

\begin{table}[hbt!]
\begin{center}
\caption{\label{t:HypothesisTest} False Alarms and Missed Detections Type I and Type II Errors, with Collision Avoidance as an Example\rev{.}}
\begin{tabular}{|r|c|c|} \hline
                    & $H_0$                 & $H_1$\\
                    & Collision             & Collision \\
                    & Not Imminent          & Imminent \\\hline
Do Not Reject $H_0$ &                       & Type II Error \\
            & Correct               & False Negative\\
System Decides           &                       & Missed Detection \\
Not Imminent           &                       & $\beta$ probability\\\hline
Reject $H_0$        & Type I Error          & \\
     & False Positive        &  Correct     \\
System Decides           & False Alarm           &       \\ 
Collision Imminent            & $\alpha$ Probability  &  \\\hline 
\end{tabular}
\end{center}
\end{table}

In some systems, like Auto GCAS, the acceptable rates of false alarms \revised{$\alpha$} are very low because of the consequence of an unnecessary interruption of the system mission is high. In other systems like elevators, the acceptable rate of \revised{a} missed detection \revised{$\beta$} (not detecting a human is in the way of the closing doors) would be very low, and false alarms would be more acceptable (there is almost always an option to take the stairs). During the development of RTA systems, studies are required to determine acceptable $\alpha$ and $\beta$ rates.


\section{The Simplex Architecture}\label{sec:simplexarch}

\begin{figure}[htb]
\begin{center}
\includegraphics[width=1\textwidth]{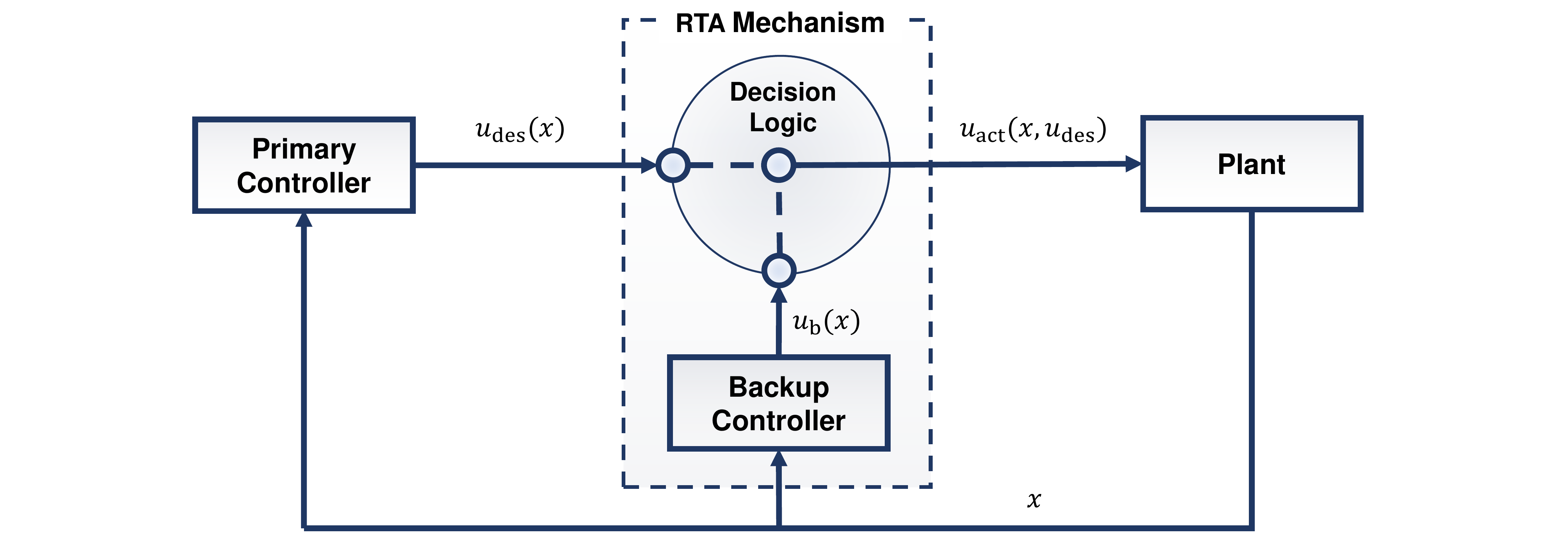}
\caption{Simplex Architecture with a physical plant, a verified safety controller, a verified decision module and switch, and an unverified complex controller.} 
\label{fig: Simplex_architechture} 
\end{center}
\end{figure}

Simplex Architecture RTA designs \cite{sha1995software,rivera1996architectural,bak11,reis2009browser} feature a \textit{monitor} or watchdog that watches for imminent violations of safety constraints, and a backup \textit{safety controller} that provides a guaranteed safe control output to remedy the unsafe situation. A critical feature here is that the backup control is applied via a switch, rather than gradually as in ASIF described next.  One way to describe Simplex RTAs is as a hybrid dynamical system \cite{goebel2012hybrid} described by Eq. \eqref{eq: switching_RTA_for_mass_spring_damper_example}. The monitor and safety controller are inserted before the plant as demonstrated in Fig. \ref{fig: Simplex_architechture}.
%
In this model, the following functional components describe specific functions and interactions that may be implemented on one or more physical components:
\begin{itemize}
\item \textit{Plant}: This functional component is the system under control and could be a spacecraft, aircraft, boat, car, or other robotic system.
\item \textit{Primary controller}: This functional component is the primary, high performance controller of the system that may not be \revised{fully} verified with traditional offline verification techniques. Depending on the application this could be a human pilot, neural network-based control, or complex control software.
\item \textit{Backup Controller}: This functional component continuously computes a safe 
control response to the state of the plant. For example, in Auto GCAS the safety controller is a predefined roll and pull maneuver. One school of thought places heavy verification burden on the safety controller, boundary monitor, and decision logic as a way to certify advanced controllers that cannot be verified to the current certification standards \cite{ASTM3269-17}. Another school of thought requires no assumptions on the safety or verification of the backup controller. That is, safety can be guaranteed through online integration of the backup controller, and the system may revert at any time to the most recently computed safe trajectory. Formal guarantees are obtainable in both cases. 
\item \textit{Boundary (a.k.a. safety) monitor}: This functional component monitors the state of the plant and the nominal controller for predetermined safety boundary violations. Examples of boundary violations might be a trajectory prediction that intersects the terrain (indicating an imminent ground collision) or a control input that would cause excessive acceleration (risking damage to the structure or components). 
\item \textit{Decision Logic}: The functional component takes inputs from all the other components pictured and determines which control output, complex or safety, to output to the plant. 
\end{itemize}

\subsection{Variations on Simplex}
\begin{figure}[htb]
\begin{center}
\includegraphics[width=\textwidth]{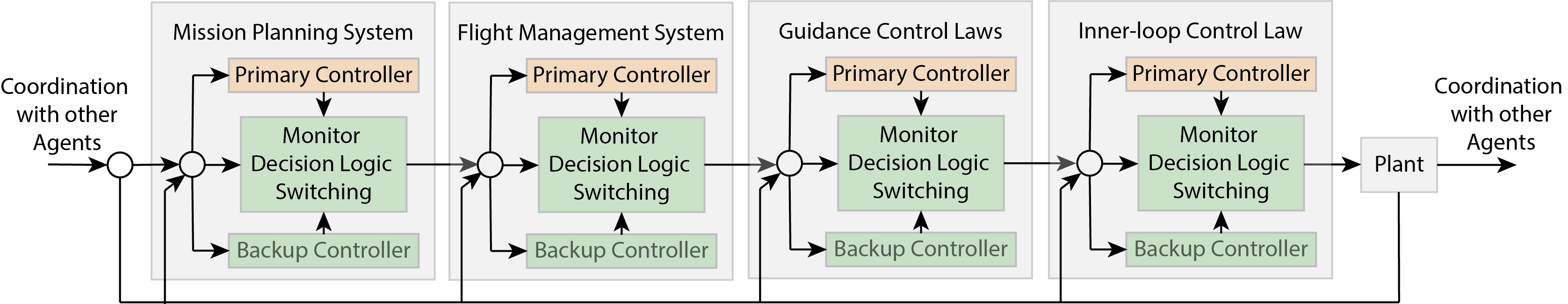}
\caption{Nested feedback RTA architecture for an Unmanned Aerial System (UAS) plant \cite{schierman2020runtime}. \rev{In a nested RTA architecture, the inner-loop control law RTA \revised{ensures aircraft stability}, the guidance control law RTA ensures safety of commands generated to follow waypoints, the flight management system RTA checks that the waypoints along the path are safe, and the mission planning system RTA ensures the safety of task allocations to meet mission goals of the plant. In  each RTA layer, a monitor watches for unsafe conditions and decision logic determines switching between the primary controller and backup controller.}} 
\label{fig:NestedRTA} 
\end{center}
\end{figure}

\begin{figure}[htb]
\begin{center}
\includegraphics[width=0.7\textwidth]{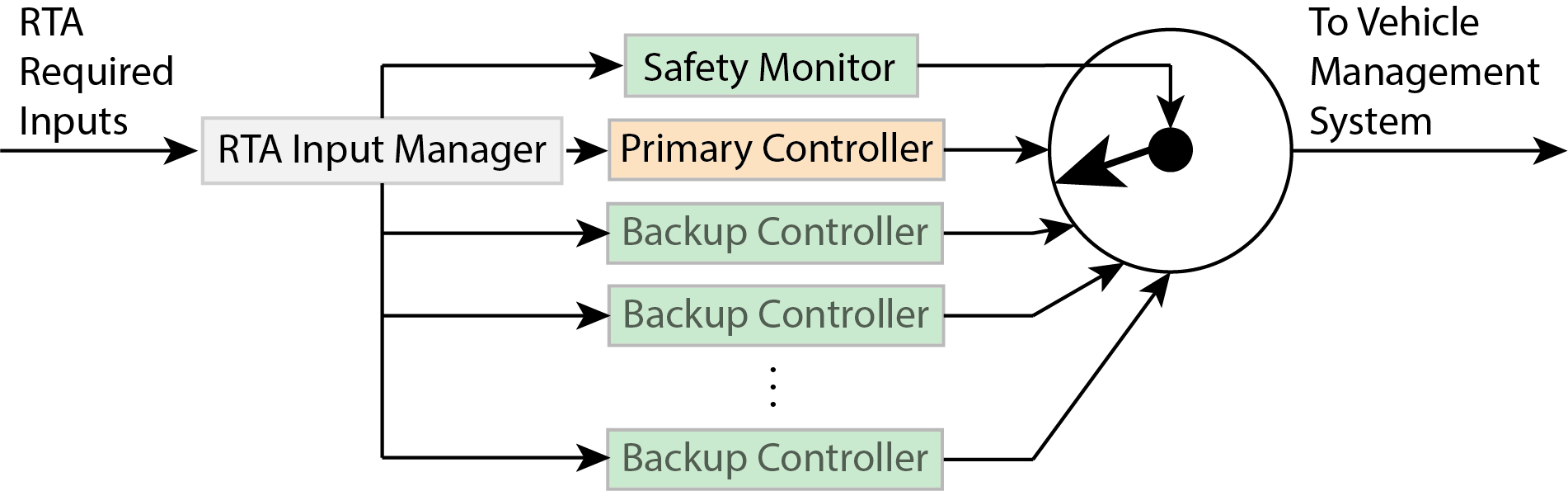}
\caption{ASTM F3269 RTA architecture with multiple recovery control functions at the same level in the control hierarchy \cite{ASTM3269-17}. \rev{An example of this might be an aircraft that avoids collisions with other aircraft, avoids collisions with the ground, and avoids flying into no-fly zones. Each of these boundaries are disjoint; however, it is possible that violations of multiple boundaries could be predicted to occur at the same time.}} 
\label{fig:MultiRecovery} 
\end{center}
\end{figure}

Variations on Simplex include \textit{nested RTA}, \textit{multi-recovery RTA}, and \textit{multi-monitor RTA} designs. \textit{Nested RTA} designs implement safety controls at each layer of the control architecture as shown in Figure 
\ref{fig:NestedRTA}. 
\textit{Multi-recovery RTA} designs have multiple reversionary controllers at the same level of the control architecture and a switch that considers all possible recovery function options.  The concept is discussed in a recent ASTM standard for the use of RTA in unmanned aircraft \cite{ASTM3269-17} and depicted in Figure \ref{fig:MultiRecovery}. 
The best approach to designing multiple recoveries is an area of active research. One approach is  to combine multiple RTA systems as an integrated solution. For example, the Automatic Integrated Collision Avoidance System (Auto ICAS) \cite{jones2017automatic,turner2019automatic} integrates ground and midair collision avoidance into a comprehensive collision avoidance system. The advantage of this integrated approach is that each system can consider possible conflicts between boundary violations and select a better solution than either might select on their own. For instance, in a situation where an aircraft flying close to the ground is on a collision course with another aircraft, a ground-aware midair collision avoidance system will filter from the possible maneuvers to select a maneuver that avoids the midair collision while at the same time ensuring none of those maneuvers will fly the aircraft into the ground. An integrated solution has the benefit of generating a better solution than either separate system might otherwise choose. However, this increase in performance also comes at a large increase in cost and schedule. 
An alternative multiple recovery controller integration approach is to use an RTA network architecture where each boundary monitor and recovery controller function is separate \cite{skoog2020leveraging}. An example of an architecture with multiple safety monitors and recovery functions is the Expandable Variable Autonomy Architecture (EVAA) RTA Network \cite{skoog2020leveraging}. 
On the one hand, this modular approach eases verification of each individual component and facilitates quick integration of new components and upgrades. On the other hand, additional stress is placed on the verification of the decision module/switching component to determine which action to take in the case where multiple safety boundaries may be violated at the same time. The EVVA \cite{skoog2020leveraging} answer to this challenge is to approach the decision like a human pilot, who responds to the most critical boundary violation first before moving onto the next violation. A disadvantage of the modular approach is that it is possible to violate a safety boundary. For instance, if a ground collision and geofence collision were both imminent, the system might decide to engage ground collision avoidance to first clear the threat of ground collision before engaging a geofence controller and violating the geofence boundary. Depending on the scenario, this may or may not be acceptable. An integrated solution could adjust the boundary of a geofence controller based on awareness of ground and midair collision avoidance, or could select a collision avoidance maneuver that also kept the aircraft within the geofence. 
%
%
%
\textit{Multi-monitor RTA} designs feature checks and balances to ensure integrity of the RTA system. These designs may include monitors beyond a boundary or safety monitor. These monitors may include a \textit{failure monitor}, an \textit{interlock monitor}, and a \textit{human supervisor}, among others \cite{hobbs2021formal}. The failure monitor watches for unreliable information coming into the RTA such as a failed or stale sensor signal and prevents the RTA from acting on incorrect state information. The interlock monitor watches for conditions where it may be safe to compute a backup control solution, but unsafe to engage a maneuver. For example, an interlock condition might occur when a robotic system is physically connected to another system and rapid maneuver may result in damage to one or both robots. Finally, there are many circumstances where a cyber-physical system may be acting with a human teammate or under supervision of a human. In these cases, a human supervisor may have roles such as turning the RTA off, changing the RTA risk or sensitivity level, or manually activating the RTA recovery control function.

\subsection{Simplex RTA Safety Monitor Approaches}
As discussed previously, approaches may be considered implicit or explicit based on whether they rely on simulating backup trajectories online.
Approaches to the design of a Simplex-based RTA safety monitor tend to fall in categories of \textit{boundary violation} (i.e. explicit), \textit{trajectory prediction} (i.e. implicit), \rev{or} \revised{\textit{reach-tube prediction} (i.e. nondeterministic)}. Boundary violation monitors activate a recovery controller after a safety boundary has been violated. These designs often feature a buffer to allow a recovery response time to complete before violating the actual safety boundary. 
\begin{figure}[htb]
\begin{center}
\includegraphics[width=0.5\textwidth]{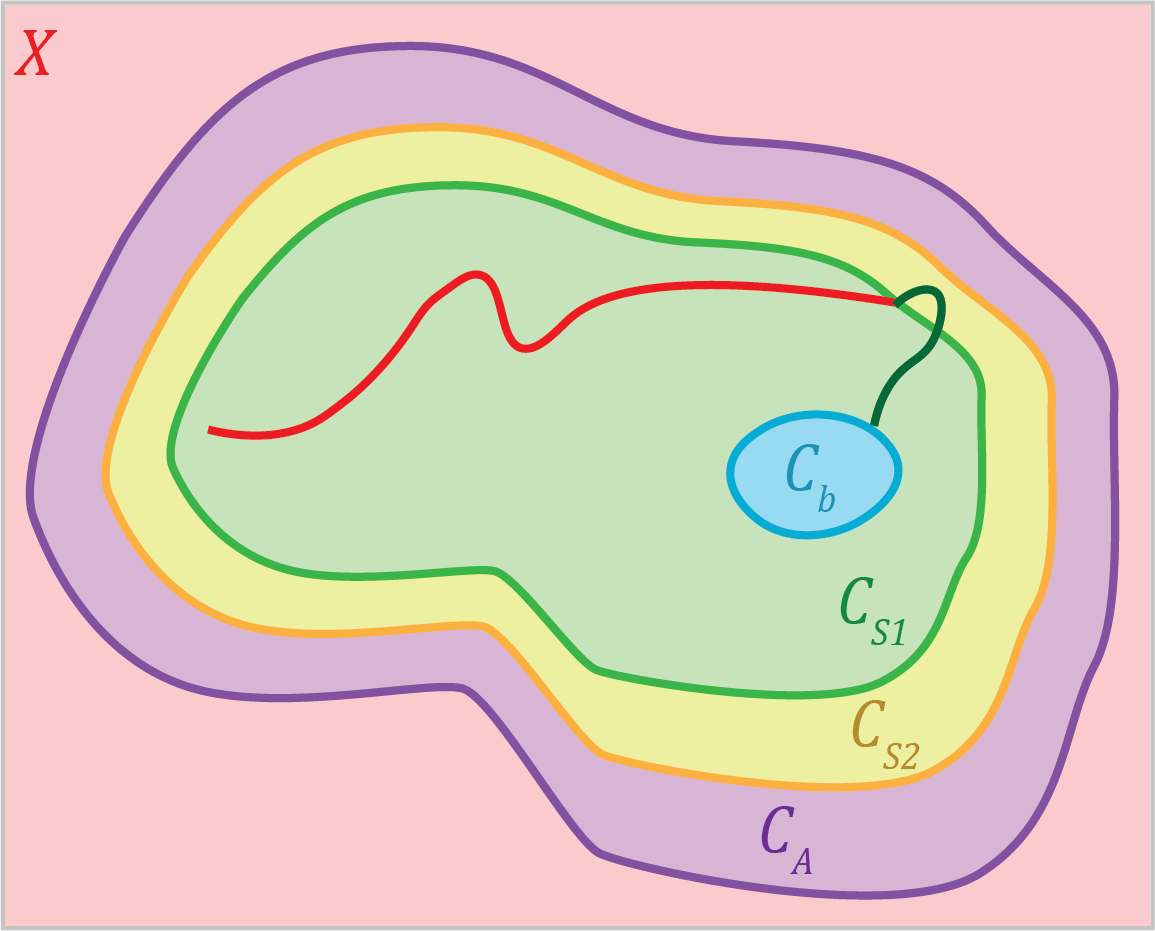}
\caption{The RTA ``Ameoba" diagram depicts the entire state space $X$, containing the set of allowable states $\Ca$ in purple, a safety buffer region $C_{S2}$, and a safe zone with without switching $C_{S1}$. The trajectory of a primary controller is depicted in red. When the trajectory crosses the switching condition boundary between $C_{S1}$ and $C_{S2}$, the RTA activates the backup controller and the green trajectory shows backup control to a safe region $\Cb$. Image inspired by \cite{schierman2020runtime}.} 
\label{fig:RTAAmoeba} 
\end{center}
\end{figure}
Trajectory prediction monitors predict the trajectory that the recovery would take if it were to activate and if that trajectory prediction intersects the safe boundary with a small buffer, the recovery maneuver will activate \cite{eller2013test,wadley2013development}. \revised{Reach-tube-based monitors study a set of possible future states, arising from uncertainty in the dynamics, and in instances where this prediction intersects the unsafe set the RTA recovery function will activate to ensure safety \cite{bak2019efficient,hobbs2018arch}.}



\subsection{Simplex RTA as a Near-Term Certification Path}

Within the last five years progress has been made towards developing certification criteria for RTA \cite{hook2016certification} based on the Simplex architecture resulting in the publication of \rev{the first} ASTM standard for the use of RTA in unmanned aircraft \cite{ASTM3269-17}. \rev{A second iteration \cite{ASTMF3269-21}} of this standard \rev{opens the door for alternative filtering RTA designs}.


\subsection{Provably Safe RTA with Black Box Backup Controllers } 
Implicit RTA systems rely on online simulations of a backup controller in order to assess safety.
\revised{An important consequence of this architectural choice is that the backup controller need not be guaranteed to always produce a safe solution in order to guarantee safety of the RTA system.
That is, in the event that the backup controller fails to compute a new safe trajectory at any time, it can follow the most recently found safe trajectory until it reaches the backup set. In this sense, the job of the backup controller is to search for and \emph{propose} candidate safe trajectories. A monitor can be used to determine whether this candidate trajectory is both feasible and safe. }
A practical consequence of this is that high performing black-box controllers can safely be used to generate backup trajectories. 
This concept has been used in practice \cite{Ariadne_Olatunde2021} and is thoroughly explored for the context of Simplex-based systems in \cite{mehmood2021_blackBoxSimplex}. 
Furthermore, the idea extends to other methods of RTA. For example, in \cite{gurriet2019scalable} an ASIF is constructed which utilizes a backup controller taking the form of a neural network approximation of an optimal control policy.

\subsection{Canonical Algorithms   } 

Included in this section are a set of canonical algorithms for Simplex-based RTA. 
In this case, we consider 
the deterministic discrete time system, 
\begin{equation} \label{eq: discrete_time_system}
    x(i+1)=F(x(i), u(i))
\end{equation}
where $x\in X \subseteq\R^n$ denotes the state, $u\in U \subset \R^m$ denotes the control input and $i\in\mathbb{Z}$ denotes the time index. 
A constraint set is defined as $\Ca\subseteq X$. 
It is assumed that a backup control law $\ub:\X\to U$ is defined and we let $\phi^{\ub}(i; x)$ be the state of \eqref{eq: discrete_time_system} reached at $i\geq0$ when beginning at state $x$ at time $i=0$ and evolving under $\ub$. 

The idea behind the algorithms in this section is to assess the safety of a \emph{probe step}. That is, at any state $\xcur\in X$, whether to accept or reject the desired control input $\ud\in U$ is based on whether the candidate next state reached under this input $\xcand:=F(\xcur, \ud)$ is safe. 
In the first case, an explicit safe set is assumed to be defined. The second case considers an implicit approach, which relies on online simulations under the backup dynamics. 
A deterministic discrete time system is assumed for clarity of the algorithms and their associated guarantees. 
However, the algorithms may be adapted to continuous time systems, to multiple backup controllers, latched implementations, etc.

\subsubsection{Explicit Simplex Filter   } 

Algorithm \ref{alg: RBSF} describes an explicit Simplex-based RTA algorithm, the  Region-Based Simplex Filter (RBSF). 
In addition to a backup control law $\ub$, it is assumed that a safe set $\Cs\subseteq\Ca$ is defined such that $\Cs$ is invariant under $\ub$.
The algorithm renders the RTA system forward invariant in $\Cs$.
This is apparent through the observations that (i) by nature of the invariance property, applying $\ub$ anywhere in $\Cs$ will lead to a next state that is still in the set (ii) the sample step allows for inputs that would cause a departure from $\Cs$ to be detected and replaced with $\ub$. 
In practice, one may choose to add conservatism to the approach by working with  any \revised{underapproximation} $\tilde{\Cs} \subseteq {\Cs}$. The region $\Cs\setminus \tilde{\Cs}$ is often referred to as a \emph{buffer}. 
A practical consideration is that the backup controller should locally attract solutions outside of $\tilde{\Cs}$ to $\tilde{\Cs}$. 
That is, it is favorable to ensure that a recovery is possible in the case where the state is perturbed outside of the safe region. 
One way to do this is to ensure that all initial conditions in the buffer will lead to $\tilde{\Cs}$ if $\ub$ is applied; \eg this is the case when $\Cs$ and $\tilde{\Cs}$ are both sublevel sets of a Lyapunov function. 


\begin{algorithm}[t]
\caption{Region-Based Simplex Filter (RBSF)}
\begin{algorithmic}[1]
\setlength\tabcolsep{0pt}
\Statex
\begin{tabulary}{\linewidth}{@{}LLp{10cm}@{}}
    \textbf{input}&:\:\:& Current State $x_{\rm curr}\in X$ \\
    &:\:\:& Desired Input $\ud \in U$ \\
    \textbf{output}&:\:\:& Safe Control Input $\ua \in U $\\
    \textbf{predefined}&:\:\:& Constraint set $\Ca \subset X$ \\
    &:\:\:& Invariant Set $\Cs\subset\Ca$ \\
    &:\:\:& Backup Control Law $\ub: \X \to U$.\\
&& 
\end{tabulary}
\Function{$ \ua=$RBSF}{$x_{\rm curr}$, $\ud$}
\State   $ x_{\rm cand}\leftarrow F(x_{\rm curr}, \ud )$
\If{$\xcand \in \Cs$}
\State \textbf{return} $\ud$ 
\Else{} 
\State \textbf{return} $\ub$
\EndIf 
\EndFunction
\end{algorithmic}
\label{alg: RBSF}
\end{algorithm}

\begin{example}
    Consider the double integrator system described by \eqref{eq: double_integrator_syst} discretized over a period of 0.1 s, and consider the constraint set \eqref{eq: double_integrator_Ca}. The backup control law $\ud=-1$ renders the set $\Cs=\{x\in\R^2\,\,|\,\,-2x_1-x_2^2\geq 0\}$ forward invariant and it is clear that $\Cs\subseteq\Ca$. Figure \ref{fig: double_int_explicit_simplex} shows a simulation of the RBSF algorithm with these parameters under the desired control $\ud=1$.  
\end{example}


\subsubsection{Implicit Simplex Filter  } 

Algorithm \ref{alg: SBSF} describes an implicit Simplex-based algorithm, the Simulation-based Simplex Filter (SBSF). 
It is assumed that a backup set $\Cb\subseteq\Ca$ is defined such that $\Cb$ is invariant under a backup control law $\ub:X\to U$. 
No assumptions are made on the structure of $\ub$, it may come from a closed form control expression, a black box function, an MPC, a motion primitive, etc. 
In contrast to the RBSF algorithm, which constrains the system to evolve in the explicitly defined invariant region, SBSF uses the smaller invariant set $\Cb$ as a means to 
access a larger implicitly defined invariant set.  
Specifically, SBSF constrains the system to the safe backward image of $\Cb$: 
\begin{equation}
    \Cs = \{x\in X \, | \, \phi^{\ub}(i; x)\in \Ca \,\, \forall i \in \{0,...,N-1\}, \,\, \phi^{\ub}(N; x)\in \Cb \}
\end{equation}
where, $\phi^{\ub}(i;x)$ is the $i^{th}$ point in an $N+1$ point trajectory obtained from simulating under the dynamics \eqref{eq: discrete_time_system} and backup controller $\ub$. 
Since $\Cs$ is forward invariant under $\ub$, the algorithm renders the RTA system forward invariant in $\Cs$. 
As in the case of the RBSF algorithm, this becomes apparent from the observations that (i) applying $\ub$ anywhere in $\Cs$ will lead to a state that is also in $\Cs$ and (ii) the sample step under $\ud$ can reliably be used to detect inputs that would cause a departure from $\Cs$. 
Conservatism can be added to this approach by using under-approximations of $\Ca$ and $\Cb$. 
Note that the above algorithm assumes a deterministic system and relies on simulating single trajectories under the backup dynamics.
This can be extended to the nondeterministic case by simulating an overapproximation of the forward reachable set under the backup dynamics and requiring that $\Cb$ is robustly forward invariant. 

\begin{algorithm}[t]
\caption{Simulation-Based Simplex Filter (SBSF)}
\begin{algorithmic}[1]
\setlength\tabcolsep{0pt}
\Statex
\begin{tabulary}{\linewidth}{@{}LLp{10cm}@{}}
    \textbf{input}&:\:\:& Current State $x_{\rm curr}\in\Rn$ \\
    &:\:\:& Desired Input $\ud \in U$ \\
    \textbf{output} &:\:\:& Safe Control Input $\ua \in U $\\
    \textbf{predefined} &:\:\:& Constraint Set $\Ca \subset X $ \\
    &:\:\:& Invariant Backup Set $\Cb\subseteq \Ca$ \\
    &:\:\:& Backup Control Law $\ub: \X \to U$.\\
&& 
\end{tabulary}
\Function{$ \ua=$SBSF}{$x_{\rm curr}$, $\ud$}
\State   $ x_{\rm cand}\leftarrow F(x_{\rm curr}, \ud )$
\State \textbf{compute:} $\phi^{\ub}(i; x_{\rm cand}) \quad \forall i \in \{0,\dots, N \}$
\If{$\phi^{\ub}(i; x_{\rm cand})\in \Ca \quad \forall i \in \{ 0, \dots, N-1 \}$ \,\,{\rm AND} \,\, $\phi^{\ub}(N; x_{\rm cand})\in \Cb$}
\State{\textbf{return} $\ud$} 
\Else{} 
\State \textbf{return} $\ub$
\EndIf 
\EndFunction
\end{algorithmic}
\label{alg: SBSF}
\end{algorithm}

\begin{example}
    Consider the double integrator system described by \eqref{eq: double_integrator_syst} discretized over a period of 0.1 s, and consider the constraint set \eqref{eq: double_integrator_Ca}. The backup control law $\ud=-1$ renders the set $\Cb=\{x\in\R^2\,\,|\,\,-x_1\geq 0,\,-x_2\geq0\}$ forward invariant and it is clear that $\Cb\subseteq\Ca$. Figure \ref{fig: double_int_implicit_simplex} shows a simulation of the SBSF algorithm with these parameters under the desired control $\ud=1$.  
\end{example}

\begin{figure*}[t!]
    \centering
    \begin{subfigure}[b]{.5\columnwidth}
\begin{center}
\includegraphics[width=0.995\columnwidth, trim=0 11.7cm 0 2.6cm, clip]{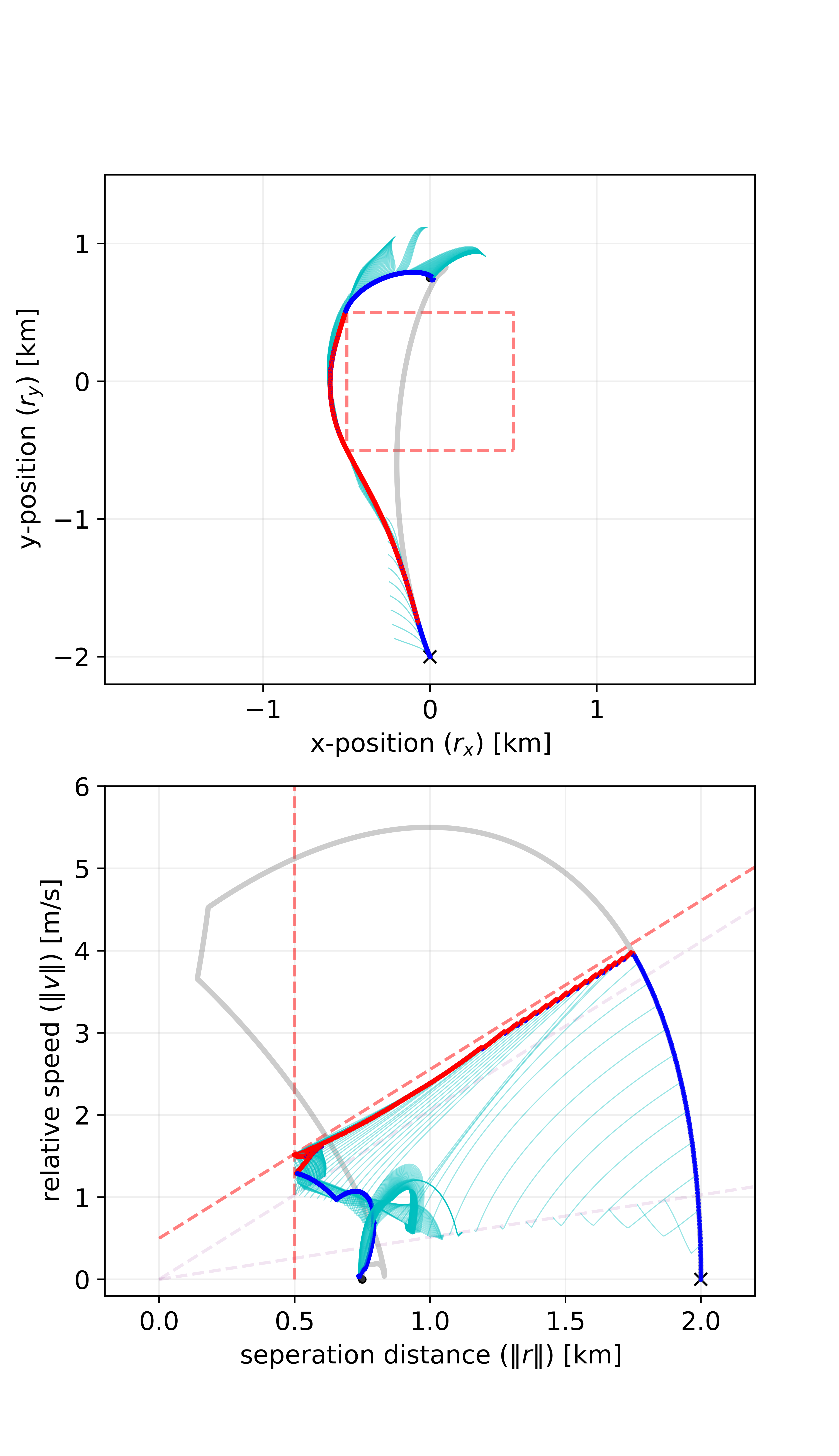}
\end{center}
    \end{subfigure}%
    ~ 
    \begin{subfigure}[b]{0.5\columnwidth}
\begin{center}
\includegraphics[width=0.995\columnwidth, trim=0 1.5cm 0 12.6cm, clip]{Figures/rta_trajectory_plot_v3.png}
\end{center}
    \end{subfigure}
\caption{Projected trajectories \rev{of a chaser spacecraft relative to a target spacecraft at the origin under Clohessy-Wiltshire-Hill dynamics, described by (\ref{eq: CWH_dynamics})} over a 2400 s period. \rev{For safety, the chaser spacecraft is required to stay at least 0.5 km away from the target in both the x and y directions (shown by dashed red lines on the left), and is required to stay below a dynamic velocity limit (shown by dashed red lines in the figure on the right) where the spacecraft's relative velocity must slow down as it gets closer.} The \rev{chaser spacecraft} trajectories without RTA are grey. The trajectories with \rev{a simulation-based Simplex filter} RTA are blue when $\ua=\ud$, and red when $\ua=\ub$. The backup trajectories (cyan) are shown at 8 s intervals.  }
\label{fig: NMT_SBSF}
\end{figure*}

\begin{example}(From \cite{mote2021_NMT})
    Consider the system \eqref{eq: CWH_dynamics} discretized in time to have a 1~s resolution, and with $\Ca=\{x\in\R^5\,|\,\Vert r\Vert_\infty-r_{\min}\geq 0, \, \kappa_1+\kappa_2\Vert r \Vert_\infty - \Vert v \Vert_\infty \geq 0 \}$ 
    where $r:=[x_1,x_2]^T$, and $v:=[x_3,x_4]^T$ are the relative position and speed vectors respectively.
    The backup controller $\ub$ is taken as the first input in the solution to a mixed integer quadratic program (MIQP); solved subject to the constraints that (i) all intermediate points lie in $\Ca$ and (ii) the endpoint lies in a set $\Cb$, that is defined from the set of natural motion trajectories lying outside of $\Ca$. In this case $x\in\Cs$ is true whenever the MIQP has a feasible solution with initial state $x$. 
    Figure~\ref{fig: NMT_SBSF} shows a simulation of the RTA system with a primary controller that is designed to drive the system to an invariant point along the $y$-axis. 
    The simulation uses parameters $m=50$ kg, $n=0.001027$ ${\rm rad/s}$, and $u_{\rm max}=0.5$ N,  $r_{\rm min}=0.5\,$km, $\kappa_1=0.5\,$m/s, and $\kappa_2=2\,n$ ${\rm s^{-1}}$. 
\end{example}

\section{Active Set Invariance Filtering} \label{sec: ASIF}


Active set invariance filtering (ASIF) methods are RTA approaches associated with a class of first-order, optimization-based algorithms. 
As with many Simplex-based techniques they are system agnostic, provably correct, and exhibit a number of robustness properties which make them attractive in practice. 
In addition, ASIFs are minimally invasive with respect to the safety constraints. 
This results in a smoother and more gradual intervention than with approaches that rely on switching to backup controllers directly. 

This section considers systems of the form
\begin{equation} \label{eq: control_affine_dynamics}
    \dot{{x}} = f({x}) + g({x}){u}. 
\end{equation}
When \eqref{eq:matt} is of the form \eqref{eq: control_affine_dynamics}, the dynamics are said to be \emph{control affine}, and it is assumed throughout the following that   $f({x}):\mathbb{R}^n\to\mathbb{R}^n$ and $g({x}):\mathbb{R}^n\to\mathbb{R}^{n\times m}$ are locally Lipshitz continuous in their inputs so that, in particular, solutions to  \eqref{eq: control_affine_dynamics} are unique when they exist.
ASIFs are constructed as quadratic programs, where the objective function is typically the $l^2$ norm of the difference between the desired input $\ud$ and actual input $\ua$, and where the constraints on the program, which are known as  \emph{barrier constraints}, take the form $BC_i(x,u) =a_i(x) u + b_i(x) \geq 0$, $i \in \{1, ..., N\}$.  
The canonical ASIF controller is given below as ASIF-QP: 
\begin{samepage}
\noindent \rule{1\columnwidth}{0.7pt}
\noindent \textbf{ASIF-QP}
\begin{equation}\label{eq: asif_QP}
    \ua(x,\ud) = \text{argmin}_{u \in U} \Vert u - \ud \Vert_2^2 
\end{equation}
\begin{equation}\label{eq: ASIF_QP_barrier_constraint}
\begin{array}{cc}
    \text{s.t.} &   BC_i(x,u)=a_i(x) u + b_i(x) \geq 0,  \qquad i=1,...,N\\
\end{array}
\end{equation}
\noindent \rule{1\columnwidth}{0.7pt}    
\end{samepage}

The ASIF-QP above guarantees system safety with respect to an operational region $\Cs$ when the properties \revised{$\vartheta^{\rm BC}_{1}$ and $\vartheta^{\rm BC}_{2}$} are satisfied by the barrier constraints:
\begin{itemize}
    \item $\vartheta^{\rm BC}_{1}$\,: if \eqref{eq: ASIF_QP_barrier_constraint} is satisfied then the system is forward invariant in some set $\Cs\subseteq\Ca$. \emph{i.e.}
    \begin{equation} \label{eq: innocuity_requirement_for_BC}
       \revised{\vartheta^{\rm BC}_{1}:\:} BC_i(x,u)\geq 0,\; \forall i\in\{1,...,N\} \implies \Cs \text{ \revised{is} forward invariant}
    \end{equation}
    where $\Cs\subseteq\Ca$.
    \item $\vartheta^{\rm BC}_{2}$\,:  it is possible to satisfy \eqref{eq: ASIF_QP_barrier_constraint} from any state in the set $\Cs$. \emph{i.e.} $\forall x \in \Cs$, $\forall i\in \{1,\dots,N\}$, 
    \begin{equation} \label{eq: viability_requirement_for_BC}
        \revised{\vartheta^{\rm BC}_{2}:\:} \sup_{u\in U}[BC_i(x,u)] \geq 0
    \end{equation}  
\end{itemize}
The first property ensures that the  RTA mechanism will bound trajectories to the set $\Cs$, as long as there exists a feasible input satisfying the barrier constraint. 
The second property ensures that for all states in $\Cs$, it is possible to find a feasible input that satisfies the barrier constraint.
Since quadratic programs can be solved in such a way that feasible solutions are found whenever they exist, guaranteeing the existence of a solution is sufficient for ensuring feasibility of the ASIF-QP.
When both properties are satisfied, 
$\Cs$ defines a safe set with respect to the ASIF-QP. 
Moreover when the barrier constraints satisfying the properties above, the ASIF-QP is \emph{minimally invasive} in the sense that the $l^2$ distance between $\ua$ and $\ud$ is minimized, subject to the barrier constraints.

As with the case of Simplex, implicit and explicit approaches emerge for ASIF based on how the safe operational region $\Cs$ is identified. 
First, \emph{explicit} ASIFs are studied. 
In this case, $\Cs$ obtained by identifying a control invariant set explicitly as the super zero level set of a smooth function. 
Next, in the case of \emph{implicit} ASIFs, a control invariant set is identified implicitly through the trajectories of a backup control law.  
A more thorough review of explicit ASIF approaches is provided in \cite{ames2019control} and implicit ASIF approaches in \cite{gurriet2020scalable}.  See also \cite{gurriet2020applied} for a review of both implicit and explicit ASIF approaches.

\subsection{Barrier Constraints from an Explicitly Defined Safe Set}

Consider a set $\Cs$ that is defined as the super zero-level set of a continuously differentiable function $h:\mathbb{R}^n \to \mathbb{R}$, \emph{i.e.}, 
\begin{align} \label{eq:practical_sets}
    \Cs &= \{x\in\mathbb{R}^n \, | \, h(x)\geq 0 \} \\ 
    \partial\Cs &=  \{x\in\Cs \, | \, h(x) = 0 \}
\end{align}
and suppose that $0$ is a regular value of $h$; \ie the gradient of $h$ does not vanish on the boundary. 
The goal is to design a barrier constraint $BC(x,u)$ that satisfies  $\vartheta^{\rm BC}_1$ and $\vartheta^{\rm BC}_2$, and consequently makes the set $\Cs$ forward invariant under the ASIF-QP. 
One method for ensuring the forward invariance of $\Cs$ involves checking subtangentality condition over the boundary of $\Cs$. Sometimes called Nagumo's condition, this result is as follows: if  
\begin{equation} \label{eq:barrier_constraint_without_buffer}
    L_f h( {x}) + L_g h( {x})  {u}(x) \geq 0
\end{equation}
holds for all $x \in \partial\Cs$, then $\Cs$ is forward invariant for the closed-loop dynamics of \eqref{eq: control_affine_dynamics} under $u(x)$, where $L_f, \, L_g$ in \eqref{eq:barrier_constraint_without_buffer} denote Lie derivatives of $h$ along $f$ and $g$, respectively. Informally, \eqref{eq:barrier_constraint_without_buffer} ensures that $\dot{h}(x) \geq 0$ for all $x \in \partial\Cs$ and, thus, any control law $u(x)$ satisfying \eqref{eq:barrier_constraint_without_buffer} ensures the forward invariance of $\Cs$ when applied to \eqref{eq: control_affine_dynamics}.


\begin{figure}[h] 
    \centering
    \includegraphics[width=0.9\textwidth]{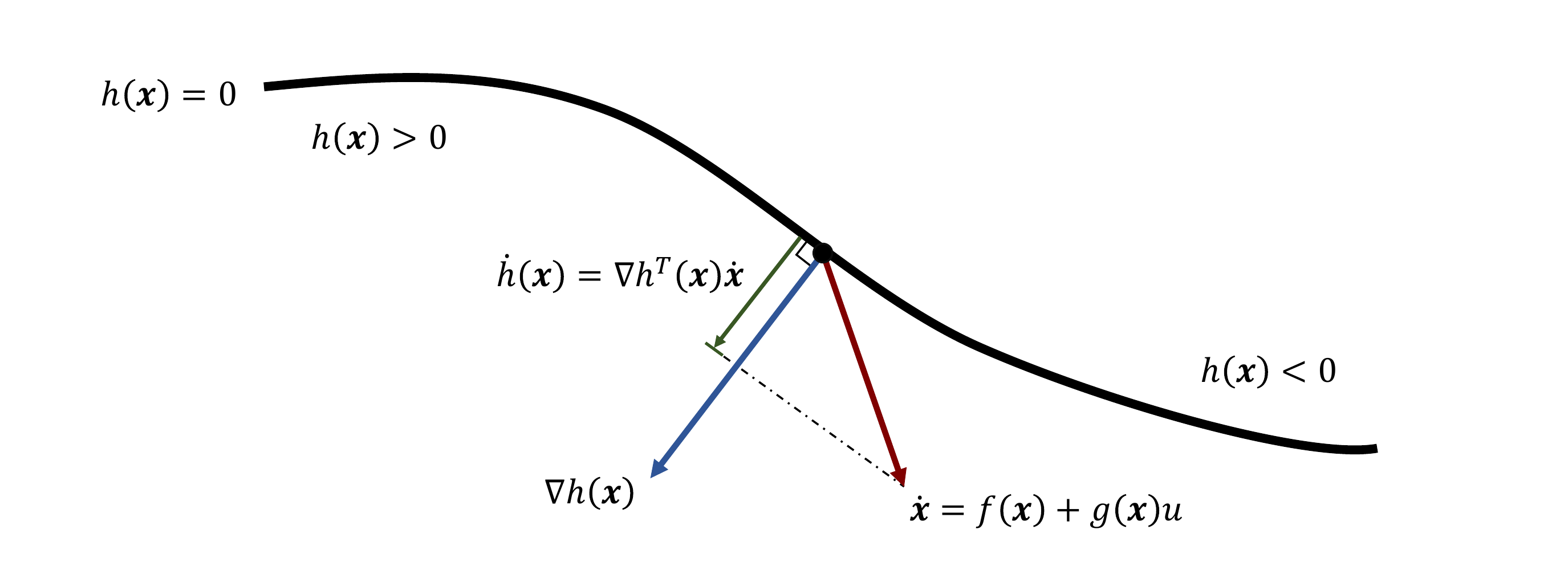}
    \caption{Barrier geometry for constraint function $h:\R^2 \to \R$.}
    \label{fig: barrier_geometry_2d}
\end{figure} 

The constraint \eqref{eq:barrier_constraint_without_buffer} is not very practical for use directly in an optimization framework as it is activated exactly on the boundary of $\Cs$, which is a region without volume. 
The solution presented in \cite{aubin2009viability} is to use similar condition that is active everywhere in $\Cs$, and that includes a strengthening term which relaxes the constraint away from the boundary: for all $x \in \Cs$
\begin{equation} \label{eq:barrier_constraint_with_buffer}
    L_f h(x) + L_g h(x) u(x) + \alpha(h(x)) \geq 0 
\end{equation}
where $\alpha:\mathbb{R}\to\mathbb{R}$ is an extended class $\kappa_\infty$ function. 
A continuous function $\alpha : \mathbb{R}\rightarrow \mathbb{R}$ is class-$\kappa_\infty$ if $\alpha$ is strictly increasing and $\alpha(0) = 0$. 
Since $\alpha(h(x))=0$ on the boundary, satisfaction of \eqref{eq:barrier_constraint_with_buffer} implies satisfaction of  \eqref{eq:barrier_constraint_without_buffer}. Hence, \eqref{eq:barrier_constraint_with_buffer} implies $\Cs$ is forward invariant for the closed-loop dynamics of \eqref{eq: control_affine_dynamics} under $u(x)$. 
For this reason, an attractive method for constructing the ASIFs \eqref{eq: asif_QP}--\eqref{eq: ASIF_QP_barrier_constraint} is to form barrier constraints using 
\begin{equation} \label{eq: expicit_barrier_constraint}
    BC(x, u) = L_f h(x) + L_g h(x) u + \alpha(h(x)).
\end{equation}
Applying the previous theoretical results, the resulting ASIF ensures the forward invariance of $\Cs$ and therefore satisfies $\vartheta^{\rm BC}_1$, as discussed above.

In order to satisfy the requirement $\vartheta^{\rm BC}_2$, the functions $h$, $\alpha$ must be chosen in such a way that ensures that there always exists $u\in\U$ satisfying the barrier constraint; equivalently,  
\begin{equation} \label{eq: viability_req_for_explicit_ASIF}
    \sup_{u\in U}[L_f h(x) + L_g h(x) u] \geq -\alpha(h(x)).
\end{equation}
must hold for all $x \in \Cs$.
Though it is on occasion possible to obtain $h,\,\alpha$ directly through intuitive reasoning, satisfying this requirement is accomplished in general
through the computation of a \emph{control invariant} subset of $\Ca$. 
Specifically, if $\Cs$ is a control invariant set, then it is always possible to find a strengthening function $\alpha(x)$ such that the viability property is satisfied \cite{gurriet2018ASIF}. 
\revised{In practice, $\alpha$ often takes the form of a linear or odd polynomial function, and can to an extent be chosen in a way that shapes the response of the RTA mechanism to have some desired behavior. For example, a common trade-off that occurs when shaping this function is in choosing between a more gradual response that activates further inside the safe set, and a more abrupt intervention activating closer to the boundary of $\Cs$.  }
Alternatively, one may bypass choosing $\alpha$ altogether with a relaxed QP formulation (see \cite{gurriet2018ASIF}, \cite{gurriet2020applied} for more details). 
Identifying a large control invariant subset $\Cs\subseteq\Ca$ is the key challenge in implementing this method. 

\begin{example}
    Consider the double integrator system described by \eqref{eq: double_integrator_syst} and the constraint set \eqref{eq: double_integrator_Ca}. A control invariant set is given by the super zero level set of $h(x) = -2x_1-x_2^2$ and given a class $\kappa_\infty$ function $\alpha$, the barrier constraint is $-2 x_2(1+u)+\alpha( -2x_1-x_2^2) \geq 0$. Note that $\forall x\in\Cs,\,\,\alpha(h(x))\geq 0$ and $\sup_{u\in U}[L_f h(x) + L_g h(x) u]=0$, hence the barrier constraint is viable. 
\end{example}

\begin{example}
    Consider the dynamics of a spacecraft in unconstrained rotation,
    \begin{equation} \label{eq: sc_rotational_dynamics }
        \dot{x} = J^{-1}(-x \times J x) + J^{-1}u
    \end{equation}
where $x\in \R^3$ represents the angular velocity, $J=\text{diag}(J_1,J_2,J_3)\in\R^{3\times 3}$ is a diagonal inertia matrix and $u\in [-1,1]^3 $. 
A control invariant set for this system is given by $\Cs = \{x\in \R^3 \, | \, h(x) \geq 0\}$ with, 
\begin{equation}
    \label{eq: example_hx_KE}
    h(x) = K - x^T J x 
\end{equation}
for any $K \in \R_+$. The fact that this set is control invariant is made apparent with the observation that the rotational kinetic energy $x^T J x$ is constant when $u=0$. 
Choosing $\alpha(x)=x$ the barrier constraint is $\nabla h^T(x)\dot{x} + h(x) \geq 0$, which simplifies to, 
\begin{equation}
    K-x^T J x - 2x^T u \geq 0 
\end{equation}
It is easy to show that this constraint is viable everywhere within $\Cs$. Namely,  $K-x^TJx \geq 0$ everywhere in $\Cs $, and one can always find $u\in U $ such that $-2x^T u \geq 0 $.

\end{example}


If there exists an extended class $\kappa_\infty$ function satisfying \eqref{eq: viability_req_for_explicit_ASIF} for all $x\in X$ then $h(x)$ is said to be a \emph{control barrier function} \cite{ames2019control, ames17_cbf_based_QPs, xu2015_robustness_of_CBFs}.
Barrier functions were first developed in the context of nonlinear and hybrid system verification \cite{prajna2004_safety_verification_of_hydrid, prajna2006barrier,maghenem2019multiple}.
The idea was first extended to the RTA problem with quadratic programs in \cite{ames2014_adaptive_cruise_control}, and it was developed more in \cite{ames17_cbf_based_QPs}. 
In recent years, there has been a surge in interest in this topic. 
The theory has been extended to relax assumptions on smoothness \cite{glotfelter2017nonsmooth}, to discrete time\cite{agrawal2017discrete} stochastic \cite{stochastic_cbfs} and nondeterministic  systems, and to enforce temporal logic specifications \cite{mohit_cbf_for_TL}\cite{lindemann2018control}\cite{yang2020continuous}.  
Likewise, the practicality of the approach is demonstrated in a variety of real world applications, 
as shown in Table \ref{t: classification_of_RTA_algs}. 
It is important to note that in order for \eqref{eq: expicit_barrier_constraint} to depend on the control input $u$, the constraint function $h(x)$ must have a relative degree of one; \ie $L_g h(x) \not = 0$. 
Safety constraint functions with higher relative degrees are addressed in the literature on  exponential control barrier functions \cite{nguyen2016exponential}, and high order control barrier functions \cite{Tan2021_higher_order_CBFs, belta2019_higher_order_CBFs}.

\begin{example}
    Consider the system $\ddot{x}=u$ 
    with $u\in U=(-\infty,\infty)$ 
    and the constraint function $h(x)= -x_1$. 
    Note that while actuation is unlimited in this case, it is not possible to enforce the barrier constraint $L_f h(x) + L_g h(x) u \geq \alpha(h(x))$ as $L_g h(x) = 0$.  
\end{example}






\begin{algorithm}[t]
\caption{Explicit Active Set Invariance Filter}
\begin{algorithmic}[1]
\setlength\tabcolsep{0pt}
\Statex
\begin{tabulary}{\linewidth}{@{}LLp{10cm}@{}}
    \textbf{input}&:\:\:& Current State $x_{\rm curr}\in X$ \\
    &:\:\:& Desired Input $\ud \in U$ \\
    \textbf{output}&:\:\:& Safe Control Input $\ua \in U $\\
    \textbf{predefined}&:\:\:& Allowable Set $\Ca \subset X$ \\
    &:\:\:& Invariant Set $\Cs\subset\Ca$ \\
    &:\:\:& Extended Class $\kappa_\infty$ Function $\alpha:\R\to\R$ \\
&& 
\end{tabulary}
\Function{$ \ua=$EASIF}{$x_{\rm curr}$, $\ud$}
\State  \textbf{solve:} \eqref{eq: asif_QP} subject to the constraint \eqref{eq: expicit_barrier_constraint}. 
\State \textbf{return} $u_{\rm act}(\xcur,\ud)$.  
\EndFunction
\end{algorithmic}
\label{alg: EASIF}
\end{algorithm}


\subsection{Barrier Constraints from an Implicitly Defined Safe Set}

As discussed in the previous section, the barrier constraint is viable when it restricts the system to a control invariant set. 
If an explicit representation of a large control invariant subset of $\Ca$ can be obtained, then with the appropriate choice of $\alpha$, a valid barrier constraint is obtained through \eqref{eq: expicit_barrier_constraint}. 
The key idea behind the implicit approach to ASIF \cite{gurriet2018ASIF, gurriet2019scalable, gurriet2020scalable} is to filter with respect to an implicitly defined control invariant set.  
Specifically, given a backup control law $\ub$ and a backup set $\Cb$ that is invariant under $\ub$, one can develop barrier constraints that activate near the boundary of the safe backward image of $\Cb$; see \eqref{eq: Safe_Backward_Image}. The idea was first proposed in \cite{gurriet2018ASIF} and further developed in \cite{gurriet2019scalable, gurriet2020scalable}.
A tutorial on the topic is provided in \cite{chen2021_backup_CBFs}, and some notable applications are listed in Table~\ref{t: classification_of_RTA_algs}. 
A key advantage to this approach is that it does not require the computation of a large control invariant set. 

Consider a smooth backup control law $\ub:\R^n \to U$ and suppose that the constraint set is given by, 
\begin{equation}
    \Ca := \{ x\in\R^n \,|\, \varphi(x)\geq 0 \}
\end{equation}
and a backup set $\Cb\subseteq\Ca$ is given by, 
\begin{equation}
    \Cb = \{ x\in \R^n \,|\, h_{\rm b}(x)\geq 0 \}.
\end{equation}
The safe backward image of $\Cb$ is 
\begin{equation}
    \Cs = \{x\in\R^n\,|\, \forall t\in[0,T_{\rm b}],\,\, \varphi(\phi^{\ub}(t;x))\geq 0\, \wedge \, h_{\rm b}(\phi^{\ub}(T_{\rm b};x))\in\mathcal{C}_{\rm b} \}.
\end{equation}
where $\phi^{\ub}(t;x)$ is the flow of \eqref{eq: control_affine_dynamics} under $\ub$, evaluated $t$ units of time from $x$. 
Given an extended class $\kappa_\infty$ function $\alpha$, it can be shown that the following constraints are sufficient for invariance in $\Cs$:
\begin{align}
    \frac{{\rm d}\varphi(t_{\rm b};x))}{{\rm d}t}
    + \alpha(\varphi( \phi(t_{\rm b};x) ) ) \geq 0 \label{eq: iasif_intermediate_BC_preexpansion}   \\ 
     \frac{{\rm d}h_{\rm b}(\phi(T_{\rm b};x))}{{\rm d}t} + \alpha(h_{\rm b}(\phi(T_{\rm b};x) ) \geq 0  \label{eq: iasif_terminal_BC_preexpansion}  
\end{align}
for all $t_{\rm b}\in[0,T_{\rm b}]$. The constraint \eqref{eq: iasif_intermediate_BC_preexpansion} enforces that the points along the backup trajectory stay in the constraint space $\Ca$, and \eqref{eq: iasif_terminal_BC_preexpansion} enforces that the endpoint of the backup trajectory stay in $\Cb$. 
Expanding the derivative terms yields the implicit ASIF barrier constraints:
\begin{align}
    \nabla \varphi (\phi^{u_b}(t_{\rm b};x)) D\phi^{u_b}(t_{\rm b};x)[f(x)+g(x)u   ]
    + \alpha(\varphi( \phi^{u_b}(t_{\rm b};x) ) ) &\geq 0 \label{eq: iasif_intermediate_BC}   \\ 
    \nabla h_{\rm b} (\phi^{u_b}(T_{\rm b};x))  D \phi^{u_b}(T_{\rm b};x)[f(x)+g(x)u] + \alpha(h_{\rm b}( \phi^{u_b}(T_{\rm b};x) ) ) &\geq 0  \label{eq: iasif_terminal_BC}  
\end{align}
for all $t_{\rm b}\in[0,T_{\rm b}]$. 
Note that since $t_{\rm b}$ lives in the interval $[0,T_{\rm b}]$, \eqref{eq: iasif_intermediate_BC} represents an uncountable number of constraints. 
In practice one may approximate the constrained set of states by numerically integrating \eqref{eq: control_affine_dynamics} forward under $\ub$ and evaluating $\phi^{\ub}(t;x)$ at discrete times $t_{\rm b}\in\{0,t_1,...,t_{\rm N-1},T_{\rm b}\}$. 
It is shown in \cite{gurriet2020scalable} how the finite set of  intermediate constraints may be tightened in such a way that is sufficient for satisfying the constraints with an infinite number of trajectory points. 
One may compute $D\phi^{\ub}(t_{\rm b}; x)$  by integrating along with  \eqref{eq: control_affine_dynamics} a \emph{sensitivity} matrix $Q(t_{\rm b},x)=D\phi^{\ub}(t_{\rm b};x)$. As explained in \cite{seywald2003desensitized}, $Q(t_{\rm b},x)$ is obtained as the solution to the following differential equation, 
\begin{equation} \label{eq: sensitivity_matrix_dynamics}
    \frac{{ \rm d}Q(t_{\rm b},x)}{{\rm d}t_{\rm b}} = Df_{\rm cl}(\phi^{\ub}(t_{\rm b};x))Q(t_{\rm b},x)
\end{equation}
with $Q(0,x)=I_{n\times n}$ and where $D f_{\rm cl}(\phi^{\ub}(t_{\rm b};x))$ is the Jacobian of the closed-loop dynamics of \eqref{eq: control_affine_dynamics} under the control law $\ub(x)$, evaluated at $\phi^{\ub}(t_{\rm b};x)$. 
In order for this term to be defined, it is necessary that the control law $\ub$ be smooth. As such, it is often necessary to consider smooth saturation functions for bounding the control law to the admissible domain $\U$. 


\begin{algorithm}[t]
\caption{Implicit Active Set Invariance Filter}
\begin{algorithmic}[1]
\setlength\tabcolsep{0pt}
\Statex
\begin{tabulary}{\linewidth}{@{}LLp{10cm}@{}}
    \textbf{input}&:\:\:& Current State $x_{\rm curr}\in\Rn$ \\
    &:\:\:& Desired Input $\ud \in U$ \\
    \textbf{output}&:\:\:& Safe Control Input $\ua \in U $\\
    \textbf{predefined}&:\:\:& Allowable Set $\Ca \subset X $ \\
    &:\:\:& Invariant Backup Set $\Cb\subseteq \Ca$ \\
    &:\:\:& Smooth Backup Control Law $\ub: \X \to U$.\\
    &:\:\:& Extended Class $\kappa_\infty$ Function $\alpha:\R\to\R$ \\
&& 
\end{tabulary}
\Function{$ \ua=$IASIF}{$x_{\rm curr}$, $\ud$}
\State \textbf{compute:} $\phi^{\ub}(t_{\rm b};\xcur) \quad \forall t_{\rm b}\in\{0,t_1,...,t_{\rm N-1},T_{\rm b}\}$ by integrating \eqref{eq: control_affine_dynamics} under $\ub$
\State \textbf{evaluate:} $D f_{\rm cl} (\phi^{\ub}(t_{\rm b};\xcur)) \quad \forall t_{\rm b}\in\{0,t_1,...,t_{\rm N-1},T_{\rm b}\}$ 
\State \textbf{compute:} $D\phi^{u_b}(t_{\rm b},\xcur)\quad \forall t_{\rm b}\in\{0,t_1,...,t_{\rm N-1},T_{\rm b}\}$ by integrating \eqref{eq: sensitivity_matrix_dynamics} 
\State \textbf{solve:} \eqref{eq: asif_QP} subject to the constraints \eqref{eq: iasif_intermediate_BC}-\eqref{eq: iasif_terminal_BC}
\State \textbf{return} $u_{\rm act}(\xcur,\ud)$.  
\EndFunction
\end{algorithmic}
\label{alg: IASIF}
\end{algorithm}

\begin{figure}
    \centering
    \includegraphics[width=1\textwidth]{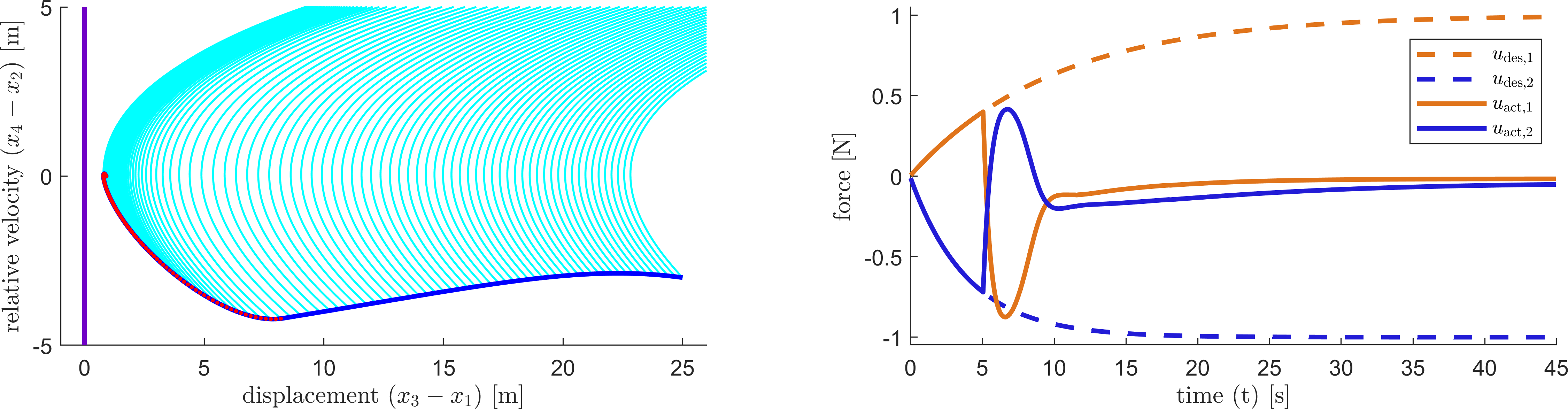}
    \caption{Simulation of Implicit ASIF algorithm for cooperative collision avoidance of two vehicles.}
    \label{fig: two_car_iasif}
\end{figure}

\begin{example}
    Considered here is finite-time safety (collision avoidance) for two vehicles having combined dynamics,  
    \begin{equation} \label{eq: iasif_example_finite_time_two_cart}
        \begin{bmatrix}
        \dot{x_1} \\ 
        \dot{x_2} \\ 
        \dot{x_3} \\ 
        \dot{x_4} 
        \end{bmatrix}
        = 
        \begin{bmatrix}
        0 & 1 & 0 & 0 \\ 
        0 & -b_1 & 0 & 0 \\ 
        0 & 0 & 0 & 1 \\ 
        0 & 0 & 0 & -b_2 
        \end{bmatrix}
        \begin{bmatrix}
        {x_1} \\ 
        {x_2} \\ 
        {x_3} \\ 
        {x_4} 
        \end{bmatrix}
        + 
        \begin{bmatrix}
        0 & 0 \\ 
        1 & 0 \\ 
        0 & 0 \\ 
        0 & 1 
        \end{bmatrix}
\begin{bmatrix}
u_1 \\ 
u_2 
\end{bmatrix}
    \end{equation}
    where $x_1,\, x_3$ denote positions and $x_2, \,$\revised{$x_4$} denote velocities for the vehicles,  $b_1=0.1$ and $b_2=0.25$. 
    The safety constraint is collision avoidance $\varphi(x)=x_3-x_1$ and the constraint set is $\Ca=\{x\in \R^4 \, | \, \varphi(x)\geq 0\} $.
    To simplify the analysis, finite-time safety is considered, with the endpoint constraint \eqref{eq: iasif_terminal_BC} being omitted. 
    The backup controller $\ub=[-1,1]^T$ is integrated over a 10 s horizon and $\alpha(x)=2x$. Figure~\ref{fig: two_car_iasif} shows the result of a simulation with $\ud(t)=[1-\exp(-0.1\,t),\,\exp(-0.25\,t)-1]^T$.
\end{example}

\begin{figure}[]
\centering
\begin{subfigure}[t]{0.5\textwidth}
\centering
\hspace{-16mm}
\includegraphics[height=6.5cm] {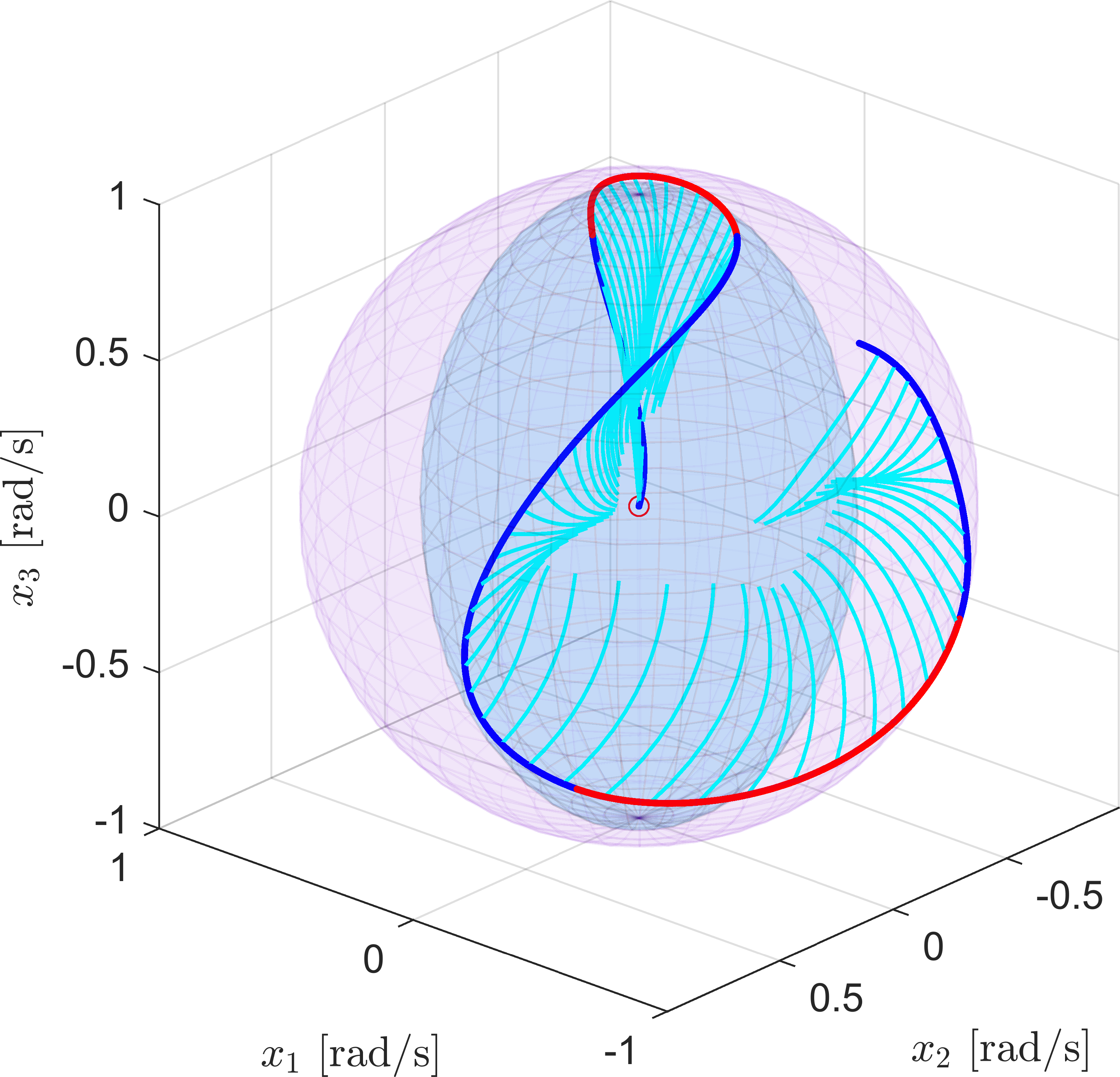}
\end{subfigure}%
\begin{subfigure}[t]{.5\textwidth}
  \centering
  \hspace{-8mm}
\includegraphics[height=6.5cm] {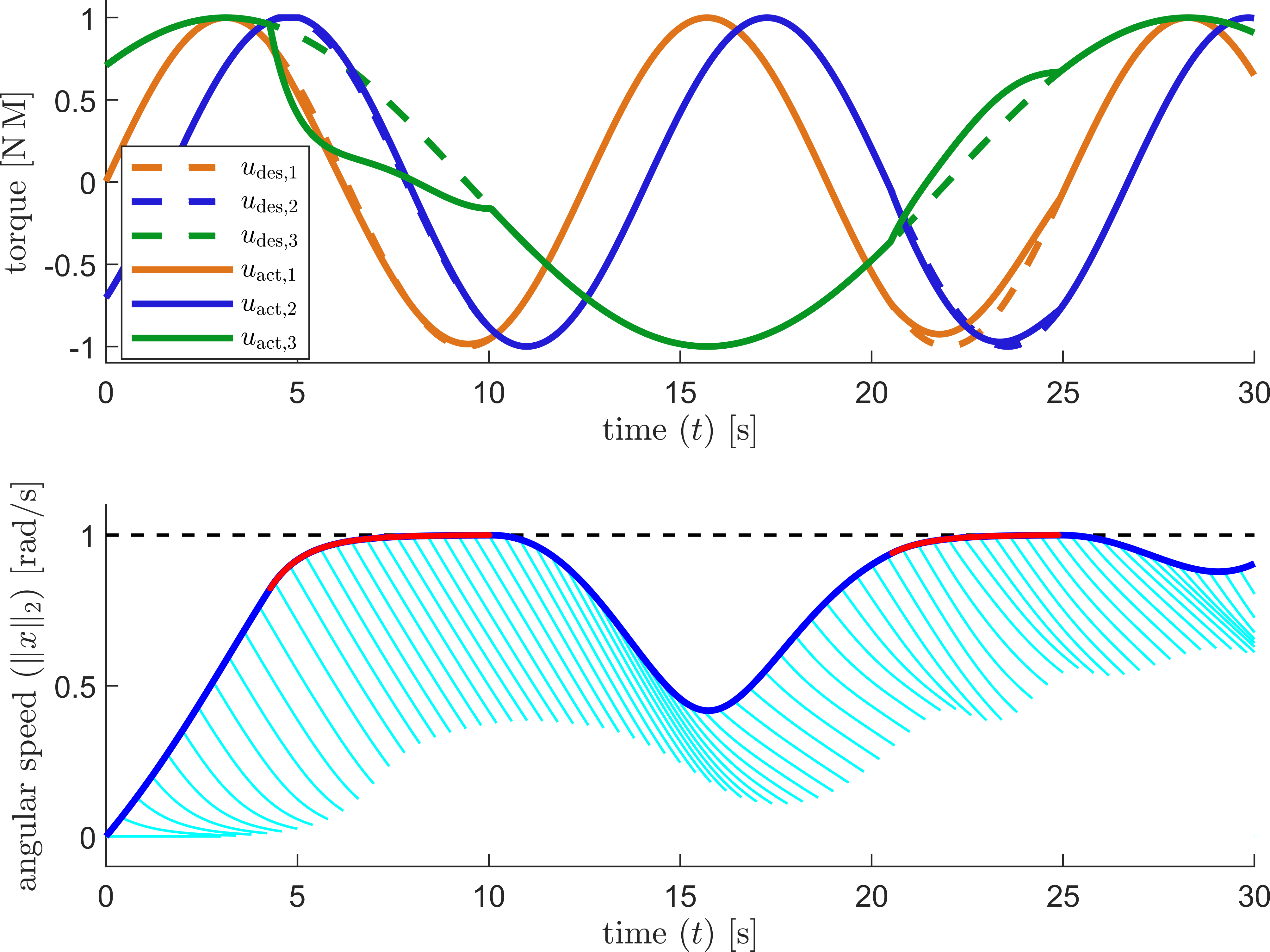}
\end{subfigure} %
\caption{Simulation of implicit ASIF algorithm for spacecraft angular velocity dynamics. The constraint set $\Ca$ is represented by the purple sphere, the backup set $\Cb$ is represented by the blue ellipsoid. The surface of this ellipsoid has constant kinetic energy. The bottom right plot projects this information into normed space, and the top right plot shows the desired and actual control values. }
  \label{fig: iasif_for_angular_speed}
\end{figure}

\begin{figure}
\centering
\begin{subfigure}[b]{.99\textwidth}
   \includegraphics[width=1\linewidth]{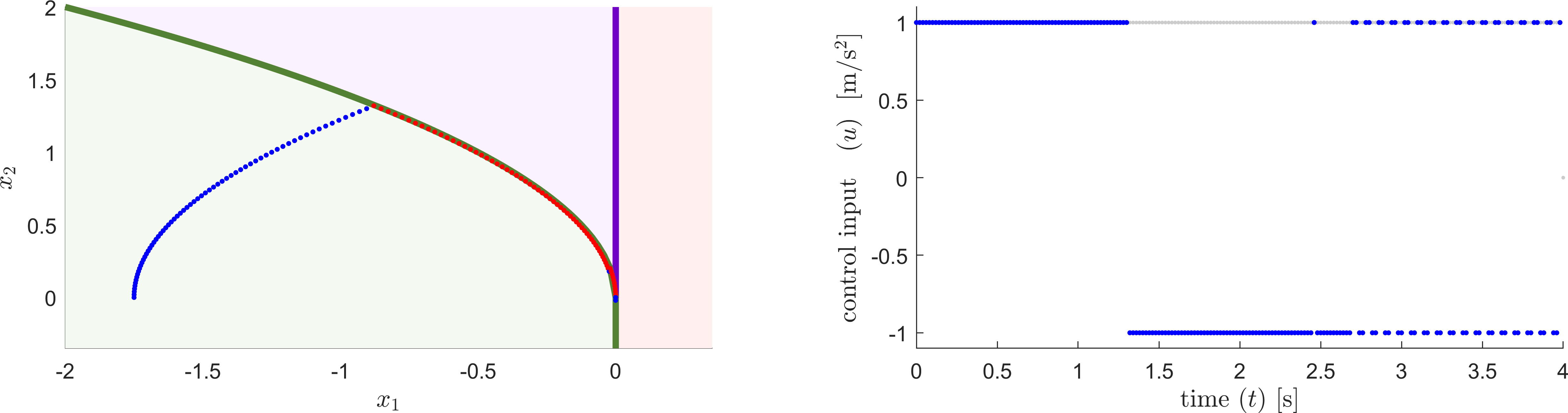}
   \caption{\revised{The explicit simplex filter switches from primary control to filtered control according to Algorithm 1.}}
   \label{fig: double_int_explicit_simplex} 
\end{subfigure}

\begin{subfigure}[b]{.99\textwidth}
   \includegraphics[width=1\linewidth]{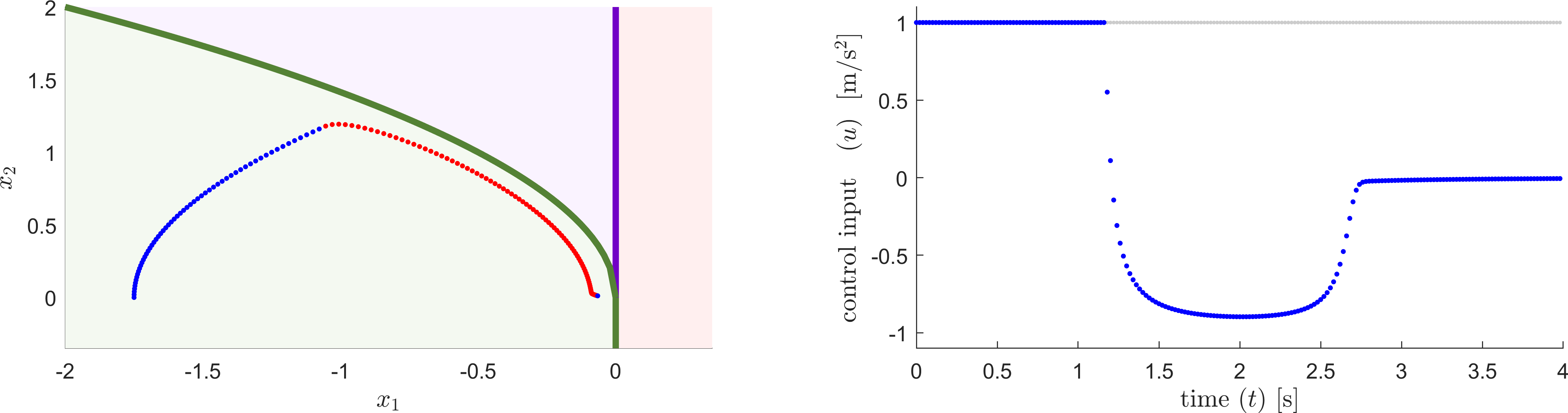}
   \caption{\revised{The explicit active set invariance filter offers a more gradual intervention that solves a quadratic program to minimize the difference of the actual control signal from the desired control signal of the primary controller, while assuring safety according to Algorithm 3.}}
   \label{fig: double_int_explicit_ASIF}
\end{subfigure}
    \caption{Comparison of explicit run time assurance algorithms on double integrator system \eqref{eq: double_integrator_syst} with constraint set \eqref{eq: double_integrator_Ca} and $U\in[-1,1]$. \revised{The viability kernel is shaded green, collision states are shaded red, and the \emph{inevitable} collision states are shaded purple. In the figures on the left, $x_1$ represents distance to the collision state boundary, $x_2$ represents velocity relative to the collision state boundary, the state under the desired control from the primary controller is blue, the state under the filtered control is red, and safety constraint boundaries are green. Figures on the right show the actual control output of the filtered signal. }}
\label{fig: comparison_of_algs_on_double_int}
\end{figure}

\begin{figure}
\centering

\begin{subfigure}[b]{.99\textwidth}
   \includegraphics[width=1\linewidth]{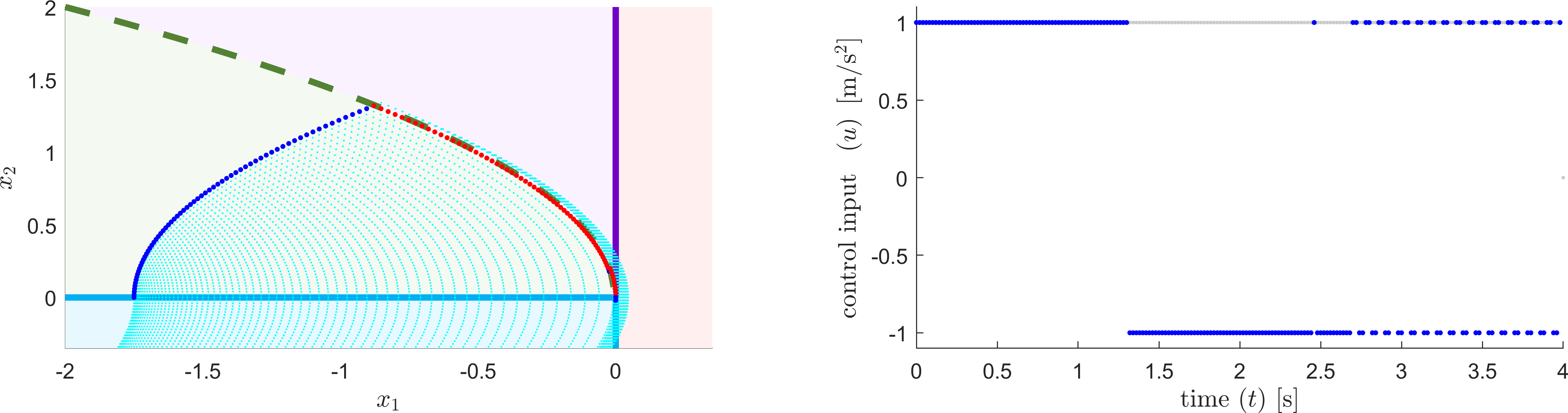}
   \caption{\revised{The implicit simplex filter switches from primary control (full acceleration) to a backup control (full deceleration\rev{)}, as predicted by the path of the system under the backup control according to Algorithm 2.}}
   \label{fig: double_int_implicit_simplex}
\end{subfigure}

\begin{subfigure}[b]{.99\textwidth}
   \includegraphics[width=1\linewidth]{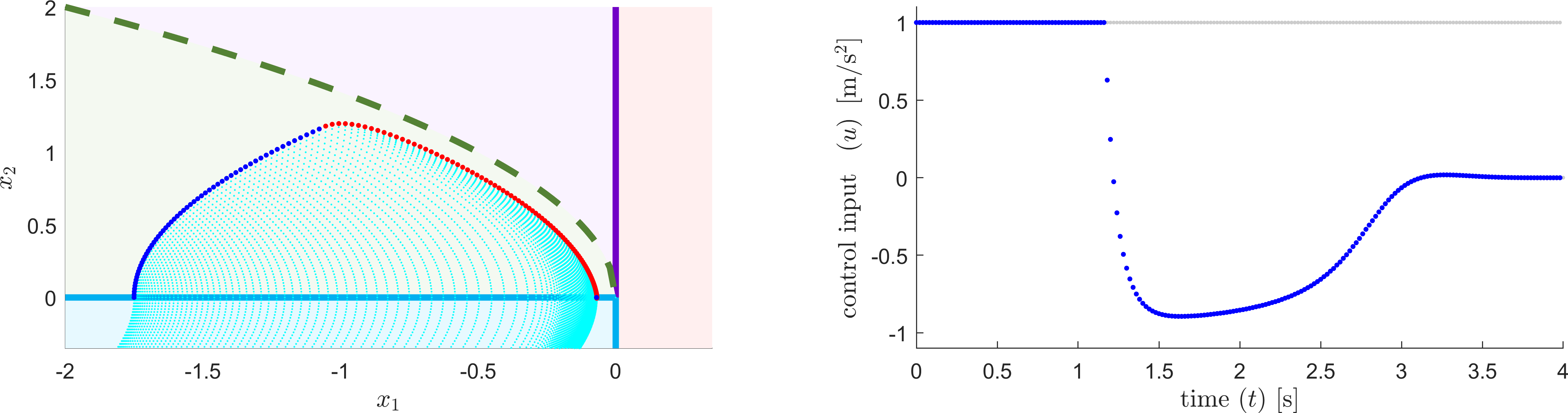}
   \caption{\revised{The implicit active set invariance filter computes a backup trajectory by integrating under a backup control law and solving the quadratic program under implicit barrier constraints according to Algorithm 4.}}
   \label{fig: double_int_implicit_ASIF}
\end{subfigure}
    \caption{Comparison of implicit run time assurance algorithms on double integrator system \eqref{eq: double_integrator_syst} with constraint set \eqref{eq: double_integrator_Ca} and $U\in[-1,1]$. \revised{The viability kernel is shaded green, collision states are shaded red, and the \emph{inevitable} collision states are shaded purple. In the figures on the left, $x_1$ represents distance to the collision state boundary, $x_2$ represents velocity relative to the collision state boundary, the state under the desired control from the primary controller is blue, the state under the filtered control is red, implicit backup trajectories are cyan, and safety constraint boundaries are green. Figures on the right show the actual control output of the filtered signal versus simulation time. }}
\label{fig: comparison_of_algs_on_double_int}
\end{figure}

\begin{example}
Consider the system given by \eqref{eq: sc_rotational_dynamics } and the constraint set $\Ca = \{x\in \R^3 \,|\, \varphi(x)\geq 0\}$ with, 
\begin{equation}
    \varphi(x) = \omega_{\rm max}^2 - x_1^2 - x_2^2 - x_3^3 
\end{equation}
where $\omega_{\rm max}\in \R_+$ represents the maximum allowable angular speed. 
It can be shown that this set is not itself forward invariant in general. 
The backup control law
\begin{equation}
    u_{\rm b} = \text{tanh}( (x\times J x) - k_{\rm d} J x )
\end{equation}
with $k_{\rm d}\geq 0$ 
is bounded to $[-1,1]$ and stabilizes system~\eqref{eq: sc_rotational_dynamics } to the origin. Furthermore, it can be shown that $\Cb=\{ x \in \R^3 \, | \, K- x^T J x \geq 0 \}$ is invariant under $u_{\rm b}(x)$ (see from previous example that this is invariant under the control law $u(x)\equiv 0$). 
The largest ellipsoid of this form that is contained in $\Ca$ is obtained by choosing $K=\omega_{\rm max}^2/\text{min}(J_1, J_2, J_3)$. 

Figure~\ref{fig: iasif_for_angular_speed} shows an illustration of the filtered trajectory under the (unsafe) primary controller $u_{\rm d}(x) = [\sin(\frac{t}{2}), \sin(\frac{t}{2}-\frac{\pi}{4}), \sin(\frac{t}{4}+\frac{\pi}{4}) ]^T$ and with  $\omega_{\rm max}=1\, m/s$, $k_d=1$, $J_1=12\,\rm{kg\,m^2},$ $J_2=12\,\rm{kg\,m^2},$ $J_3 = 5\,\rm{kg\,m^2}$ and the backup horizon $t_{\rm b}\in \{0, 0.05, ..., 3\}$. The trajectory is colored blue where $\ud=\ua$ and red where the RTA mechanism is active. 
\end{example}


\subsection{Additional Design Considerations}

Recall that when the properties $\vartheta^{\rm BC}_1$ and $\vartheta^{\rm BC}_2$ are satisfied, a set $\Cs$ is forward invariant under the RTA control law $\ua$ given by the ASIF-QP. 
While guarantees exist under the assumptions of the model, and indeed the formulations presented exhibit some robustness properties, it is desirable in practice to consider the behavior of the system in the case where an unmodelled effect perturbs the state outside of $\Cs$. 
A simple solution is to make $\Cs$ a locally attractive set by switching to a stabilizing backup controller whenever $x\not\in\Cs$. 
Alternatively, for explicit ASIF, \emph{if $h$ is a concave function}, then the barrier constraint will cause $\ua$ to attract solutions to the interior of $\Cs$ whenever the ASIF-QP is feasible. A backup controller should be present for the case where a feasible solution is not found. 
While outside of the scope of this article, it is interesting to note that attracting solutions outside of $\Cs$ to $\Cs$ doesn't necessarily correspond to the best response of the physical system. For instance \cite{mote2020collision} (Section V.B) presents a method for minimizing a function of \emph{damage} in the presence of an inevitable collision.

\section{Discussion and Comparison of Approaches on Double Integrator System}

Figure \ref{fig: comparison_of_algs_on_double_int} shows a comparison of the four algorithms presented in this article on  the double integrator system given by \eqref{eq: double_integrator_syst} with $u\in[-1,1]$,  $\Ca=\{x\in\X\,\,|\,\,-x_1\geq0\}$, and a controller update period of 10 Hz. For the case of the Simplex algorithms, the continuous dynamics are discretized to fit this period. 
The simulations begin at state $x(0)=[-1.75,0]^T$ and have desired input $\ud=1$. 
Here it can be seen that both Simplex-based algorithms exhibit near identical behavior, and likewise both ASIF algorithms exhibit similar behavior. 
In spite of the implicit methods \revised{lacking} explicit knowledge of the boundary, all cases exhibit similar operational regions. 
Note that the backup trajectory under the implicit Simplex filter violates the safety constraint. This is by design, as the trajectories are integrated forward after taking a theoretical probe step. \revised{A buffer could be added to the switching region to assure safety}. 
The ASIF approaches are slightly more conservative due to the presence of the $\alpha$ term in the barrier constraints. 
This function may be modified to shape the response.

While the Simplex and ASIF methods are most distinctive in terms of behavior, the implicit and explicit methods are most distinctive in terms of requirements.
For instance, the key challenge in using the implicit methods in the design of the backup controller and the identification of the backup set. 
Once these have been identified the it is relatively straightforward to implement either approach. 
Likewise, the key challenge in implementing the explicit algorithms is identifying an invariant set.
Importantly, the explicit ASIF approach is the only algorithm that does not require any backup controller to be defined.

\section{Assurance in the Presence of Uncertainty}

While CPS may be \emph{modeled} using dynamical systems as in \eqref{eq:matt}--\eqref{eq:mark}, real world systems rarely ---if ever--- obey the dynamics of their model precisely. Furthermore, deterministic models do not capture the effects of external disturbances, which are universal in real-world systems. 

One way to \revised{address} model error is to construct a higher fidelity system model, \emph{i.e.} a higher order parameterized model that can be tuned in system identification. However, producing higher fidelity modes can become an exhaustive procedure and can be unfruitful when a newly suggested complex model performs less accurately in comparison to its simpler predecessor.  These hindrances have attracted the attention of numerous fields of study, \eg, robust control \cite{zhou1998essentials}, Sim2Real \cite{hofer2021sim2real}.

A second way to address model error and external disturbances is to construct a \emph{nondeterministic} system model, \ie a dynamical model that accounts for environmental, internal disturbances, errors, by incorporating one or more noise terms. In continuous-time, one such a model is represented by
\begin{equation} \label{eq:nondet}
    \dot{x} = f(x) + g_1(x)u + g_2(x)w,
\end{equation}
where $x \in \X$ and $u \in \U$ maintain their definitions as the system state and control input, and where now the term $w(t) \in W \subset \R^p$ is included to represent a bounded nondeterministic input signal to the system. 
Employing a nondeterministic system model as in \eqref{eq:nondet} reduces the design burden involved in forming an accurate representation of real world phenomena. When designed correctly, a nondeterministic model \emph{will} in general describe the real world behavior accurately.

This section presents run time assurance mechanisms in the context of uncertain systems, as in \eqref{eq:nondet}. First, an \emph{explicit} RTA formulation is presented in which \emph{robust control barrier functions} are employed to assure system trajectories evolve in a given safe set.  Next, an implicit solution to this problem is presented, where run time safety is assessed through the online computation of a reachable sets.


\subsection{Explicit Run Time Assurance for Uncertain Systems}


In this setting of \eqref{eq:nondet}, an ASIF provides a control policy $\ua$ that renders a given safe set $\Cs = \{x\in \X \,\vert\, h(x) \geq 0\}$ forward invariant, regardless to disturbance input $w(t)$ chosen.
For this reason, ASIFs are naturally constructed from \emph{robust control barrier functions}, as discussed next.

Robust control barrier functions extend the general barrier function theory presented above to the nondeterministic setting of \eqref{eq:nondet}. A continuously differentiable function $h : \R^n \rightarrow \R$ is a robust control barrier function for \eqref{eq:nondet} if there exists a class-$\K$ function $\alpha : \R \rightarrow \R$ such that for all $x\in \Cs := \{x\in \X \,\vert\, h(x) \geq 0\}$ there exists a $u \in \R^m$ satisfying 
\begin{equation}\label{eq:barrier_constraint2}
    BC(x, u, w) = L_f h(x) + L_{g_1} h(x)u + L_{g_2} h(x)w + \alpha(h(x)) \geq 0 
\end{equation}
for all $w \in W$, where $ L_f h(x),$ $ L_{g_1} h(x)$ and $L_{g_2}h(x)$ are naturally taken to be the lie derivatives of $h$ along $f$, $g_1$ and $g_2$. 
The barrier constraint \eqref{eq:barrier_constraint2} is a linear inequality on the variable $u$ for any $x \in \R^n$ and any $w \in W$, and when a candidate controller $\ud(x)$ satisfies \eqref{eq:barrier_constraint2} for all $x \in \Cs$ and all $w \in W$, the set  $\Cs$ will be robustly forward invariant for the closed-loop dynamics of \eqref{eq:nondet} under $\ud$. 

When instead $\ud(x)$ does not satisfy \eqref{eq:barrier_constraint2}, an ASIF can be employed to assure the forward invariance of $\Cs$.  One such filter, given below as RASIF-QP, is posed as quadratic program where the constraints in this program are constructed from \eqref{eq:barrier_constraint2}.

\begin{samepage}
\noindent \rule{1\columnwidth}{0.7pt}
\noindent \textbf{RASIF-QP}
\begin{equation}
    \ua(x) = \argmin_{u \in \U} || u - \ud(x)||_2^2 
\end{equation}
\begin{equation}
\begin{array}{cl}
    \text{s.t.} &   L_f h(x) + L_{g_1} h(x) u + L_{g_2}h(x)w + \alpha(h(x)) \geq 0  \\
    &  \text{for all $w \in W$}  \\
\end{array}
\end{equation}
\noindent \rule{1\columnwidth}{0.7pt}
\end{samepage}

When $h$ is a robust control barrier function for \eqref{eq:nondet}, the RCBF-QP is always feasible and the set $\Cs$ is robustly forward invariant for closed-loop dynamics of \eqref{eq:nondet} under $\ua$.
However, this program contains an infinite number of linear constraints and, thus, is not always practically implementable.  In the special instance  where the disturbance set $W$ is a polytope, the program RCBF-QP can be reduced to include only a finite number of linear constraints.  Assume, for instance, that 
\begin{equation}
    W = \textbf{Conv}\{w^1, \cdots, w^q\}
\end{equation}
for some $w^1, \cdots, w^q \in \R^p$, where \textbf{Conv} denotes the convex hull operator.  Then an explicit ASIF is constructed using

\begin{samepage}
\noindent \rule{1\columnwidth}{0.7pt}
\noindent \textbf{RASIF-QP} (polytope W)
\begin{equation}
    \ua(x) = \argmin_{u \in \R^m} || u - \ud(x)||_2^2 
\end{equation}
\begin{equation}
\begin{array}{cl}
    \text{s.t.} &   L_f h(x) + L_{g_1} h(x) u + L_{g_2}h(x)w + \alpha(h(x)) \geq 0  \\
    &  \text{for all $w \in \{w^1, \cdots, w^q\}$}  \\
\end{array}
\end{equation}
\noindent \rule{1\columnwidth}{0.7pt}
\end{samepage}
and this program contains only $q$ linear constraints.


\subsection{Implicit Active Set Invariance for Uncertain Systems}
In this section, an implicit RTA formulation is introduced for the nondeterministic setting of \eqref{eq:nondet}. As in the case of deterministic systems, the implicit ASIF allows the system to leave the safe set $\Cs$ when safety is verified \emph{a priori} via the assessment of a known safe backup control policy.  

In the deterministic setting of \eqref{eq: control_affine_dynamics}, the safety of a given state $x \in \X$ with respect to the backup controller $\ub$ is assessed via a system simulation; however, in the nondeterministic setting of \eqref{eq:nondet}, \revised{it is} necessary to assess the effects of the disturbance, perhaps through, \eg, computing reachable sets \cite{abate2019monitor, abate2020enforcing, abate2021verification}.  For initial set $A \subseteq \X$ and time $t \geq 0$, the time-$t$ reachable set of \eqref{eq:nondet} under $\ub$ is 
\begin{equation}
\label{reach1}
     R(t; A) :=
     \big{\{}\Phi^{\ub}(t;\, x, \mathbf{w}) \in \X \,\big{|}\, x\in A,\, \mathbf{w} : [0, t) \rightarrow \W\big{\}}
\end{equation}  
where $\Phi^{\ub}(t;\, x,\, \mathbf{w})$ denotes the state of \eqref{eq:nondet} reached at time $t$ when beginning at state $x\in \X$ at time $0$ and evolving subject to the backup control input $\ub$ and the disturbance signal $\mathbf{w} : [0, t] \to \W$.  In particular, an implicit ASIF ensures $R(T_{\rm b}; x) \subseteq \Cb$ for all states $x$ along the system trajectory, where $T_{\rm b} \geq 0$ maintains its definition and function as the horizon time of the backup controller. 

In general, it can be difficult to compute reachable sets in closed form. However numerous efficient computational methods exists for over-approximating reachable sets, with computational speeds suitable for, \eg, computing reachable sets in the control-loop and enforcing safe behavior online from these predictions \cite{llanes2021Safety, llanes2021Safety2, abate2020enforcing, abate2021verification}.  Of particular interest to this work, the mixed monotonicity property of dynamical systems can be applied to compute a hyperrectangular over-approximation of $R(t; x)$ using a single simulation of a related $2n$-dimensional deterministic embedding system; a technique that have been applied in domains including transportation system \cite{7799445}, and biological systems \cite{smith2006discrete}.  Further details on the mixed monotonicity property are provided in the sidebar titled ``Mixed  Monotonicity  for  Efficient  Reachability" in this work.

In the remainder of this section, we present one possible implicit ASIF construction, where over-approximations of reachable sets are computed online using the mixed monotonicity property.  The basic idea is as follows: when the system is at state $x$, the ASIF computes a hyperrectangular overapproximation of $R(T_{\rm b}; x)$ using the mixed monotonicity procedure.  A barrier condition is then enforced for each vertex of the approximation, as to ensure $R(T_{\rm b}; x) \subseteq \Cb$ for all states $x$ along the system trajectory.  In this way safety is ensured.  Specific details are as follows.

For a finite set $\S \subset \R$, and some fixed parameter $p > 0$, the \emph{Log-Sum-Exponential} (LSE) of $\S$ is given by 
\begin{equation}\label{eq:lse}
    \mbox{LSE}(\S) = -\frac{1}{p} \log \sum_{s \in \S} \mbox{exp}(-p\cdot s).
\end{equation}
The LSE has several useful properties: namely, $\mbox{LSE}(\S, p)$ is differentiable with respect to the elements of $\S$, and $\mbox{LSE}(\S, p)$ approximates $\min \S$, \emph{i.e.},
    \begin{equation}\label{eq:minbound}
        \min \S -\frac{n}{p} \log 2\leq \mbox{LSE}(\S, p) < \min \S
    \end{equation}
for all $p > 0$, and this approximation can be made {arbitrarily tight} by choosing $p$ large enough. We use the LSE, in this section, to approximate the minimum evaluation of $h$ over a hyperrectangular subset of the statespace. Define the following: 
\begin{align}
\label{eq:LSEh}
\mbox{LSE}_{h}(a) 
    &:= \mbox{LSE}(\,\{h(z)\;\vert\; z \in \corn{a}\,\}, p)\\
\label{eq:gammafunct}
    \gamma(t;\, x) 
    &:= \mbox{LSE}_{h}(\Phi^{E}(t; x))\\
\label{eq:Psi}
\Psi^{\rm}(x) &:=\sup_{0 \leq \tau\leq T_{\rm b}} \gamma^{\rm}(\tau;\, x)  
\end{align}
where $\corn{x, y}$ in \eqref{eq:LSEh} denotes the set of $2^{n}$ corners of a rectangle $[x, y]$ and is given by
\begin{equation}
    \corn{x, y} := \{z \in \X  \,\vert\, z_i \in \{x_i, y_i\} \}.
\end{equation}
From the above definitions we have that $\Psi^{\rm}(x) \geq 0$ when there exists a $t \in [0, T_{\rm b}]$ so that $[\Phi^{E}(t; x)] \subseteq \Cb$ and in this case  $R(T_{\rm b}; x) \subseteq \Cb$ as $\Cb$ is assumed robustly forward invariant for \eqref{eq:nondet} under the backup controller $\ub$.  Thus, an implicit ASIF for \eqref{eq:nondet} is constructed using Algorithm \ref{alg:MMASIF}. See \cite{abate2020enforcing, llanes2021Safety, llanes2021Safety2} for further details.

\begin{algorithm}[t]
\caption{Implicit ASIF for Nondeterministic Control Systems}
\begin{algorithmic}[1]
\setlength\tabcolsep{0pt}
\Statex
\begin{tabulary}{\linewidth}{@{}LLp{10cm}@{}}
    \textbf{input}&:\:\:& Current State $x_{\rm curr}\in X$ \\
    &:\:\:& Desired Input $\ud \in U$ \\
    \textbf{output}&:\:\:& Safe Control Input $\ua \in U $\\
    \textbf{predefined}&:\:\:& Constraint set $\Ca \subset X$ \\
    &:\:\:& Invariant Set $\Cs\subset\Ca$ \\
    &:\:\:& Backup Control Law $\ub: \X \to U$.\\
&
\end{tabulary}
\Function{$\ua(x)=$ASIF}{$x_{\rm curr}$,\, $\ud$}
\State \textbf{compute:}  
\State $u^* = \argmin_{u \in \R^m} ||u - \ud||_2^2$
\State s.t. $\frac{\partial \Psi}{\partial x}(x_{\rm curr}) (f(x_{\rm curr}) + g_1(x_{\rm curr})u + g_2(x_{\rm curr})w)\geq -\alpha(\Psi(x_{\rm curr}))$
\State \hspace{0.45cm} $\forall w \in \corn{ \underline{w},\, \overline{w} }$
\If{Program feasible} \textbf{return} $u^*$
\Else{} \textbf{return} $\ub(x_{\rm curr})$
\EndIf
\EndFunction
\end{algorithmic}
\label{alg:MMASIF}
\end{algorithm}




\section{Verification of Run Time Assurance Algorithms and Architectures}
A key question in trusting RTA systems is ``who checks the checker?" In many applications, RTA systems serve a safety critical role and therefore must be rigorously verified. One can imagine a second RTA to check the RTA, and another third to check that RTA until it is turtles all the way down. Proponents of the Simplex architecture advocate simple switching logic and preplanned backup control actions that reduce the verification burden \cite{ASTM3269-17}. Other approaches may rely on mathematical proofs for verification \cite{munoz2015daidalus}. Nearly all approaches see the RTA as a modular component within the closed-loop control system that can be verified sub-component. It is worth noting that this verification can be done on the algorithms or the actual software implementation. This section focuses on verification of RTA algorithms and architectures in the early design phases using formal and compositional reasoning. 

\subsection{Formal Specification and Analysis of RTA System Requirements, Architecture and Design}
Formal methods tools from computer science for automated verification of hardware and software \cite{Baier08} can be brought to the problem of verifying RTA. In addition to the ability to verify software implementations, formal methods can be used to aid in modeling and automated analysis of algorithm and decision logic with mathematical rigor. At one level, Lyapunov proofs can be used to guarantee safety of control systems. Formal methods can supplement these proofs, or perhaps be used where Lyapunov proofs do not exist. Three goals of formal methods are specification, analysis, and synthesis. 

\textit{Formal specification} facilitates a common, unambiguous understanding of the system between users, designers, programmers, and testers. ``Specification is difficult, unglamorous, and arguably the biggest bottleneck facing verification and validation of aerospace, and other, autonomous systems"  \cite{rozier2016specification}. RTA systems and formal specification and analysis have been applied successfully in a variety of applications \cite{avram2017nonlinear,reis2009browser,bodson1994analytic,rivera1996architectural,sha1998evolving,seto1998simplex,sha01,bak2009system,bak11,bak2010hybrid,alur1995algorithmic,kaynar2010theory,mitra2007verification,aiello2010run,sorokowski2015small,hook2018initial, GrossThesis,hobbs2020elicitation,hobbs2019formal}. Formal RTA requirements specifications have also been developed and analyzed \cite{lee1998monitoring,lee1999runtime, Torens15}.

\textit{Formal analysis} helps to find errors and increase reliability of the system. Formal analysis can be divided \cite{wooldridge1998agents} into axiomatic (\eg. Hoare logic \cite{hoare1969axiomatic}, theorem proving \cite{davis1962machine}), semantic approaches (\eg. model checking \cite{Baier08}), and static analysis methods (\eg. abstract interpretation \cite{cousot1992abstract}). Many formal analysis approaches focus on the satisfiability modulo theories (SMT) where sets of variables are substituted with binary-valued functions of non-binary values called predicates \cite{de2011satisfiability}. An example predicate is an inequality that is evaluated as true or false, such as $x+y\leq c$ with variables $x$ and $y$, and constant $c$. These predicates can be combined in a variety of logics such as first order and temporal logics such as linear temporal logic (LTL), computational tree logic (CTL), timed computational tree logic (TCTL), metric temporal logic (MTL), and signal temporal logic (STL) \cite{clarke1981design,koymans1990specifying,Baier08,maler2004monitoring}.
%
%
Formal verification of linear temporal logic specifications has been applied to PID attitude control of spacecraft \cite{GrossSpace}, a Simplex RTA system for spacecraft attitude \cite{gross2017run,GrossRTA,GrossThesis}, switching logic for automatic maneuvering \cite{hobbs2019formal}, an LTL specification monitor automaton \cite{abate2019monitor}, and many other applications.

\textit{Formal synthesis} integrates formal methods into the development process to convert a design to implementation with some level of automation. While many controller designs are synthesized in some fashion, formal synthesis typically involves automatic generation of a controller, or a run time assurance filter from formal specifications \cite{konighofer2017shield}.

\subsection{Compositional Verification}

\textit{Compositional verification}, sometimes referred to as a ``divide and conquer" approach to verification \cite{puasuareanu2008learning}, helps to verify or test portions of algorithms or systems architectures separately and then make conclusions about the system as a whole. This verification approach is helpful in modern engineering practices where engineering tasks are divided over large teams \cite{peled2013software}. Compositional verification includes statements such as: if property $p_1$ for subsystem $A$ and property $p_2$ for subsystem $B$ both hold, it implies property $p$ holds for the entire system. A compositional verification approach can be used to apply formal methods to verify components of an RTA system, especially in Simplex-based RTA systems where multiple simple components (monitors, backup controllers, and switching functions) work together to ensure safety \cite{gross2017run,gross2016incremental,GrossRTA,GrossThesis, hobbs2020elicitation,hobbs2021formal}. 

\begin{figure}[htb]
\begin{center}
\includegraphics[width=0.7\textwidth]{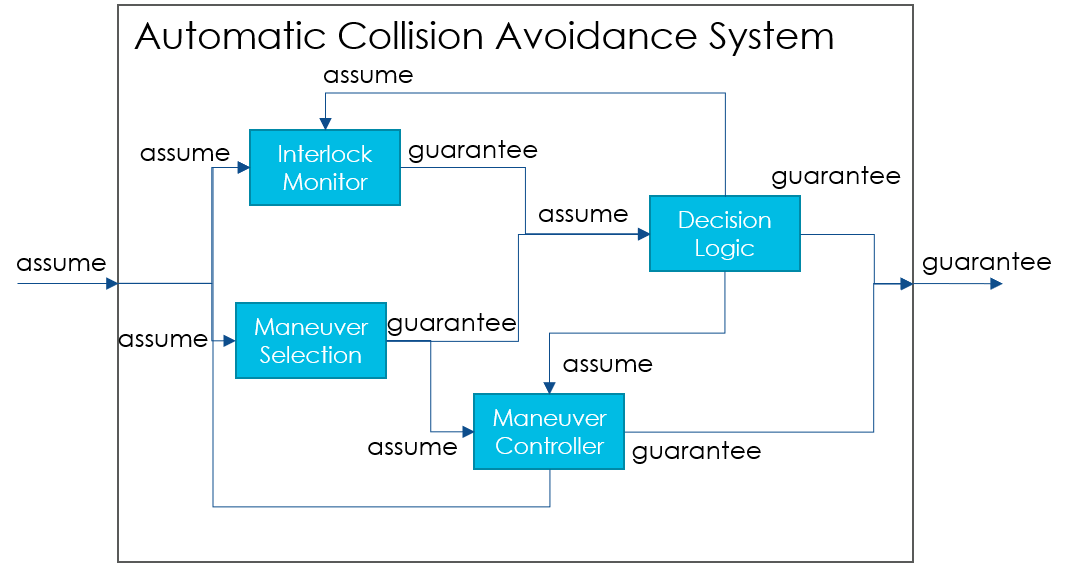}
\caption{Depiction of assume-guarantee reasoning for an automatic collision avoidance system, were assumptions properties are guaranteed by a component based on assumptions about their environment. Guarantees on the output of one component translate to assumptions on the input of another component until larger properties of the overall system are proven.} 
\label{fig:AssumeGuarantee} 
\end{center}
\end{figure}

Compositional verification often relies on \textit{assume-guarantee reasoning} of the architecture \cite{whalen2012your} or the hybrid switching \cite{henzinger2001assume}. Assume-guarantee contracts make guarantees on the output of a component based on assumptions about the input. Assumptions are expectations that a component has about it's environment, while guarantees are behavior properties provided by the component. Verification techniques are applied to individual components with explicit, formal, assumptions and guarantees, as depicted in Figure \ref{fig:AssumeGuarantee}.

\section{Conclusions}
RTA Systems are growing in popularity as a way to assure safety by altering unsafe control inputs from a primary controller. The assurance mechanism of RTA systems is constructed agnostic to the underlying structure of the primary controller, whether the primary control comes from a human operator, an advanced control approach, or an autonomous control approach. By effectively decoupling the enforcement of safety constraints from performance-related objectives, RTA offers a number of useful advantages over traditional (offline) verification.

This article provided a tutorial on developing RTA systems with particular emphasis on implicit and explicit safety definitions as well as Simplex architecture and active set invariance filtering RTA approaches. Implicit (or trajectory-based) safety definitions computed finite-time trajectories under a backup controller online, while explicit (or region-based) approaches identified trusted regions offline via the construction of either a large forward invariant or control invariant set. Simplex Architecture RTA designs \textit{monitor} for imminent violations of safety constraints, and switch  from the primary controller to a backup controller when an imminent unsafe condition is detected. By contrast, backup control intervenes gradually in ASIF approaches, rather than with a hard switch.

Given the increasing complexity of modern control systems, RTA in its various forms is one tool to assure safety and a potential path to certification in safety critical domains.



\setcounter{equation}{0}
\renewcommand{\theequation}{S\arabic{equation}}
\setcounter{table}{0}
\renewcommand{\thetable}{S\arabic{table}}
\setcounter{figure}{0}
\renewcommand{\thefigure}{S\arabic{figure}}


\section[Run Time Assurance]{Sidebar: Run Time Assurance}\label{sidebar-RTA Aliases}
\rev{Run Time Assurance (RTA) systems guarantee safety of increasingly complex and intelligent control system designs by monitoring the state of the system and intervening when necessary. RTA acts as a safety filter that sits between the primary controller and the plant to assure the state of the system adheres to safety properties. The design of the RTA safety mechanism is decoupled from the design of the primary controller, allowing for RTA to focus on safety while the primary controller optimizes performance. Historically, verification techniques for novel control approaches trail years to decades behind the development of the approaches themselves \cite{hobbs2020elicitation}. RTA provides a path to introduce the use of these advanced controllers faster, even in safety-critical applications. This article provides a tutorial on many ways to design RTA, including explicit and implicit monitoring, Simplex and Active Set Invariance Filter (ASIF) intervention approaches, and uncertainty with a variety of examples. Explicit monitors define forward invariant or control invariant safe sets offline that are enforced online at run time, while implicit monitors project the a finite-time trajectory of a backup controller online. Simplex interventions switch from the primary controller to a backup controller, while ASIF interventions gradually optimally modify the control signal subject to safety constraints.  A conceptual and a mathematical description are provided for each. Where possible, the Automatic Ground Collision Avoidance System  (Auto GCAS), an implicit Simplex RTA design, is used as an example to illustrate several RTA concepts.  The article assumes an introductory understanding of control theory and state space concepts. }

\section[RTA Aliases]{Sidebar: RTA Aliases}\label{sidebar-RTA Aliases}

RTA provides a conceptually simple and highly effective approach to enforcing safety constraints that has frequently been explored, utilized, and independently reinvented in a wide variety of areas.
As such, RTA and its applications go by many names in the literature, including: active set invariance, active collision avoidance, advanced driver assistance, active safety, safety filtering, emergency braking, sandboxing, conflict resolution, and constraint-based planning. 
The primary controller goes by several names in the literature including {advanced controller} \cite{schierman2018runtime}, {advanced system} \cite{schierman2015runtime}, {experimental control module} \cite{seto1998simplex}, {baseline controller} \cite{seto1998simplex}, {unverified controller} \cite{clark2013study,gross2017run}, {nominal controller} \cite{hobbs2021evolving}, {complex subsystem} \cite{bak2009system,clark2013study}, and {complex controller} \cite{bak11}. 
Finally, the backup controller goes by many names in the literature, including the recovery mechanism, {reversionary controller} \cite{schierman2018runtime}, {reversionary system} \cite{schierman2015runtime}, {safety control module} \cite{seto1998simplex}, {verified controller}\cite{gross2017run}, {backup controller} \cite{hobbs2021evolving}, {safety remediation controller} \cite{bak2009system}, {recovery controller} \cite{swihart2011automatic,ASTM3269-17,hook2018initial,clark2013study}, and {safety controller} \cite{bak11}. 

\revised{Part of the different naming conventions comes from the different intentions and the resulting implementation of the RTA intervention, which are grouped here as: recovery and reversionary controllers. \textit{Recovery Controllers} intervene in emergency situations and return control authority to the primary controller as soon as possible. Recovery controllers are not intended to perform the wide range of tasks a dynamical system is designed to perform, but instead focuses solely on recovery maneuvers from an unsafe condition (e.g. collision avoidance, adherence to keep out zones, or maintaining state limits). By contrast, \textit{reversionary controllers} detect inappropriate behavior of the primary controller and revert to a simpler, verified controller to complete the rest of the tasks. The reversionary controller may sacrifice performance for assurance. This article focuses on the design of recovery controller-based RTA systems.}

%


\section[Shielded Learning]{Sidebar: Shielded Learning}\label{sidebar-ShieldedLearning}
No discussion of RTA for complex and autonomous control systems would be complete without discussing the emerging challenges of using machine learning, and in particular, reinforcement learning (RL), where a dynamical system learns from experience collected over many episodes \cite{sutton2018reinforcement}. The use of RL has attracted attention for its ability to tackle complex tasks with high-dimensional state spaces by world experts in Go \cite{silver2016mastering,silver2017mastering} and StarCraft II \cite{vinyals2019grandmaster}. In addition, it is becoming popular in robotics research to operate in unknown environments and learn state representations for many tasks \cite{ibarz2021train}.  Both control theory and RL share a common concept of a system \textit{state} (usually $x\in \mathcal{X}$ in control theory and $s\in\mathcal{S}$ in RL). A control input ($u\in\mathcal{U}$ in control theory) is formulated as an \textit{action} $a$ or $A\in \mathcal{A}$ in RL, and the actions are determined by a \textit{policy} (i.e. controller) that maps states to actions. State-action sequences ($S_0,A_0,S_1, A_1,...S_n,A_n$) are called \textit{trajectories} in RL. Both fields have a notion of partial observability. A measurement $y$ may come from sensors in control theory, or an \textit{observation} $O$ in RL may be used to estimate the system state (\eg. $\hat{x}\in\mathcal{X}$). In control theory, a controller is optimized to minimize a cost function, while in RL, a policy is optimized to maximize a \textit{reward function}. In both control theory and RL, \textit{dynamics} describe how the state evaluates in the environment. RL approaches may be \textit{model-based}, where the agent learns a model of the environment, or \textit{model-free} where dynamics are not explicitly modeled and the action is learned directly through interactions with the environment.  

A challenge in RL, especially when it takes place on a physical agent, is ensuring safety of that agent. Safe RL is defined as the process of learning a policy that maximizes expected return while ensuring adherence to safety constraints \cite{garcia2015comprehensive}. Safe RL usually employs either \textit{reward shaping} or \textit{shielding} (\emph{i.e.} RTA) to incorporate safety in the training process. 
Reward shaping factors safety into the reward function to reduce risk. However, it is difficult to properly weight safety rewards and reward shaping can sometimes have unintended impacts on performance. In addition, reward shaping does not guarantee safety during operations. %
By contrast, RTA can be used to filter the actions of the RL algorithms to ensure safety.
A variety of RTA approaches have been explored in the literature including human-like intervention \cite{saunders2018trial}, Lyapunov-based approaches \cite{perkins2002lyapunov, chow2018lyapunov}, barrier functions \cite{cheng2019end}, and formal verification of safety constraints \cite{murugesan2019formal,fulton2018safe}. Determining the appropriate way to include RTA in the training process is still an area of active research. 


\section[The Case for Plan B]{Sidebar: The Case for Plan B}

A fundamental concept underlying RTA is: always maintaining a safe ``Plan B" during mission execution, with the purpose of executing ``Plan B" as soon as the nominal activity appears to be dangerous. Considered from that perspective, maintaining a ``Plan B" makes sense and hundreds of examples can be found in everyday life. For example, financial advisers always recommend their clients keep a few weeks' worth of salary in a savings account so as to be ready and face unexpected financial expenses, such as the medical costs following an accident, the replacement of a commuting vehicle following a collision, a few weeks at a hotel if the home is damaged by a hurricane, and so on. Defensive driving is another example, whereby a driver is trained to always be aware of her surroundings and be ready to take appropriate action in case of unexpected events. 
Currently U.S. regulations specify three levels of ``circuit breakers," that are used to halt trading in stock exchanges when the S\&P 500 index drops below certain critical levels. 
One of the oldest examples of an engineered ``Plan B" can be found in Greek mythological the story of Ariadne's thread \cite{AriadnesThread, Philosophyinlabyrinths, MythandCreativity}. 
In the story, Ariadne, daughter of Minos the king of Crete, conceived of using a thread as a ``Plan B" mechanism that would allow her lover to exit Daedalus' Labyrinth after killing the Minotaur. 
Similar themes appear in folktales such as ``Le Petit Poucet" (Hop-o'-My-Thumb) \cite{LittleThumb, LittleThumbEnglish} and ``Hansel and Gretel." 

One may also observe this paradigm in nature. 
It is now known that acacia trees, whose leaves are one common form of food for giraffes and other herbivorous fauna in Africa, are capable of exercising a kind of ``Plan B" making these leaves lethal by raising tannin-C in their leaves to deadly levels in a matter of minutes if they are over-grazed \cite{killertree}. 
Likewise, rodents very often create burrows with emergency exits or ``bolt holes" to escape predators, flooding or other causes that may block their main entrances \cite{1023071375126}. 
The sympathetic nervous system in mammals 
acts as a physiological ``Plan B" by preparing the body for a ``fight or flight" response. Upon detection of a threat, glucose is transmitted from storage sites, blood is shifted to the organs that are most essential for physical exertion, heart rate is increased, adrenaline is released, and cognition is sharpened \cite{sapolsky1990stress}.

\section[Safety, Reliability, and Security]{Sidebar: Safety, Reliability, and Security}\label{sidebar-safrelsec}

The primary function of an RTA system is to ensure safety (and in some cases security), not reliability. Safety and reliability are two different concepts. Reliability is freedom from errors and failures, while safety is freedom from harm. \textit{Reliability} of a system is its ability to perform consistently over time. For example, physical components may fatigue and degrade over time, and reliability focuses on how long the performance is consistent and failure is unlikely \revised{\cite{wiegers2013software,leveson2011engineering}}. Reliability of software, which doesn't degrade over time (unless it changes over time, in the case of intelligent control that continues to learn online), is based on it's freedom from design errors. In a related sense, \textit{security} is freedom from harm from others. The difference in safety and security is in intent, severity, and likelihood of harm.  





\section[To RTA or not to RTA? That is the Question]{Sidebar: To RTA or not to RTA? That is the Question}\label{sidebar-RTAorNoRTA}

RTA is not a panacea for controller safety. Employing RTA incurs a cost-benefit trade-off. Implementing RTA can increase system cost, complexity, the amount of testing or evidence required for certification, the number of hardware and software components that can fail, and the attack surface when cyber security is a concern. Designers must ask themselves critical questions before deciding to use an RTA system. First, \textit{could a simple controller meet design requirements?} If a simple control design is sufficient, traditional simulation and testing can provide adequate confidence, or the primary controller can be exhaustively verified, an RTA system should not be used. Second, \textit{is the system safety or mission critical?} If the system is not safety or mission critical, an RTA could be an unnecessary design complication.

If the system under consideration is safety or mission critical and the control functions can only be achieved by a complex or neural network control scheme, and certification of that high performance controller is not possible via traditional simulation, test, or exhaustive analysis, then an RTA system may be appropriate.

\section[Reference and Command Governors]{Sidebar: Reference and Command Governors} 
\begin{figure}[htb]
\begin{center}
\includegraphics[width=0.7\textwidth]{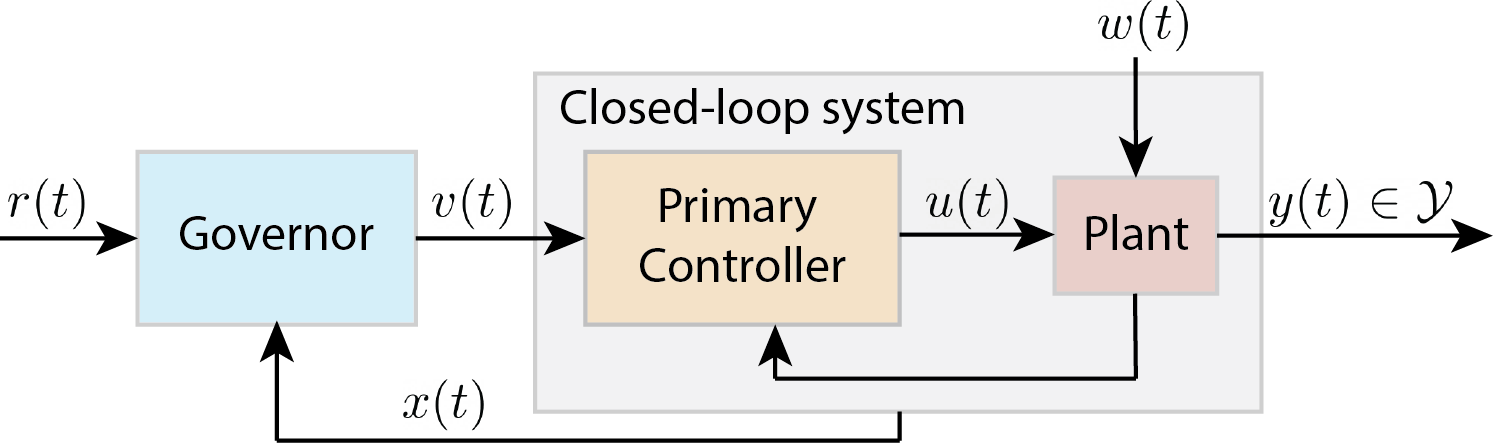}
\caption{Depiction of a reference governor which modifies the input to the primary controller, rather than the primary controller output to assure safety.} 
\label{fig:ReferenceGovernor} 
\end{center}
\end{figure}

While this article focuses entirely on RTA designs that monitor the primary controller and intervene before the control signal reaches the plant, it is worth noting that other approaches, such as \textit{reference and command governors} can be used for safety assurance. Instead of modifying the control signal of the primary controller before it reaches the plant, reference and governors wrap around the closed loop system and change the reference signal \cite{garone2017reference}. The reference governor acts as a pre-filter on the desired reference command $r(t)$ and state measurement or estimate $x(t)$ under noise $w(t)$, resulting in a modified reference command $v(t)$. A command or reference governor architecture is depicted in Figure \ref{fig:ReferenceGovernor}.

\section[The Robotarium: An RTA Enabled Remote-Access Swarm Robotics Testbed]{Sidebar: The Robotarium: An RTA Enabled Remote-Access Swarm Robotics Testbed   }

The Robotarium \cite{pickem2016safe, pickem2017robotarium, wilson2020robotarium, robotarium_2021} is a remotely accessible swarm robotics testbed with the stated mission to \emph{democratize robotics by providing remote access to a state-of-the-art multi-robot research facility}.
The lab consists of a 12ft$\times$14ft custom arena and a number of differential drive robots, known as GritsBots. The testbed exhibits features such as Vicon-based real time motion-capture, wireless inductive charging, video capture of experiments, and an above mounted projector that allows users to project time-varying environmental projections onto the testbed during experiments. 
Users may submit code to control the robots through MATLAB or Python scripts, that are submitted through the Robotarium website. 

The open-access nature of the lab presents unique challenges in terms of safety. 
Experiments must remain as faithful as possible to the user specified behavior while being provably safe, so as to prevent damage to the lab equipment. 
Offline verification of the user code would be a prohibitively difficult and time consuming task for the operators of the lab, and would place an unnecessary burden of correctness on its users. 
Consequently, an RTA solution to safety is utilized, and user submitted algorithms are treated as black box functions that can be probed at run time. 
Specifically, control barrier function based quadratic programs (explicit ASIF) are used to prevent collisions between the various robots involved in an experiment. 
An explicit ASIF-based approach is attractive in this case as it is minimally invasive to the desired inputs and provides a smooth overriding behavior. 
Since the planning is conducted under a single-integrator dynamics model, explicit identification of viable sets is fully-tractable problem offline.

\section[Mixed Monotonicity for Efficient Reachability]{Sidebar: Mixed Monotonicity for Efficient Reachability}

A dynamical system, possibly subject to a nondeterministic disturbance input, is mixed monotone when there exists a related decomposition function that separates the system dynamics into cooperative and competitive state interactions. Mixed monotonicity applies to continuous-time systems \cite{ coogan2020mixed, sontag2006nonmonotone}, discrete-time systems \cite{smith2006discrete}, as well as systems with disturbances \cite{meyer2019tira, coogan2015efficient}, and it generalizes the \emph{monotonicity} property of dynamical systems for which trajectories maintain a partial order over states \cite{smith2008monotone, angeli2003monotone}.

For an $n$-dimensional mixed monotone system with a disturbance input, it is possible to construct a $2n$-dimensional monotone embedding system from the decomposition function. This enables one to apply the powerful theory of monotone dynamical systems to the embedding system and can be used to compute useful properties of the initial system.  In particular, such approaches are useful for \revised{efficiently} computing reachable sets and robustly forward invariant sets for systems with disturbances\revised{, techniques which have been successfully demonstrated in online safety applications \cite{llanes2021Safety, llanes2021Safety2, abate2020enforcing, abate2021verification}}. 

A dynamical system
\begin{equation}\label{NAS}
    \dot{x} = F(x, w),
\end{equation}
with state $x \in \X \subseteq \R^n$ and disturbance $w \in \W \subseteq \R^m$, is a \emph{mixed monotone system} when there exists a related \emph{decomposition function} $d : \X \times \W \times \X \times \W \rightarrow \mathbb{R}^n$ so that for all $x, \widehat{x} \in \X$ and all $w, \widehat{w} \in \W$ the following hold:
\begin{itemize}
    \item $d(x, w, x, w) = F(x, w)$.
    \item $\frac{\partial d_i}{\partial x_j}(x, w, \widehat{x},  \widehat{w}) \geq 0$ for all $i, j$ with $i \neq j$.
    \item $\frac{\partial d_i}{\partial  \widehat{x}_j}(x, w,  \widehat{x}, \widehat{w}) \leq 0$ for all $i, j$.
    \item$\frac{\partial d_i}{\partial w_k}(x, w,  \widehat{x},  \widehat{w}) \geq 0$ and $\frac{\partial d_i}{\partial  \widehat{w}_k}(x, w, \widehat{x},  \widehat{w}) \leq 0$ for all $i, k$. 
\end{itemize}
Large classes of systems have been shown to be mixed monotone, with decomposition functions constructed from, \eg, bounds on the systems Jacobian matrix \cite{meyer2019tira} of domain specific knowledge \cite{abate2020computing, coogan2016stability}.  In certain instances decomposition functions can also be computed by solving an optimization problem \cite{abate2020tight}.

An important feature of mixed monotone systems is that hyperrectangular over-approximations of reachable sets can be efficiently computed using a single simulation of a related $2n$ dimensional embedding system constructed from the decomposition function. Assume $\X$ is an extended hyperrectangle and $\W := [\underline{w}, \overline{w}]$ is a hyperrectangle.  When \eqref{NAS} is mixed monotone with decomposition function $d$,
\begin{equation}\label{eq:embedding}
\begin{bmatrix}
  \dot{x}\\
  \dot{ \widehat{x}} 
\end{bmatrix}
  = E(x,\, \widehat{x})
  := 
\begin{bmatrix}
  d (x,\, \underline{w},\,  \widehat{x},\,\overline{w})\\
  d ( \widehat{x},\,\overline{w},\, x,\, \underline{w}) 
\end{bmatrix}
\end{equation}
is the \textit{embedding system} relative to $d$, and $\Phi^{E}( t; a)$ denotes the state of \eqref{eq:embedding} reached at time $t \geq 0$ when beginning at state $a \in \X\times\X$ at time $0$.  
For any hyperrectangular set of states $[\underline{x}, \overline{x}] \subset \X$ the following is true: if $\Phi^{E}( \tau; (\underline{x}, \overline{x}))\in \X\times \X$ for all $0\leq \tau\leq t$, then 
\begin{equation}\label{eq:MM_reach}
    R(t; [\underline{x}, \overline{x}]) \subseteq [\Phi^{E}_{1:n}( t;\, (\underline{x}, \overline{x})), \Phi^{E}_{n+1:2n}( t;\, (\underline{x}, \overline{x}))].
\end{equation}
The results of \eqref{eq:MM_reach} provide an efficient procedure for over-approximating reachable sets for systems with disturbances; in particular, a hyperrectangular over-approximation of $R(t; [\underline{x}, \overline{x}])$ is \revised{computable} from a single simulation of the embedding system \eqref{eq:embedding}, where the bottom and top corners of the approximation are the first $n$ and last $n$ coordinates of $\Phi^{E}( t ; (\underline{x}, \overline{x}))$.  This result is demonstrated in the following numerical example.

\begin{figure*}[t!]
\centering
    \begin{subfigure}{0.31\textwidth}
    \scalebox{.9}{
%
%
\begin{tikzpicture}

\begin{axis}[%
width=4.8cm,
height=3.5cm,
at={(0cm,0cm)},
scale only axis,
xmin=-1,
xmax=5,
xtick={-1,  0,  1,  2,  3,  4,  5},
xlabel style={font=\color{white!15!black}},
xlabel={$x_1$},
ymin=-1,
ymax=3,
ytick={-1,  0,  1,  2,  3},
ylabel style={font=\color{white!15!black}},
ylabel={$x_2$},
axis background/.style={fill=white},
axis x line*=bottom,
axis y line*=left,
xmajorgrids,
ymajorgrids,
axis on top
]

\addplot[area legend, line width=1.2pt, draw=black, fill=red, fill opacity=0.7, forget plot]
table[row sep=crcr] {%
x	y\\
-0.5	-0.5\\
-0.4	-0.5\\
-0.3	-0.5\\
-0.2	-0.5\\
-0.1	-0.5\\
0	-0.5\\
0.1	-0.5\\
0.2	-0.5\\
0.3	-0.5\\
0.4	-0.5\\
0.5	-0.5\\
0.5	-0.4\\
0.5	-0.3\\
0.5	-0.2\\
0.5	-0.1\\
0.5	0\\
0.5	0.1\\
0.5	0.2\\
0.5	0.3\\
0.5	0.4\\
0.5	0.5\\
0.4	0.5\\
0.3	0.5\\
0.2	0.5\\
0.1	0.5\\
0	0.5\\
-0.1	0.5\\
-0.2	0.5\\
-0.3	0.5\\
-0.4	0.5\\
-0.5	0.5\\
-0.5	0.4\\
-0.5	0.3\\
-0.5	0.2\\
-0.5	0.1\\
-0.5	0\\
-0.5	-0.1\\
-0.5	-0.2\\
-0.5	-0.3\\
-0.5	-0.4\\
-0.5	-0.5\\
}--cycle;

\addplot[area legend, line width=1.2pt, draw=black, fill=green, fill opacity=0.1, forget plot]
table[row sep=crcr] {%
x	y\\
1.504	0.0010000000000008\\
1.77607204650367	0.0010000000000008\\
2.04814409300734	0.0010000000000008\\
2.32021613951101	0.0010000000000008\\
2.59228818601468	0.0010000000000008\\
2.86436023251835	0.0010000000000008\\
3.13643227902202	0.0010000000000008\\
3.40850432552569	0.0010000000000008\\
3.68057637202936	0.0010000000000008\\
3.95264841853303	0.0010000000000008\\
4.2247204650367	0.0010000000000008\\
4.2247204650367	0.246387472371358\\
4.2247204650367	0.491774944742715\\
4.2247204650367	0.737162417114072\\
4.2247204650367	0.982549889485429\\
4.2247204650367	1.22793736185679\\
4.2247204650367	1.47332483422814\\
4.2247204650367	1.7187123065995\\
4.2247204650367	1.96409977897086\\
4.2247204650367	2.20948725134221\\
4.2247204650367	2.45487472371357\\
3.95264841853303	2.45487472371357\\
3.68057637202936	2.45487472371357\\
3.40850432552569	2.45487472371357\\
3.13643227902202	2.45487472371357\\
2.86436023251835	2.45487472371357\\
2.59228818601468	2.45487472371357\\
2.32021613951101	2.45487472371357\\
2.04814409300734	2.45487472371357\\
1.77607204650367	2.45487472371357\\
1.504	2.45487472371357\\
1.504	2.20948725134221\\
1.504	1.96409977897086\\
1.504	1.7187123065995\\
1.504	1.47332483422814\\
1.504	1.22793736185679\\
1.504	0.982549889485429\\
1.504	0.737162417114072\\
1.504	0.491774944742715\\
1.504	0.246387472371358\\
1.504	0.0010000000000008\\
}--cycle;

\addplot[area legend, line width=1.2pt, draw=black, fill=green, fill opacity=0.9, forget plot]
table[row sep=crcr] {%
x	y\\
1.68302138159568	0.121993086103313\\
1.76015774985729	0.209747379074114\\
1.8429050733731	0.298957361637399\\
1.93139932981606	0.389642913027619\\
2.02577962349615	0.481824229462162\\
2.12618826055129	0.575521829962475\\
2.23277082609746	0.670756562297627\\
2.34567626339393	0.767549609053254\\
2.4650569550808	0.865922493828845\\
2.59106880654812	0.965897087566484\\
2.72387133149772	1.06749561501416\\
2.74763306207465	1.1557763559593\\
2.79307395399336	1.25383784823433\\
2.86219108885708	1.36215532600483\\
2.95718437634688	1.48123629123835\\
3.08047619831661	1.61162285120856\\
3.23473316894653	1.75389425634589\\
3.42289025857682	1.90866965784314\\
3.64817756071541	2.07661110655015\\
3.91415001807577	2.25842681707799\\
4.2247204650367	2.45487472371357\\
3.94846751450065	2.31371978529659\\
3.68337188566274	2.17484265475353\\
3.42916022694129	2.038209473659\\
3.18556578792349	1.90378696895352\\
2.95232825032407	1.77154244129062\\
2.72919356364142	1.64144375364846\\
2.51591378536907	1.51345932019937\\
2.31224692562502	1.38755809543062\\
2.1179567960661	1.26370956351008\\
1.93281286295861	1.14188372789053\\
1.79691353306218	0.991056371990598\\
1.69217333313875	0.85257654190603\\
1.61573189596617	0.725818279977681\\
1.56499526278975	0.610196243191891\\
1.53761152462038	0.505162864037498\\
1.53144887813123	0.410205739530193\\
1.54457584037302	0.32484522807017\\
1.57524339571181	0.248632235802881\\
1.62186887401088	0.181146175940976\\
1.68302138159568	0.121993086103313\\
}--cycle;
\addplot[only marks, mark=*, mark options={}, mark size=1.5000pt, color=black, fill=black] table[row sep=crcr]{%
x	y\\
1.504	0.0010000000000008\\
};
\addplot[only marks, mark=*, mark options={}, mark size=1.5000pt, color=black, fill=black] table[row sep=crcr]{%
x	y\\
4.2247204650367	2.45487472371357\\
};
\end{axis}

\begin{axis}[%
width=5.899cm,
height=4.301cm,
at={(-0.767cm,-0.478cm)},
scale only axis,
xmin=0,
xmax=1,
ymin=0,
ymax=1,
axis line style={draw=none},
ticks=none,
axis x line*=bottom,
axis y line*=left,
legend style={legend cell align=left, align=left, draw=white!15!black}
]
\end{axis}
\end{tikzpicture}%
        }
        \caption{ }
        \label{fig:ex1_1}
    \end{subfigure}
    ~\hspace{1cm}~
    \begin{subfigure}{0.31\textwidth}
    \scalebox{.9}{
        \input{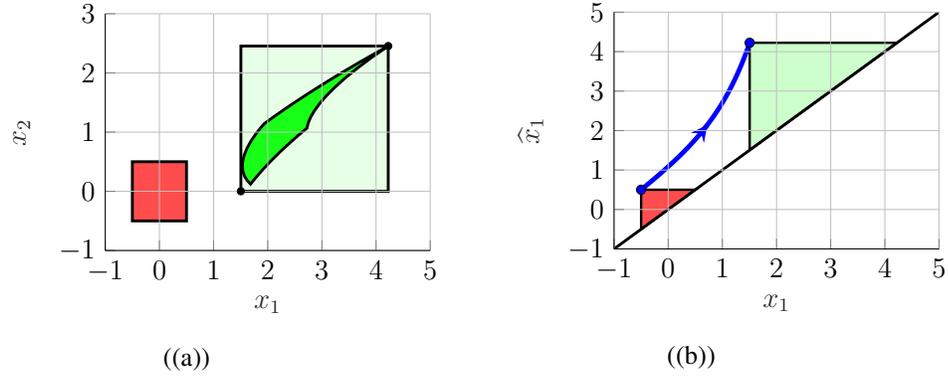}
        }
        \caption{ }
        \label{fig:ex1_2}
    \end{subfigure}
    \caption{ 
    Approximating reachable sets of \eqref{eq3} from the set of initial conditions $\X_0 = [-1/2, 1/2]\times[-1/2, 1/2]$. (a) $\X_0$ is shown in red. $R^F(1; \X_0)$ is shown in green.  The hyperrectangular over-approximation of $R^F(1; \X_0)$, which is computed from the embedding system \eqref{eq:embedding} as described in \eqref{eq:MM_reach}, is shown in light green. (b)  Visualization of the bounding procedure from \eqref{eq:MM_reach}. 
        The trajectory of \eqref{eq:embedding} that yields Figure \ref{fig:ex1_1} is shown in blue, where $\Phi^E$ is projected to the $x_1, \widehat{x}_1$ plane.  
        The southeast cones corresponding to $\X_0$ and the hyperrectangular over-approximation of $R^F(1; \X_0)$ are shown in red and green, respectively.
    }
    \label{fig56}
\end{figure*}

\begin{example}
Consider the system
\begin{equation}\label{eq3}
\begin{bmatrix}
\dot{x}_1 \\ \dot{x}_2
\end{bmatrix} = 
F(x) = 
\begin{bmatrix}
x_2^2 + 2 \\ x_1
\end{bmatrix}
\end{equation}
with state space $\X = \R^2$.  The system \eqref{eq3} is mixed monotone with respect to the \emph{decomposition function} $d: \X \times \X \rightarrow \R^2$ given by
\begin{equation}\label{eq4}
\begin{split}
     d_1(x, \widehat{x}) &= 
    \begin{cases}
    x_2^2 + 2
    & \text{if } x_2 \geq \max\{0, -\widehat{x}_2\}, \\
    \vspace{-.4cm} \\
    \widehat{x}_2^2 + 2
     & \text{if } \widehat{x}_2 \leq \min\{0, -x_2\}, \\ 
    \vspace{-.4cm} \\
    2
     & \text{if } x_2 \leq 0 \leq \widehat{x}_2,
    \end{cases} \\
    d_2(x, \widehat{x}) &= x_1.
\end{split}
\end{equation}
Reachable sets for \eqref{eq3} are now approximated from a single simulation of the embedding system \eqref{eq:embedding}, as described in \eqref{eq:MM_reach}. An example is shown in Figure \ref{fig56} where $R(1; \X_0)$ is approximated for $\X_0 = [-1/2, 1/2]^2$.
\end{example}

\bibliographystyle{IEEEtran}
\bibliography{references}


\section{Author Biographies}

\begin{biography}[{\includegraphics[width=1in,height=1.25in,clip,keepaspectratio]{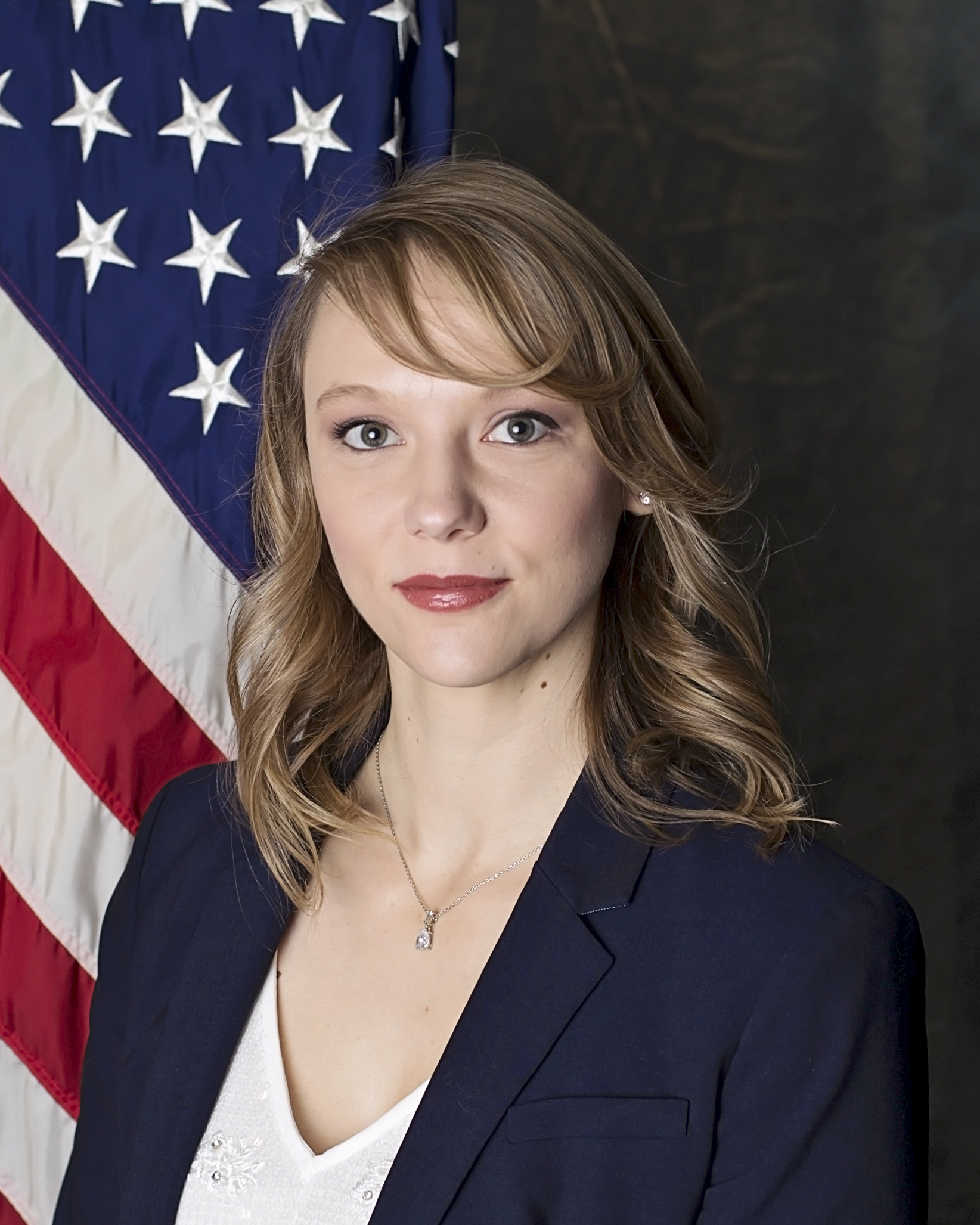}}]{Kerianne Hobbs} is the Safe Autonomy Lead at the Autonomy Capability Team (ACT3) at the Air Force Research Laboratory. There she investigates rigorous specification, analysis, and bounding techniques to enable certification of autonomous and learning controllers for aircraft and spacecraft applications. Her previous experience includes work in automatic collision avoidance at AFRL from 2011-2014, and Autonomy Verification and Validation research from 2012-2020. Kerianne has a BS in Aerospace Engineering from Embry-Riddle Aeronautical University, an MS in Astronautical Engineering from the Air Force Institute of Technology, and a Ph.D. in Aerospace Engineering from the Georgia Institute of Technology.
\end{biography}

\begin{biography}[{\includegraphics[width=1in,height=1.25in,clip,keepaspectratio]{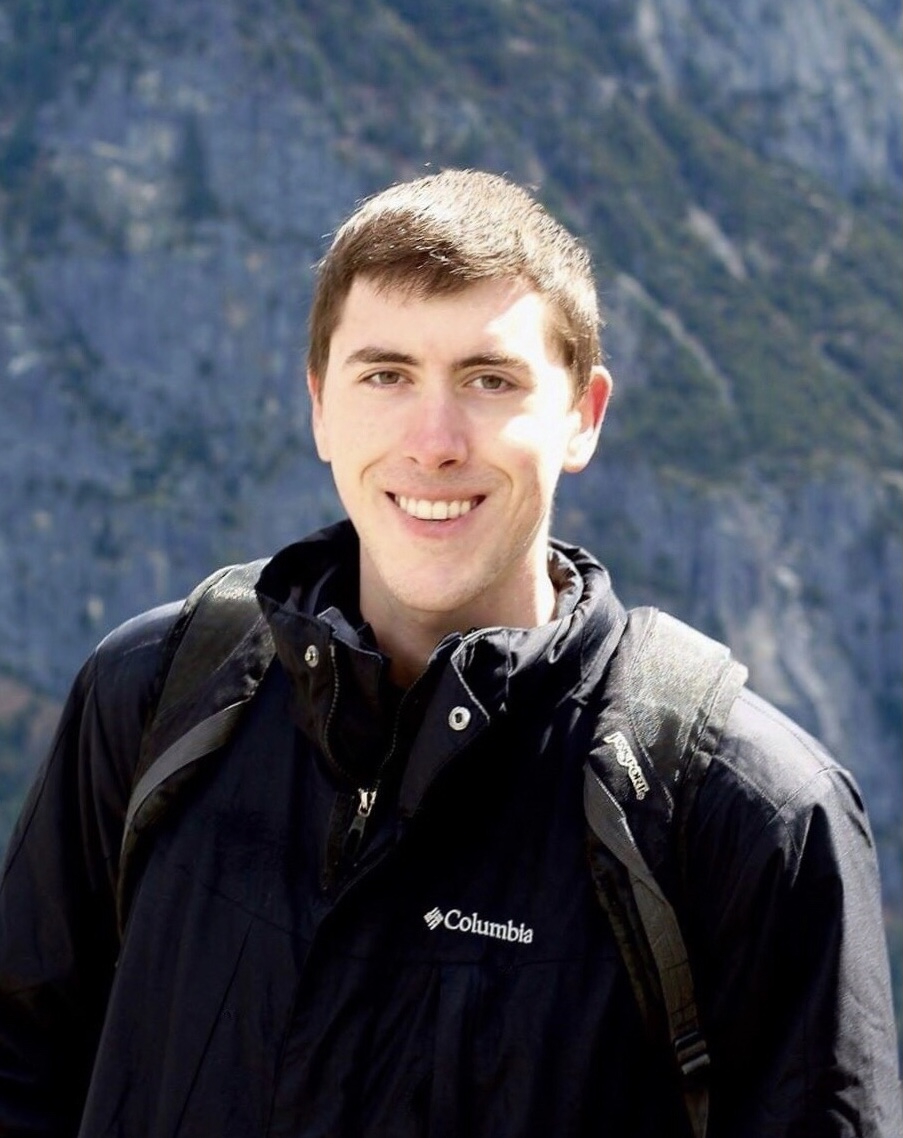}}]{Mark L. Mote}
 is a co-founder of Pytheia. He received his Ph.D. in Robotics from the Georgia Institute of Technology in 2021, and his B.S. and M.S. degrees in Aerospace Engineering from the Georgia Institute of Technology in 2015 and 2018. His current research interests are in safe autonomy, perception, optimization, trajectory planning, and run time assurance for safety-critical systems. 
\end{biography}

\begin{biography}[{\includegraphics[width=1in,height=1.25in,clip,keepaspectratio]{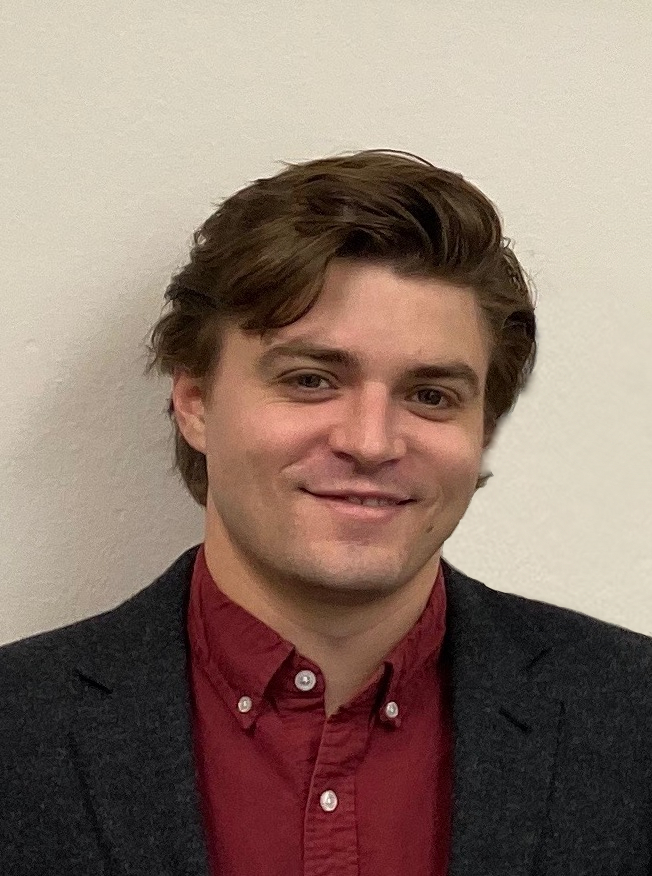}}]{Matthew C. L. Abate} (Student Member, IEEE) received a B.A. degree in engineering sciences and a B.E degree in mechanical engineering from Dartmouth College, Hanover, NH, USA, in 2017. He received a M.S. degree in Mechanical Engineering from Georgia Tech, Atlanta, GA, USA, in 2018, and a M.S. degree in Electrical and Computer Engineering from the same institution in 2020.  He is currently pursuing a Ph.D. degree in Robotics at Georgia Tech, supervised by Dr.'s Samuel Coogan and Eric Feron. His research interest is in the area of safety and verification for uncertain dynamical systems.
\end{biography}

\begin{biography}[{\includegraphics[width=1in,height=1.25in,clip,keepaspectratio]{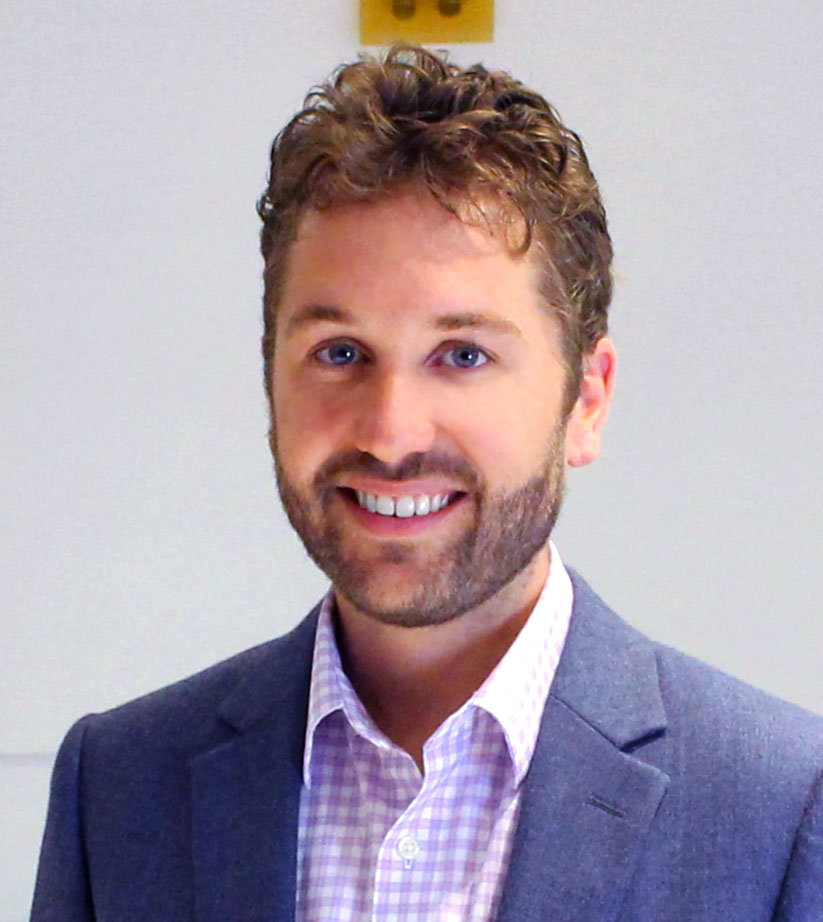}}]{Sam Coogan}
is assistant professor at Georgia Tech in the School of Electrical and Computer Engineering and the School of Civil and Environmental Engineering. Prior to joining Georgia Tech in 2017, he was an assistant professor in the Electrical Engineering Department at UCLA. He received the B.S. degree in Electrical Engineering from Georgia Tech and the M.S. and Ph.D. degrees in Electrical Engineering from the University of California, Berkeley. His research is in the area of
dynamical systems and autonomy and focuses on developing scalable tools for verification and control of networked, cyber-physical systems with an emphasis on transportation systems. He received the Donald P Eckman Award from the American Automatic Control Council in 2020, a Young Investigator Award from the Air Force Office of Scientific Research in 2019, a CAREER award from NSF in 2018, and the Outstanding Paper Award for the IEEE Transactions on Control of Network Systems in 2017.
\end{biography}

\begin{biography}[{\includegraphics[width=1in,height=1.25in,clip,keepaspectratio]{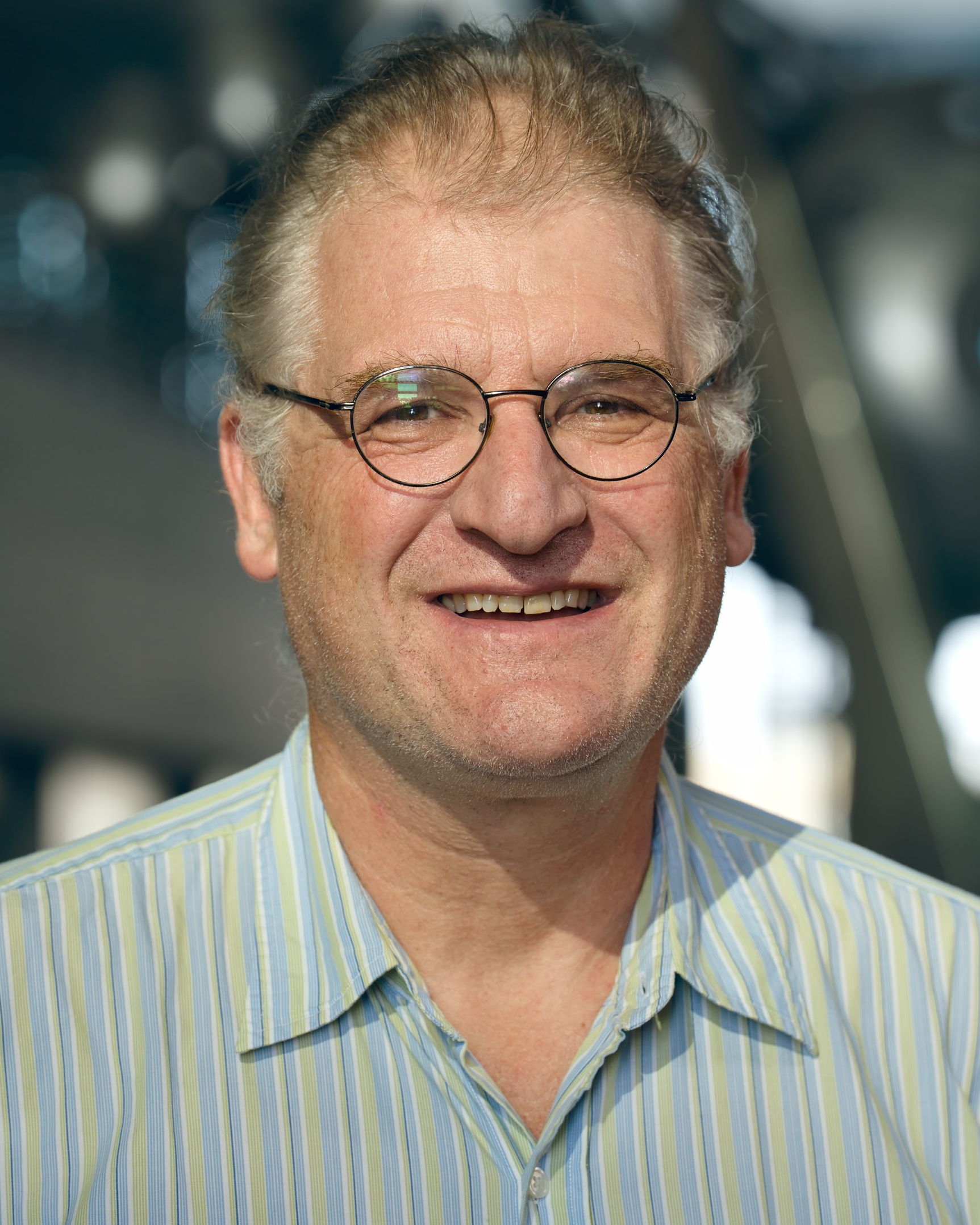}}]{Eric Feron}
is professor of Electrical, Computer, and Mechanical Engineering at King Abdullah University of Science and Technology (KAUST), Saudi Arabia. Eric Feron's interests are with aerospace systems, ranging from aircraft to space systems, with an emphasis on robotic autonomy and artificial intelligence. 
Prior to his KAUST appointment, Eric Feron was a professor of Aeronautics and Astronautics at the Massachusetts Institute of Technology (MIT) and a professor of Aerospace Engineering at the Georgia Institute of Technology (Georgia Tech), United States. Eric Feron also works on occasion with the Ecole Nationale de l'Aviation Civile (ENAC) and the Institut Supérieur de l'Aéronautique et de l'Espace (ISAE-Supaéro), France, where he was born.
\end{biography} 

\end{document}